\documentclass[%
	pdftex,%              PDFTex verwenden da wir ausschliesslich ein PDF erzeugen.
	a4paper,%             Wir verwenden A4 Papier.
	oneside,%             Ein-/Zweiseitiger Druck.
	12pt,%                Grosse Schrift, besser geeignet für A4.
	headsepline,%         Linie nach Kopfzeile.
]{scrartcl}
\pdfoutput=1
\newcommand{\dd}{\mathrm{d}}
\newcommand{\DD}{\mathrm{D}}

\newcommand{\R}{\mathbb{R}}

\newcommand{\fracpd}[2]{\frac{\partial #1}{\partial #2}} % partielles Differential
\newcommand{\fracppd}[2]{\frac{\partial^2 #1}{\partial #2^2}}  % zweite partielle
\newcommand{\fracd}[2]{\frac{d #1}{d #2}} %  Differential
\newcommand{\fracD}[2]{\frac{D #1}{D #2}} %  Differential mit großen D
\newcommand{\fracdd}[2]{\frac{d^2 #1}{d #2^2}}  % zweite partielle
\newcommand{\fracDD}[2]{\frac{D^2 #1}{D #2^2}}  % zweite partielle D
\newcommand{\const}{{\text{const}}}
\newcommand{\norm}[1]{\left|\left|#1\right|\right|}
\newcommand{\sose}{{\overset{!}{=}}}
\newcommand{\refb}[1]{(\ref{#1})}
\newcommand{\crs}[3]{\genfrac{\{}{\}}{0pt}{}{ #1}{ #2 #3}}
\newcommand{\g}{g_{\mu\nu}}
\newcommand{\bd}{\textbf{d}}
\newcommand{\bomega}{\boldsymbol{\omega}}

\renewcommand\theequation{\thesection.\arabic{equation}}  %Nummerierung der Gleichung in der Form:
\renewcommand{\baselinestretch}{1.5}\normalsize    %Zeilenabstand

\let\iint\relax				% amsmath und wasysym benutzen beide \iint. Das behebt die Fehlermeldungen

\usepackage{color}
\usepackage[latin1]{inputenc}
\usepackage[T1]{fontenc}
\newcommand{\changefont}[3]{
\fontfamily{#1} \fontseries{#2} \fontshape{#3} \selectfont}
\usepackage{makeidx}
\usepackage[ngerman]{babel}
\usepackage[babel,german=quotes]{csquotes}
\usepackage{array}
\usepackage{graphicx}
\usepackage{amsmath}
\usepackage{amssymb}
\usepackage{dsfont}
\usepackage{rotating}
\usepackage{fancyhdr}
\usepackage[savemem]{listings}
\usepackage{epsfig}
\usepackage{subfigure}
\usepackage{picins}
\usepackage{caption}
\usepackage{pstricks}
\usepackage[numbers,square]{natbib}
\usepackage{bibgerm} % deutsches Literaturverzeichnis
\usepackage{multirow}
\definecolor{darkblue}{rgb}{0,0,.5}
\definecolor{darkgreen}{rgb}{0,0.5,0}
\definecolor{darkred}{rgb}{0.5,0,0}
\usepackage[%	
	pdftitle={Masterarbeit},%                       							 Titel des PDF Dokuments.
	pdfauthor={Thomas Schönenbach},%             									 Autor des PDF Dokuments.
	pdfsubject={Pseudokomplexe ART angewandt auf die Kerr-Lösung},%               Thema des PDF Dokuments.
	pdfcreator={MiKTeX, LaTeX with hyperref and KOMA-Script},% 		 Erzeuger des PDF Dokuments.
	pdfpagemode=UseOutlines,%                               		   Inhaltsverzeichnis anzeigen beim Öffnen%
	pdfdisplaydoctitle=true,%                               		   Dokumenttitel statt Dateiname anzeigen.
	pdflang=de,%                                            		   Sprache des Dokuments.
  hyperfootnotes=false,%																				 Fußnoten kriegen keine Hyperlinks
	plainpages=false,%																						 Soll Warnings vermindern
	pdfpagelabels,
	colorlinks=true,
	breaklinks=true,
	linkcolor=darkblue,
	menucolor=darkblue,
	urlcolor=darkred,
	citecolor=darkgreen,
  citebordercolor=darkgreen
]{hyperref}

\graphicspath{{Grafiken/}}
\setcapindent{1em}
\makeindex

\begin{document}
\changefont{cmr}{m}{n}

\clearpage
%
% Titelseite
%
\begin{titlepage}
\setlength{\oddsidemargin}{0cm} %nicht links / rechts ausgerichtet
\pagenumbering{alph}
\vspace*{2cm}
\begin{center}
	\LARGE
	\textbf{Die Kerr-Metrik in pseudokomplexer Allgemeiner Relativitätstheorie} \\
	\vspace{1cm}
	\large 
%	\textbf{Untertitel der Masterarbeit} \\
	\vspace{2cm}
	\normalsize
	\textbf{Masterarbeit von Thomas Schönenbach}  \\ \vspace{2cm} \textbf{März 2011} \\ \vspace{2cm}
\end{center}
\begin{center}
Betreuer:\\
 Prof. Dr. Dr. h.c. mult. Walter Greiner \\
Prof. Dr. Peter O. Hess
\end{center}

\end{titlepage}

\clearpage 											%Blankoseite um richtige Nummerierung zu 
\thispagestyle{empty}					%garantieren
\quad 
\clearpage

\thispagestyle{empty}
\pagenumbering{roman}
\tableofcontents
\clearpage
\pagenumbering{arabic}
\setcounter{page}{1}

\clearpage

\clearpage
%
% Einleitung
%
%evtl ohne Nummerierung
\thispagestyle{empty}
\pagenumbering{roman}

\section*{Vorwort}

Ziel dieser Arbeit ist einerseits eine Aufbereitung und Überarbeitung des Papers über die pseudokomplexe Allgemeine Relativitätstheorie \cite{Hess:2008wd} 
(siehe Kapitel \ref{sec:pseudoart}), welche in gemeinsamer Arbeit mit Gunther Caspar und in ständiger Korrespondenz mit Prof. Dr. Dr. h.c. mult. Walter Greiner und Prof. Dr. Peter O. Hess stattfand, und andererseits die Anwendung des pseudokomplexen Formalismus um eine Lösung der Einsteingleichung zu finden, die der Kerr-Lösung 
entspricht (siehe Kapitel \ref{sec:ergebnisse}). In den ersten beiden Kapiteln wird eine Einführung in die Gebiete der Allgemeinen Relativitätstheorie und der pseudokomplexen Zahlen gegeben. Sie beschränken sich dabei auf bereits bekanntes Wissen. \\
Im Rahmen der Arbeit wird mit den so genannten "`Natürlichen Einheiten"' ($\hbar = 1$ und $c=1$) gerechnet. Da wir uns auf dem Gebiet der ART befinden, benutzen wir
auch die von Einstein eingeführte Summenkonvention, wobei über doppelt auftretende Indizes in einem Ausdruck impliziert summiert wird, z.B.:
\begin{equation*}
 \dd s^2 = \sum_{\mu,\nu=0}^4 g_{\mu\nu} \dd x^\mu \dd x^\nu =  g_{\mu\nu} \dd x^\mu \dd x^\nu \qquad .
\end{equation*}
 Weiter benutzen wir die Minkowski-Metrik mit Signatur $(+,-,-,-)$, also
 \begin{equation*}
	\eta_{\mu\nu} = \begin{pmatrix} 1  & 0 & 0  & 0 \\ 0 & -1 & 0 & 0 \\ 0 & 0 & -1 & 0 \\ 0 & 0 & 0 & -1
                          \end{pmatrix} \qquad .
\end{equation*}

% Motivation
% Nomenklatur erläutern (Ableitungen Summenkonvention Signatur usw)
% gemeinsame Erarbeitung bestimmter Themen
% 
% 
% 

%%%%%%%%%%%%%%%%%%%%%%%%%%%%%%%%%%%%%%%%%%%%%%%%%%%%%%%%%%%%%%%%%%%%%%%%%%%%%%%%%%%%%%%%%%%%%%%%%%%%%%%%%%%%%%%%%%%%%%%%%%%%%%%%%%%%%%%%%%%%%%%
%\setcounter{footnote}{0}
\setcounter{equation}{0}
\clearpage
\pagenumbering{arabic}
\setcounter{page}{1}
%
%################    EOF    #######################
\clearpage

\clearpage
%
% Kapitel über die Standard-ART
%

\section{Allgemeine Relativitätstheorie}\label{sec:art}
Dieses Kapitels gibt eine kurze Einführung in die Allgemeine Relativitätstheorie (\textbf{ART}). Es wird sich dabei an dem Buch von Adler, Bazin und Schiffer \cite{adler:1975} orientieren.\\%, einige der neueren Tests der ART sind in \cite{fliessbach:2006} beschrieben. \\
Einstein selbst sah drei Grundprinzipien als Pfeiler der ART \cite{Einstein:1918}:
\begin{itemize}
	\item \textbf{Relativitäts- oder Kovarianzprinzip}: Physikalische Naturgesetze lassen sich durch kovariante Gleichungen beschreiben. Diese Gleichungen gelten forminvariant in beliebig bewegten Bezugssystemen.
	\item \textbf{Äquivalenzprinzip}: Träge und Schwere Masse sind äquivalent. Ein Beobachter in einem geschlossenen Kasten kann nicht zwischen gleichmäßiger Beschleunigung und gravitativer Anziehung unterscheiden.
	\item \textbf{Machsches Prinzip}: Das Gravitationsfeld (die Krümmung des Raumes) wird vollständig durch die Massenansammlungen im Raum festgelegt. Durch die Äquivalenz von Masse und Energie legt also der Energie-Impuls-Tensor $T_{\mu\nu}$ die Krümmung des Raumes - beschrieben durch den metrischen Tensor $g_{\mu\nu}$ - fest.
\end{itemize}

Neben diesen Prinzipien brauchen wir eine mathematische Basis für die ART. Diese Basis bilden zum einen Tensoren und die zugehörigen Rechenvorschriften und zum anderen eine semi-riemannsche Mannigfaltigkeit, die Raumzeit, in der diese Tensoren operieren. Eine semi-riemannsche Mannigfaltigkeit besitzt eine Metrik, die nicht positiv definit ist\footnote{Wäre die Metrik positiv definit, würde für alle Vektoren $u_\mu$ gelten, dass $u^\mu u_\mu = g^{\mu\mu}u_\mu u_\mu > 0$ ist.}. Wir benötigen hier eine nicht definite Metrik, um zwischen raumartigen, zeitartigen und Null-Vektoren unterscheiden zu können. Die Raumzeit lässt sich lokal durch den $\R^4$ mit der Minkowski-Metrik beschreiben.

%evtl noch einführen was riemann mannigfaltigkeiten sind und dass art in solchen stattfindet

\subsection{Mathematische Grundlagen der ART}
\subsubsection*{Tensoren}
In der Physik werden Tensoren durch ihr Transformationsverhalten zwischen verschiedenen Koordinatensystemen charakterisiert. Wir betrachten zwei Koordinatensysteme, einmal mit gestrichenen $\bar{x}^\mu$ 
und einmal mit ungestrichenen Koordinaten $x^\nu$, zwischen denen es stetige Koordinatentransformationen gibt
\begin{eqnarray}
 \bar{x}^\mu &=& f^\mu(x^0,x^1,x^2,x^3) \notag\\
 x^\nu &=& h^\nu(\bar{x}^0,\bar{x}^1,\bar{x}^2,\bar{x}^3)  \qquad . \notag
\end{eqnarray}
Eine physikalische Größe, die in beiden Systemen unverändert bleibt, nennen wir einen Tensor nullter Stufe, beziehungsweise \textit{Skalar}. Ein Beispiel für einen Skalar ist das Quadrat des Linienelements $\dd s$
\begin{equation}
 \dd s^2 =  g_{\mu\nu} \dd x^\mu \dd x^\nu  \qquad .
\label{eq:laengenelement}
\end{equation}
Es bleibt in allen Koordinatensystemen gleich\footnote{Hier wurde bereits die Einsteinsche Summenkonvention benutzt, bei der über doppelt auftretende Indizes in einem Ausdruck summiert wird.}.  \\
Das nächst kompliziertere Objekt - ein Tensor erster Stufe - wird auch als \textit{Vektor} bezeichnet. Hier unterscheiden wir zwischen ko- und kontravarianten Vektoren, die sich einerseits durch die unterschiedliche Schreibweise
und andererseits durch ihr unterschiedliches Transformationsverhalten auszeichnen. Ein kontravarianter Vektor $\xi^\mu$ ist ein vier-komponentiges Objekt, welches sich wie folgt transformiert
\begin{equation*}
 \bar{\xi}^\mu = \fracpd{\bar{x}^\mu}{x^\nu} \xi^\nu  \qquad .
\end{equation*}
Entsprechend dazu transformiert sich ein kovarianter Vektor $\zeta_\mu$
\begin{equation*}
 \bar{\zeta}_\mu = \fracpd{x^\nu}{\bar{x}^\mu} \zeta_\nu  \qquad .
\end{equation*}
Tensoren höherer Stufe können wie Vektoren ko- und kontravariante Indizes tragen. Ihr Transformationsverhalten hängt dann davon ab, welche Indizes ko- und welche kontravariant sind. Ein \textit{Tensor} $n$-ter 
Stufe $T^{\alpha_1 \cdots \alpha_a}_{\beta_1 \cdots \beta_b}$ hat $n = a +b$ Indizes, von denen $a$ kontra- und $b$ kovariant sind. Er transformiert sich wie folgt
\begin{equation*}
 \bar{T}^{k_1 \cdots k_a}_{l_1 \cdots l_b} = \fracpd{\bar{x}^{k_1}}{x^{\alpha_1}} \cdots \fracpd{\bar{x}^{k_a}}{x^{\alpha_a}} \fracpd{x^{\beta_1}}{\bar{x}^{l_1}} \cdots \fracpd{x^{\beta_b}}{\bar{x}^{l_b}} T^{\alpha_1 \cdots \alpha_a}_{\beta_1 \cdots \beta_b}  \qquad .
\end{equation*}
Eine Möglichkeit aus einem Tensor einen Tensor geringerer Stufe zu bilden ist die \textit{Kontraktion}. Dabei setzt man zwei Indizes, je einen ko- und einen kontravarianten, gleich und summiert
gemäß Einsteinscher Summenkonvention über diesen Index. Daraus resultiert ein Tensor dessen Stufe um 2 im Vergleich zum Ausgangstensor reduziert wurde. Ein Beispiel hierfür ist
die Spur einer Matrix ${A^\mu}_\nu$
\begin{equation*}
 \text{sp}({A^\mu}_\nu) = {A^\mu}_\mu = \sum_\mu  {A^\mu}_\mu \qquad .
\end{equation*}
%%%%%%%%%%%%%%%%%%%%%%%%%%%%%%%%%%%%%%%%%%vvv  
Bisher haben wir Tensoren zwar in verschiedenen Koordinaten dargestellt, aber immer an ein und dem selben Punkt im Raum betrachtet. In der Physik ist es von Interesse zu wissen, wie sich ein Tensor an 
verschiedenen Punkten im Raum verhält. Ein wichtiges Beispiel dafür ist die Konstanz eines Vektors. Aus der klassischen Mechanik sind wir es gewohnt einen Vektor als konstant zu bezeichnen, wenn er an jedem
Raumpunkt die selben Komponenten enthält. Dies gilt allerdings nur in euklidischen Räumen, im Allgemeinen ist die Länge eines Vektors $l^2 = \xi_\mu\xi^\mu = g_{\mu\nu} \xi^\mu \xi^\nu$. Sobald der metrische Tensor $g_{\mu\nu}$ nicht mehr an 
jedem Punkt im Raum gleich ist, lässt sich der klassische Begriff der Konstanz nicht mehr anwenden. Ein "`konstanter"' Vektor hätte dann an verschiedenen Punkten verschiedene Längen. \\
Eine andere Möglichkeit einzusehen, dass der gewöhnliche Begriff für die Konstanz eines Vektors unzureichend ist, besteht darin einen konstanten Vektor $\xi^\mu$ in einem anderen Koordinatensystem zu betrachten
\begin{equation*}
 \bar{\xi}^\mu = \fracpd{\bar{x}^\mu}{x^\nu} \xi^\nu \qquad .
\end{equation*}
Nur für lineare Koordinatentransformationen bleibt der Vektor konstant. Es gilt nun also eine Möglichkeit zu finden, den Begriff der Konstanz zu erweitern. Dazu betrachten wir die Änderung von $\bar{\xi}^\mu$
entlang einer parametrisierten Kurve, die wir mit einem Parameter $p$ beschreiben können
\begin{equation*}
 \fracd{\bar{\xi}^\mu}{p} = \frac{\partial^2 \bar{x}^\mu}{\partial x^\nu \partial x^\tau} \fracd{x^\tau}{p} \xi^\nu = \frac{\partial^2 \bar{x}^\mu}{\partial x^\nu \partial x^\tau} \fracd{x^\tau}{\bar{x}^\sigma} \fracd{\bar{x}^\sigma}{p} \fracpd{x^\nu}{\bar{x}^\gamma} \bar{\xi}^\gamma  \qquad .
\end{equation*}
An dieser Stelle ist es praktisch die Abkürzung
\begin{equation*}
 \bar{\Gamma}^\mu_{\sigma\gamma} = \frac{\partial^2 \bar{x}^\mu}{\partial x^\nu \partial x^\tau} \fracd{x^\tau}{\bar{x}^\sigma}  \fracpd{x^\nu}{\bar{x}^\gamma}
\end{equation*}
einzuführen\footnote{Hierbei handelt es sich noch nicht um die Christoffelsymbole zweiter Art, welche in der Literatur häufig mit $\Gamma^\mu_{\sigma\gamma}$ bezeichnet werden, sondern um affine Verbindungen.}. Damit haben wir dann
\begin{equation}
 \fracd{\bar{\xi}^\mu}{p} =  \bar{\Gamma}^\mu_{\sigma\gamma} \fracd{\bar{x}^\sigma}{p}  \bar{\xi}^\gamma  \qquad .
 \label{eq:translaw}
\end{equation}
Dieser Ausdruck wird in differentieller Form als Verschiebungsgesetz ("`transplantation law"') für Vektoren bezeichnet. \\
Das gerade aufgestellte Verschiebungsgesetz gilt auch in Räumen, die keinen metrischen Tensor $g_{\mu\nu}$ haben. Für uns sind aber gerade die metrischen Räume interessant und hier lässt sich ein wichtiger
Zusammenhang zwischen $g_{\mu\nu}$ und $\Gamma^\mu_{\nu\sigma}$ aufstellen. Dazu betrachten wir die Forderung, dass das Skalarprodukt zweier Vektoren $g_{\mu\nu} \xi^\mu \eta^\nu$ konstant bleibt, wenn wir sie entlang einer mit
$s$ parametrisierten Kurve verschieben
\begin{align}
 &\fracd{}{s} (g_{\mu\nu} \xi^\mu \eta^\nu) \sose 0 \notag \\
 \Leftrightarrow ~& \fracpd{g_{\mu\nu}}{x^\sigma} \fracd{x^\sigma}{s} \xi^\mu\eta^\nu + g_{\mu\nu} \fracd{\xi^\mu}{s} \eta^\nu + g_{\mu\nu} \xi^\mu \fracd{\eta^\nu}{s} = 0  \notag \\
 \Leftrightarrow ~& \fracpd{g_{\mu\nu}}{x^\sigma} + g_{\gamma\nu} \Gamma^\gamma_{\sigma\mu} + g_{\mu\gamma} \Gamma^\gamma_{\sigma\nu} = 0  \qquad .
 \label{eq:metgamma} 
\end{align}
In der letzten Zeile wurde noch ausgenutzt, dass man doppelt auftretende Indizes beliebig benennen kann und dass $\xi$ und $\eta$ beliebige Vektoren sind. Nun können wir uns die Symmetrie der ${\Gamma^\gamma}_{\nu\mu}$
und des metrischen Tensors zu Nutze machen und die Indizes in Gleichung  \refb{eq:metgamma} zyklisch permutieren. Aus den drei daraus resultierenden Gleichungen ergibt sich dann der Zusammenhang \cite{adler:1975}
\begin{equation*}
 {\Gamma^\gamma}_{\nu\mu} = -\frac{1}{2} g^{\sigma\gamma} \left( \fracpd{g_{\nu\sigma}}{x^\mu} + \fracpd{g_{\sigma\mu}}{x^\nu} - \fracpd{g_{\mu\nu}}{x^\sigma}\right) = -\frac{1}{2} g^{\sigma\gamma} \left[ \nu\mu,\sigma \right]
\end{equation*}
Die Größe $-\frac{1}{2} g^{\sigma\gamma}$ wird Christoffelsymbol erster Art genannt. Von großer Bedeutung für unsere weiteren Rechnungen sind die \textit{Christoffelsymbole zweiter Art}
\begin{equation}
 \crs{\sigma}{\mu}{\nu} = - {\Gamma^\sigma}_{\mu\nu} \qquad .
\label{eq:christoffel2}
\end{equation}
Das oben eingeführte Verschiebungsgesetz für Vektoren beschreibt in metrischen Räumen eine verallgemeinerte \textit{Parallelverschiebung}, also die Änderung des Vektors $\xi$, wenn wir ihn 
parallel zu sich selbst verschieben
\begin{equation}
 \dd \xi^\mu = - \crs{\mu}{\nu}{\sigma} \dd x^\nu  \xi^\sigma  \qquad .
 \label{eq:paraversch}
\end{equation}
%%%%%%%%%%%%%%%%%%%%%%%%%%%%%%%%%%%%%%%%%%%%%%%%%%%%%%%%%%%%%%%%%%%%%%%%%%%%%%%%%%%%%%%%%%%%%%%%%%%%%%%%%%%%
\subsubsection*{Kovariante Ableitung}
Wir können jetzt genauer untersuchen, wie sich Tensoren an verschiedenen Punkten im Raum verhalten. Dazu schauen wir uns zunächst die Differenz zwischen einem kontravarianten Vektor
$\xi^\mu(x^\nu+\dd x^\nu)$ an der Stelle $x^\nu + \dd x^\nu$ und dem dazu parallel verschobenen Vektor $\xi^{'\mu}(x^\nu+\dd x^\nu)$ an. Der erste Term lässt sich mit Hilfe einer Taylorentwicklung umschreiben
\begin{equation*}
  \xi^\mu(x^\nu+\dd x^\nu) = \xi^\mu(x^\nu) +  \fracpd{\xi^\mu}{x^\nu} \dd x^\nu + \mathcal{O}\left((\dd x^\nu)^2\right) \qquad .
\end{equation*}
Für den zweiten Teil verwenden wir das Parallelverschiebungsgesetz \refb{eq:paraversch}
\begin{equation*}
\xi^{'\mu}(x^\nu+\dd x^\nu) = \xi^\mu(x^\nu) - \crs{\mu}{\gamma}{\sigma} \xi^\gamma \dd x^\sigma \qquad .
\end{equation*}
Damit ist die Differenz also
\begin{equation*}
  \xi^\mu(x^\nu+\dd x^\nu) - \xi^{'\mu}(x^\nu +\dd x^\nu) = \left[ \fracpd{\xi^\mu}{x^\sigma} +  \crs{\mu}{\sigma}{\gamma} \xi^\gamma\right] \dd x^\sigma + \mathcal{O}\left((\dd x^\nu)^2\right) \qquad .
\end{equation*}
Nun vernachlässigen wir die Terme höherer Ordnung und stellen fest, dass auf der linken Seite die Differenz zweier Vektoren steht. Also muss die rechte Seite der Gleichung auch ein Vektor sein. 
Nach dem Quotiententheorem\footnote{Angenommen das Produkt eines Tensors mit einer beliebig indizierten Matrix $T^{k_1 \cdots k_a}_{l_1 \cdots l_b}$ ist wieder ein Tensor, so ist $T^{k_1 \cdots k_a}_{l_1 \cdots l_b}$ auch ein Tensor \cite{adler:1975}.} 
handelt es sich bei $\left[ \fracpd{\xi^\mu}{x^\sigma} +  \crs{\mu}{\sigma}{\gamma} \xi^\gamma\right]$ somit um einen Tensor. Diesen nennen wir die \textit{Kovariante Ableitung} eines kontravarianten Vektors. 
Die Kovariante Ableitung erhält in der Regel eine eigene Abkürzung. Wir werden
\begin{equation}
 {\xi^\mu}_{||\sigma} = {\xi^\mu}_{|\sigma} + \crs{\mu}{\sigma}{\gamma} \xi^\gamma
\label{eq:kovdiffkontra}
\end{equation}
verwenden, wobei die normale Ableitung als $\fracpd{\xi}{x^\sigma} = \xi_{|\sigma}$ notiert ist (In der Literatur wird die Kovariante Ableitung auch häufig mit einem Semikolon geschrieben ${\xi^\mu}_{;k}$). Die Kovariante Ableitung eines kovarianten Vektors ist (hier ohne Ableitung)
\begin{equation}
  \xi_{\mu||\sigma} = \xi_{\mu|\sigma} - \crs{\gamma}{\mu}{\sigma} \xi_\gamma \qquad .
\label{eq:kovdiffko}
\end{equation}
Eine wichtige Eigenschaft der Kovarianten Ableitung ist, dass sie im Allgemeinen nicht wie die normale Ableitung vertauscht
\begin{equation*}
 {\xi^\mu}_{||\sigma||\nu} \neq {\xi^\mu}_{||\nu||\sigma} \qquad .
\end{equation*}

%%%%%%%%%%%%%%%%%%%%%%%%%%%%%%%%%%%%%%%%%%%%%%%%%%%%%%%%%%%%%%%%%%%%%%%%%
\subsubsection*{Geodäten}
In der Newtonschen Mechanik haben wir gelernt, dass sich ein kräftefreies Teilchen entlang einer Geraden bewegt \cite{Greiner:2003}. In Euklidischen Räumen können wir eine solche Gerade dadurch charakterisieren, dass wir an ihr einen Tangentialvektor verschieben können und dieser dabei immer parallel zu sich selbst bleibt. In gekrümmten Räumen können wir dazu die verallgemeinerte Parallelverschiebung \refb{eq:paraversch} benutzen. Ein möglicher Tangentialvektor an eine beliebige mit $s$ parametrisierte Kurve ist $\fracd{x^\mu}{s}$. Damit haben wir
\begin{align}
 &\fracd{}{s} \fracd{x^\mu}{s} = - \crs{\mu}{\sigma}{\gamma} \fracd{x^\sigma}{s}  \fracd{x^\gamma}{s} \notag \\
\Leftrightarrow~ & \fracdd{x^\mu}{s} +\crs{\mu}{\sigma}{\gamma} \fracd{x^\sigma}{s}  \fracd{x^\gamma}{s} = 0 \qquad .
\label{eq:geodaet}
\end{align}
Eine andere Art diese Geodätengleichung abzuleiten, ist die Verwendung des Variationsprinzips. So wie eine Gerade die kürzeste Strecke zwischen zwei Punkten im Euklidischen Raum beschreibt, ist eine Geodäte die kürzeste (mathematisch gesehen extremale) Verbindung zwischen zwei Punkten im gekrümmten Raum. Die Bogenlänge dieser Verbindung können wir mit 
\begin{equation*}
 s = \int \dd s = \int \fracd{s}{s}   \dd s = \int  \sqrt{g_{\mu\sigma} \fracd{x^\mu}{s} \fracd{x^\sigma}{s}} \dd s
\end{equation*}
berechnen. Sie wird extremal, wenn ihre Variation $\delta s$ verschwindet
\begin{equation}
 \delta \int \sqrt{g_{\mu\sigma} \fracd{x^\mu}{s} \fracd{x^\sigma}{s} } \dd s = 0 \qquad .
\label{eq:variationspr}
\end{equation}
Die daraus resultierenden Euler-Lagrange-Gleichungen sind
\begin{equation*}
 \fracd{}{s} \left( \fracpd{}{\dot{x}^\gamma} \sqrt{g_{\mu\sigma} \dot{x}^\mu \dot{x}^\sigma} \right) = \fracpd{}{x^\gamma} \sqrt{g_{\mu\sigma} \dot{x}^\mu \dot{x}^\sigma } \qquad .
\end{equation*}
Hier wurde noch die Abkürzung $ \dot{x}^\mu = \fracd{x^\mu}{s}$ eingeführt. Unter Berücksichtigung, dass $ \sqrt{g_{\mu\sigma} \dot{x}^\mu \dot{x}^\sigma }$ entlang der Kurve konstant ist, wird diese Gleichung zu
\begin{equation*}
 \fracd{}{s} \left( g_{\gamma\sigma} \fracd{x^\sigma}{s} \right) = \frac{1}{2} \fracpd{g_{\mu\sigma}}{x^\gamma} \fracd{x^\mu}{s} \fracd{x^\sigma}{s} \qquad .
\end{equation*}
Ausgeschrieben und unter Ausnutzung der Symmetrie von $g_{\mu\sigma}$ wird daraus
\begin{equation*}
 g_{\gamma\sigma} \fracdd{x^\sigma}{s} + \frac{1}{2}  \left( \fracpd{g_{\gamma\sigma}}{x^\mu} + \fracpd{g_{\mu\gamma}}{x^\sigma} -  \fracpd{g_{\mu\sigma}}{x^\gamma} \right) \fracd{x^\mu}{s} \fracd{x^\sigma}{s} = 0  \qquad .
\end{equation*}
Multipliziert man nun noch mit dem inversen metrischen Tensor $g^{\gamma\sigma}$ und benutzt die Definition für die Christoffelsymbole, so erhält man wieder die Geodätengleichung \refb{eq:geodaet} \cite{adler:1975}.

%Im Folgenden werden wir auf die einzelnen Prinzipien eingehen und den genauen Zusammenhang zwischen $g_{\mu\nu}$ und $T_{\mu\nu}$ in Form der Einstein-Gleichungen darstellen.
%%%%%%%%%%%%%%%%%%%%%%%%%%%%%%%%%%%%%%%%%%%%%%%%%%%%%%%%%%%%%%%%%%%%%%%%%%%%%%%%%%%%%%%%%%%%%%%%%%%%%%%%%%%%
\subsection{Die Einsteingleichungen}
Die Einsteingleichungen bilden den Kern der Allgemeinen Relativitätstheorie. Es gibt einige sinnvolle Forderungen, die man an die Gleichungen stellen kann und die dabei helfen sie aufzustellen. Adler, Bazin und Schiffer \cite{adler:1975} stützen sich auf die vier Forderungen
\begin{enumerate}
 \item Die Gleichungen müssen in Tensorform vorliegen. Dies entspricht dem Relativitätsprinzip, was wir am Anfang dieses Kapitels bereits erwähnt haben.
 \item Die Komponenten des metrischen Tensors sollen in zweiter Ordnung in die Feldgleichungen eingehen, in Analogie zur klassischen Laplace-Gleichung für das Gravitationspotential.
 \item Im Grenzfall eines komplett materiefreien Raumes müssen die Gleichungen die Lorentzmetrik als globale Lösung zulassen. 
 \item Um die Eindeutigkeit der Lösung zu garantieren, müssen die Gleichungen linear in den zweiten Ableitungen der Metrik sein.
\end{enumerate}
Hinzu kommt noch die in der Physik gängige Maxime, Gleichungen so einfach wie möglich zu halten.\\

Zunächst betrachten wir den Fall, dass keinerlei Materie im Raum vorliegt. Der dritte Punkt sagt uns dann, dass wir diesen Raum durch die Lorentzmetrik beschreiben können. 
In dieser Metrik geht die kovariante Ableitung in die gewöhnliche Ableitung über, denn die Metrik bleibt überall konstant und ihre Ableitungen und damit die Christoffelsymbole verschwinden.
Für einen beliebigen Vektor $\xi$ gilt dann
\begin{equation*}
 {\xi^\mu}_{||\nu} = {\xi^\mu}_{|\nu} \qquad .
\end{equation*}
Damit ist aber auch klar, dass kovariante Ableitungen in diesem Fall vertauschen
\begin{equation*}
  {\xi^\mu}_{||\nu||\eta} = {\xi^\mu}_{|\nu|\eta} = {\xi^\mu}_{|\eta|\nu} = {\xi^\mu}_{||\eta||\nu} \qquad .
\end{equation*}
 Bei dieser Gleichung handelt es sich nun um eine Tensorgleichung und sie gilt somit in allen Koordinatensystemen, solange man sich im masselosen Raum befindet.
Durch ein wenig Umformarbeit und Umbenennung von Summationsindizes lässt sich dieser Ausdruck in Abhängigkeit der Christoffelsymbole und ihrer Ableitungen darstellen
\begin{align}
    {\xi^\mu}_{||\nu||\eta} - {\xi^\mu}_{||\eta||\nu} &= \left[ \crs{\mu}{\nu}{\sigma}_{|\eta} - \crs{\mu}{\sigma}{\eta}_{|\nu} + \crs{\mu}{\tau}{\eta} \crs{\tau}{\nu}{\sigma}  -\crs{\mu}{\tau}{\nu} \crs{\tau}{\eta}{\sigma} \right] \xi^\sigma \notag \\
 &= {\mathcal{R}^\mu}_{\sigma\nu\eta} \xi^\sigma \qquad .
\label{eq:riemanntensor}
\end{align}
Bei dem hier eingeführten Tensor ${\mathcal{R}^\mu}_{\sigma\nu\eta}$ handelt es sich um den \textit{Riemannschen Krümmungstensor}. Er gibt ein Maß für die Krümmung des Raumes an und verschwindet im masselosen
Raum. Die Gleichungen
\begin{equation*}
 {\mathcal{R}^\mu}_{\sigma\nu\eta}  = 0
\end{equation*}
erfüllt schon alle vier Forderungen, die wir an die Einsteingleichungen gestellt haben. Sie gelten allerdings nur für den freien Raum und lassen sich noch durch Kontraktion zweier Indizes vereinfachen
\begin{equation}
 {\mathcal{R}^\mu}_{\sigma\mu\eta} = \mathcal{R}_{\sigma\eta} \qquad .
\label{eq:riccitensor}
\end{equation}
 Der daraus resultierende Tensor $\mathcal{R}_{\sigma\eta}$ wird \textit{Ricci-Tensor} genannt. Die beiden Indizes, über die summiert wird, ergeben sich aus Symmetrieeigenschaften des Krümmungstensors, auf die wir hier nicht weiter eingehen. Für den freien Raum sind die Einsteingleichungen dann also
\begin{equation}
 \mathcal{R}_{\sigma\eta} = 0 \qquad .
\label{eq:einsteinfrei}
\end{equation}
Sie lassen sich für schwache Felder auf die klassische Laplacegleichung für das Gravitationspotential \cite{adler:1975}
\begin{equation*}
 \sum_{i=1}^3 \phi_{|i|i} = 0
\end{equation*}
reduzieren. \\
Jetzt wollen wir die Einsteingleichungen aber auch für den nichtleeren Raum aufstellen können. Im Klassischen wird aus der Laplacegleichung die Poissongleichung
\begin{equation*}
  \sum_{i=1}^3 \phi_{|i|i} = 4\pi \kappa \rho \qquad ,
\end{equation*}
 wobei hier $\kappa$ die Newtonsche Gravitationskonstante und  $\rho$ die Materiedichte ist. Erweitert auf den Tensorformalismus suchen wir nun eine Tensor-Gleichung, die die Krümmung des Raumes einerseits mit der Massen- und durch $E= m c^2$ auch Energieverteilung andererseits in Verbindung bringt. Für den masselosen Raum muss sie sich auf \refb{eq:einsteinfrei} reduzieren und für schwache Gravitation die klassische Poissongleichung ergeben. Als Kandidat für die Beschreibung der Energie bietet sich der Energie-Impuls-Tensor $T_{\mu\nu}$ an, der (bis auf gravitative Anteile) die gesamte Energieverteilung beschreibt. Eine einfache Proportionalität zwischen dem Ricci- und dem Energie-Impuls-Tensor kommt allerdings nicht für die Einsteingleichungen in Frage, da der Energie-Impuls-Tensor im Gegensatz zum Ricci-Tensor divergenzfrei ist. Hier lässt sich der \textit{Einstein-Tensor} $G_{\mu\nu} = \mathcal{R}_{\mu\nu} - \frac{1}{2} g_{\mu\nu} {\mathcal{R}^\alpha}_\alpha$ benutzen, sodass wir die Einsteingleichungen als 
\begin{equation}
 \mathcal{R}_{\mu\nu} - \frac{1}{2} g_{\mu\nu} {\mathcal{R}^\alpha}_\alpha = C T_{\mu\nu}
 \label{eq:einsteinmat}
\end{equation}
ansetzen können. Die Konstante hier ist durch den klassischen Grenzfall der Gleichungen bestimmt und beträgt $C = - 8\pi \kappa$ \cite{adler:1975}. \\
Die Gleichungen \refb{eq:einsteinmat} stellen den formalen mathematischen Ausdruck des Machschen Prinzips dar. Sie beschreiben den Zusammenhang zwischen den Massen, repräsentiert durch den Energie-Impuls-Tensor, und der Krümmung des Raumes. Für die konkrete Berechnung von Teilchentrajektorien verwendet man die Geodätengleichung. Sie lässt sich aus den Einsteingleichungen aber ableiten.

%%%%%%%%%%%%%%%%%%%%%%%%%%%%%%%%%%%%%%%%%%%%%%%%%%%%%%%%%%%%%%%%%%%%%%%%%%%%%%%%%%%%%%%%%%%%%%%%%%%%%%%%%%%%%%%%%%%%%%%%%%%%%%%%%%%%%%%%%%
\subsubsection{Ableitung der Einsteingleichungen aus dem Variationsprinzip}
In der Physik spielt das Variationsprinzip von der klassischen Mechanik bis hin zu Quantenfeldtheorien eine entscheidende Rolle. Aus diesem Grund gehen wir hier 
auf die Herleitung der Einsteingleichungen unter Ausnutzung des Variationsprinzips ein. Eine kleine Änderung des Variationsprinzips wird uns in Kapitel \ref{sec:pseudoart} 
als Grundlage für die pseudokomplexe ART dienen. Ausgehend von der Variation der Wirkung
 \begin{equation}
  \delta S  = \delta \int R \sqrt{-g} ~\dd^4 x = 0
\label{eq:varprinadler}
 \end{equation}
lassen sich die Einsteingleichungen für den freien Raum \refb{eq:einsteinfrei} ableiten. Die Rechnungen dazu lassen sich z.B. in \cite{adler:1975} finden.  

%%%%%%%%%%%%%%%%%%%%%%%%%%%%%%%%%%%%%%%%%%%%%%%%%%%%%%%%%%%%%%%%%%%%%%%%%%%%%%%%%%%%%%%%%%%%%%%%%%%%%%%%%%%%%%%%%%%%%%%%%%%%%%%%%%%%%%%%%%
\clearpage
\subsection{Lösungen der Einsteingleichungen} \label{sec:loesungenklassisch}
Obwohl die gerade angesprochenen Einsteingleichungen aufgrund ihrer Nichtlinearität sehr komplex sind, wurde bereits 1916 von Schwarzschild die erste exakte Lösung für den freien Raum gefunden \cite{schwarzschild:1916}.
\subsubsection{Die Schwarzschildlösung}
 Durch Ausnutzen der Radialsymmetrie und Zeitunabhängigkeit lässt sich der Ansatz für die Schwarzschild-Metrik wie folgt formulieren
\begin{equation}
 g^{\text{Schw}}_{\mu\nu} = \begin{pmatrix} e^{\nu(r)} & 0 & 0  & 0 \\ 0 & -e^{\lambda(r)} & 0 & 0 \\ 0 & 0 & -r^2 & 0 \\ 0 & 0 & 0 & -r^2 \sin^2\vartheta
                          \end{pmatrix}  \qquad .
\label{eq:schwarzschild}
\end{equation}
Man verwendet hier die Koordinaten $(t,r,\vartheta,\varphi)$, welche den Kugelkoordinaten zusammen mit der Zeit entsprechen. Es bleiben nur noch die Funktionen
 $\nu(r)$ und $\lambda(r)$ unbestimmt, alle anderen Teile der Metrik sind durch die Symmetrie schon festgelegt \cite{adler:1975}. \\
Nun ist es aufwendig alle Christoffelsymbole aus dem Ansatz der Metrik zu berechnen. Wesentlich einfacher geht es, wenn  man das Variationsprinzip  \refb{eq:variationspr} verwendet, um die Geodätengleichung \refb{eq:geodaet} abzuleiten. Aus ihr können wir dann die einzelnen Christoffelsymbole ablesen. Mit diesen wiederum lässt sich der Ricci-Tensor aufstellen und damit die freien Einsteingleichungen \refb{eq:einsteinfrei}. Diese Rechnungen sind trotz allem noch recht aufwendig und deshalb werden wir uns hier mit ihrem Ergebnis begnügen \cite{adler:1975}:
\begin{align}
 \mathcal{R}_{00} &= -\frac{e^{\nu-\lambda}}{2} \left( \nu'' + \frac{\nu'^2}{2} - \frac{\lambda'\nu'}{2} + \frac{2\nu'}{r} \right) = 0 \notag \\
 \mathcal{R}_{11} &= \frac{1}{2} \left( \nu'' + \frac{\nu'^2 }{2} - \frac{\lambda'\nu'}{2}  - \frac{2\lambda'}{r} \right) = 0 \notag \\
 \mathcal{R}_{22} &=  e^{-\lambda} - r \lambda' e^{-\lambda}  -1 + r e^{-\lambda} \left( \frac{\lambda' + \nu'}{2}  \right)  = 0 \notag \\
 \mathcal{R}_{33} &=  \sin^2(\vartheta) \left[ e^{-\lambda} - r \lambda'e^{-\lambda} - 1 + r e^{-\lambda}  \left(\frac{\nu' + \lambda'}{2} \right)  \right] = 0 \qquad .
 \label{eq:schwarzschildricci}
\end{align}
Der Strich bedeutet eine Ableitung nach der Variablen $r$. Zunächst bemerken wir, dass die letzten beiden Gleichungen in \refb{eq:schwarzschildricci} bis auf den Faktor $\sin^2(\vartheta)$ identisch sind. Die ersten beiden Gleichungen können wir verwenden, um $\nu$ und $\lambda$ zu bestimmen, dazu ziehen wir die zweite von der ersten ab und erhalten
\begin{equation}
 \nu' + \lambda' = 0  \qquad .
 \label{eq:nulambdaschwarzschild}
\end{equation}
Also ist $\nu = -\lambda + \const$ . Diese Konstante kann aber durch eine Umdefinierung der Zeitkoordinate absorbiert werden. Dadurch wird auch gewährleistet, 
dass die Lösung für große Abstände asymptotisch in den flachen Raum übergeht. Dieses Resultat können wir in die dritte Gleichung einsetzen und erhalten
\begin{equation}
 e^{-\lambda} -r \lambda' e^{-\lambda} -1 = 0 ~\qquad \Leftrightarrow \qquad ~ \left(r e^{-\lambda}\right)' = 1 \qquad .
\end{equation}
Diese Gleichung kann man direkt lösen und erhält
\begin{equation}
 e^{-\lambda} = 1 - \frac{r_S}{r} \qquad .
\label{eq:r22klassisch}
\end{equation}
Die Konstante $r_S$ ist hier erst einmal eine Integrationskonstante. Sie lässt sich durch Grenzbetrachtungen dann mit dem Schwarzschildradius $2\kappa M$ identifizieren \cite{adler:1975}.
Setzt man \refb{eq:nulambdaschwarzschild} in die zweite Gleichung ein, bleibt
\begin{equation}
 \lambda'' - \lambda'^2 + \frac{2 \lambda'}{r} = 0 ~\qquad \Leftrightarrow \qquad ~ \left(r e^{-\lambda}\right)'' = 0 \qquad ,
 \label{eq:r11klassisch}
\end{equation}
was konsistent mit der gerade gefundenen Lösung ist. Damit haben wir die Schwarzschildmetrik
\begin{equation}
 g^{\text{Schw}}_{\mu\nu} = \begin{pmatrix} 1 - \frac{r_S}{r} & 0 & 0  & 0 \\ 0 & - \dfrac{1}{1- \frac{r_S}{r}} & 0 & 0 \\ 0 & 0 & -r^2 & 0 \\ 0 & 0 & 0 & -r^2 \sin^2\vartheta
                          \end{pmatrix} \qquad .
\label{eq:schwarzschildmetrik}
\end{equation}

\subsubsection{Die Kerr-Lösung} \label{sec:kerrklassisch}
Deutlich länger als bei der Schwarzschildlösung hat es gedauert um eine komplette analytische Lösung zu finden, die eine rotierende Zentralmasse beschreibt. Erst 1963 gelang es Kerr \cite{kerr:1963} eine solche Lösung aufzustellen. Die hier vorgestellte Ableitung der Kerr-Lösung aus \cite{adler:1975} basiert auf der Schwarzschildlösung - genauer gesagt auf einer bestimmten Darstellung der Schwarzschildlösung. Zunächst wird mit einer Koordinatentransformation $\bar{x}^0 = x^0 +r_S \ln\left| \dfrac{r}{r_S} - 1\right|$ die Schwarzschildlösung auf die so genannte {\bf Eddington-Form} \cite{Eddington:1924} 
\begin{equation}
 \dd s^2 = (\dd \bar{x}^0)^2 - (\dd r)^2 - r^2 (\dd \vartheta^2 + \sin^2\vartheta \dd \varphi^2) - \frac{r_S}{r} (\dd \bar{x}^0 + \dd r)^2
 \label{eq:eddington}
\end{equation}
gebracht. Dieser Ansatz funktioniert leider nicht für den Fall, wenn $\nu \neq -\lambda$ ist (siehe Kapitel \ref{sec:ergebnissekerr1}). Die Eddington-Form stellt also das Linienelement des flachen Raums mit einem Zustatzterm, der gemischt in der zeitlichen und radialen Koordinate ist, dar. Umgeschrieben in kartesische Koordinaten nimmt sie die folgende Form an
\begin{equation}
 \dd s^2 = (\dd \bar{x}^0)^2 - (\dd x)^2 - (\dd y)^2 - (\dd z)^2 - \frac{r_S}{r} \left( \dd \bar{x}^0 + \frac{x \dd x + y \dd y + z \dd z}{r} \right)^2 \qquad .
\label{eq:eddingtoncart}
\end{equation}
In diesen Koordinaten lässt sich der metrische Tensor also als
\begin{equation}
 \g = \eta_{\mu\nu} - r_S l_\mu l_\nu
\label{eq:degmetric}
\end{equation}
mit $l_\mu = \dfrac{1}{\sqrt{r}}\left(1, \dfrac{x}{r}, \dfrac{y}{r}, \dfrac{z}{r} \right)$ schreiben. Eine bemerkenswerte Eigenschaft der so definierten Vektoren $l_\mu$ ist $l_\mu l_\nu \eta^{\mu\nu} = 0$. Eine Metrik der Form \refb{eq:degmetric} wird auch als entartet (engl. "`degenerate"') bezeichnet. Die Tatsache, dass die Schwarzschildmetrik ein Spezialfall einer solchen entarteten Metrik ist, dient in \cite{adler:1975} als Motivation ausgehend von \refb{eq:degmetric} nach Verallgemeinerungen der Schwarzschildlösung zu suchen. \\
Betrachtet man nun die Einsteingleichungen \refb{eq:einsteinfrei} in Abhängigkeit von $r_S$, so lassen sie sich als Polynome in $r_S$ auffassen. Da $r_S$ aber zu diesem Zeitpunkt nur eine in \refb{eq:degmetric} eingeführte (beliebige) Konstante darstellt, müssen die Einsteingleichungen für jede Potenz von $r_S$ erfüllt sein. Es ergibt sich daraus ein Satz von vier mal zehn Gleichungen 
\begin{align}
\eta^{\alpha \rho} [\mu \nu,\rho]_{|\alpha} = 0 &\qquad \mathcal{O}(r_S) \notag \\
R_S(l^\alpha l^\rho[\mu\nu,\rho])_{|\alpha} - \eta^{\alpha \rho} \eta^{\beta \lambda} [\beta \mu, \sigma][\alpha \nu, \lambda] = 0 &\qquad \mathcal{O}(r_S^2) \notag \\
l^\beta l^\lambda \eta^{\alpha \sigma} [\beta \mu, \sigma] [\alpha \nu, \lambda] + l^\alpha l^\lambda \eta^{\beta \sigma} [\beta \mu,\lambda][\alpha \nu,\sigma] = 0 &\qquad \mathcal{O}(r_S^3)  \notag \\
l^\alpha l^\sigma l^\beta l^\lambda [\beta \mu,\sigma] [\alpha \nu,\lambda] = 0 &\qquad \mathcal{O}(r_S^4)  \quad .
\label{eq:polynomeinstein} 
\end{align}
 Einige dieser Gleichungen sind direkt erfüllt. Adler et al. \cite{adler:1975} zeigen, dass diese sich auf eine einzige Gleichung reduzieren lassen
\begin{equation}
 - \Box^2 (l_\mu l_\nu) = [(L+A)l_\mu]_{|\nu} + [(L+A)l_\nu]_{|\mu} \qquad ,
\end{equation}
wobei $L= - {l^\alpha}_{|\alpha}$ und $A$ über $l^\alpha {l^\nu}_{|\alpha} =: -A l^\nu$ definiert sind. Durch die Annahme einer stationären Lösung, gelingt es Adler et al. \cite{adler:1975} die Gleichung so weit umzuformen, dass am Ende eine Differentialgleichung für eine komplexe Funktion $\gamma$ bleibt. Die dazu notwendigen Rechnungen sind recht aufwendig und würden an dieser Stelle zu weit führen. Es bleibt 
\begin{equation}
 \nabla^2 \gamma  = 0 \quad,\quad (\vec{\nabla} \omega)^2 = 1 \quad ,\quad \omega := \frac{1}{\gamma} \qquad .
\label{eq:laplaceeikonal}
\end{equation}
Die erste Gleichung ist gerade die Laplacegleichung, bei der zweiten handelt es sich um die Eikonal-Gleichung, welche u.a. in der Optik eine Rolle spielt \cite{adler:1975}. Die Lösungen dieser Gleichungen zusammen mit Randbedingungen bestimmen die Metrik komplett. Die $l_\mu$ aus \refb{eq:degmetric} lassen sich über $l_\mu = l_0 (1,\vec{\lambda})$ und 
\begin{equation}
 \vec{\lambda} = \frac{\vec{\nabla}\omega + \vec{\nabla}\omega^* - i (\vec{\nabla}\omega \times \vec{\nabla} \omega^*)}{1 + \vec{\nabla}\omega \cdot \vec{\nabla}\omega^*} \qquad l_0^2 = \text{Re}(\gamma)
  \label{eq:loesungdeg}
\end{equation}
bestimmen. \\
Adler et al. \cite{adler:1975} nehmen nun die Gleichungen \refb{eq:laplaceeikonal}, um zunächst die Schwarzschildmetrik in der Eddington-Form herzuleiten. Setzt man mit
\begin{equation}
 \gamma = \frac{1}{r} = \frac{1}{\sqrt{x^2 + y^2 +z^2}}
\end{equation}
an, so erhält man 
\begin{equation}
 l_0^2 = \frac{1}{r} \qquad \lambda_1 = \frac{x}{r}\qquad \lambda_2 = \frac{y}{r}\qquad \lambda_3 = \frac{z}{r} \qquad ,
\end{equation}
was die Metrik in der Form \refb{eq:eddingtoncart} liefert %\footnote{Es ist klar, dass diese Lösung herauskommen muss, da die Eddington-Form ein Spezialfall von \refb{eq:degmetric} ist. Das Einsetzen dient an dieser Stelle nur der Veranschaulichung.}
. \\
Nun kann man nach den einfachsten Verallgemeinerungen der Schwarzschildmetrik suchen. Eine Möglichkeit der Verallgemeinerung stellt eine Verschiebung des Ursprungs dar. Allerdings bringt eine Verschiebung der Form
\begin{equation}
	\gamma = \frac{1}{\sqrt{(x-a_1)^2 + (y-a_2)^2 + (z-a_3)^2}}
\end{equation}
keine neue physikalische Lösung, sondern nur eine um den Ursprung verschobene Lösung \cite{adler:1975}. Anders verhält es sich, wenn man eine imaginäre Koordinatenverschiebung zulässt
\begin{equation}
	\gamma = \frac{1}{\sqrt{x^2 + y^2 + (z-ia)^2}} \qquad .
\end{equation}
Es ist nun einfacher $\omega = 1/\gamma$ zu betrachten, um die Lösungen \refb{eq:loesungdeg} zu erhalten. Wir schreiben $\omega = \rho + i \sigma$ und bekommen damit
\begin{equation}
	\rho + i \sigma = \sqrt{r^2 -a^2 -2iaz} \qquad .
\end{equation}
Quadriert man diese Gleichung und trennt nach Real- und Imaginärteil auf, so bleiben die zwei Gleichungen
\begin{equation}
	\rho^2 - \sigma^2 = r^2 - a^2 \qquad \sigma = - \frac{az}{\rho}
\end{equation}
übrig. Kombiniert ergeben sie eine quadratische Gleichung für $\rho^2$
\begin{equation}
	\rho^4 - \rho^2(r^2 -a^2) -a^2z^2 = 0 \qquad ,
\end{equation}
deren Lösung durch
\begin{equation}
	\rho^2 = \frac{r^2-a^2}{2} + \sqrt{\frac{(r^2-a^2)^4}{4} +a^2z^2}
	\label{eq:rhokerr}
\end{equation}
gegeben ist. Dabei nehmen wir nur die positive Wurzel, sodass im Grenzfall $r >> a$ die Variablen $\rho$ und $r$ einander entsprechen. Jetzt können wir $l_0^2 = \text{Re}(\gamma)$ direkt bestimmen und erhalten
\begin{equation}
	l_0^2  = \frac{\rho^3}{\rho^4 + a^2z^2} \qquad .
	\label{eq:lnullkerr}
\end{equation}
Ebenso bekommen wir über \refb{eq:loesungdeg} die Komponenten von $\vec{\lambda}$ nach \cite{adler:1975}
\begin{align*}
 \lambda_1 &= \frac{\rho x  + ay}{a^2 + \rho^2} \\ 
 \lambda_2 &= \frac{\rho y  - ax}{a^2 + \rho^2} \\ 
 \lambda_3 &= \frac{z}{\rho}  \qquad .
\end{align*}
Damit können wir das Linienelement nun vollständig aufschreiben
\begin{align}
	\dd s^2  &= (\dd x^0)^2 - (\dd x)^2 - (\dd y)^2 - (\dd z)^2 - \frac{r_S \rho}{\rho^4 + a^2z^2}  \notag \\
	 &\times \left[\dd x^0 + \frac{\rho}{a^2 + \rho^2}(x\dd x + y \dd y) + \frac{a}{a^2 + \rho^2}(y \dd x - x \dd y) + \frac{z}{\rho} \dd z \right]^2  \qquad .
	\label{eq:kerrnachadler}
\end{align}
In dieser Form wurde es erstmals auch von Kerr \cite{kerr:1963} aufgestellt. Durch geeignete Koordinatentransformationen \cite{adler:1975} lässt sich das Längenelement auf eine Form bringen, die die Interpretation als Längenelement außerhalb einer rotierenden Masse deutlicher macht
\begin{align}
 \dd s ^2 =& \left( 1 - \frac{r_S \rho}{\rho^2 + a^2 \cos^2\vartheta} \right) \dd \hat{t}^2 - \frac{\rho^2 + a^2 \cos^2\vartheta}{\rho^2 + a^2 - r_S\rho} \dd \rho^2 - \left( \rho^2 + a^2\cos^2\vartheta \right)\dd \vartheta^2 \notag \\
  &  - \left[ (\rho^2 +a^2) \sin^2\vartheta + \frac{r_S \rho a^2 \sin^4\vartheta}{\rho^2 + a^2 \cos^2 \vartheta} \right] \dd \hat{\varphi}^2 - 2 \frac{r_S \rho a \sin^2\vartheta}{\rho^2 +a^2 \cos^2\vartheta}\dd \hat{t} \dd \hat{\varphi}   \qquad .
  \label{eq:boyerlindquist}
\end{align}
Dieses Längenelement wurde zuerst von Boyer und Lindquist \cite{boyerlindquist:1967} aufgestellt. Hier tritt die Ähnlichkeit zu einem rotierenden flachen Raum \begin{equation*}
 	\dd s^2 = \left( 1- \omega^2 r^2\right) \dd t^2 - \left( \dd r^2 + r^2 \dd\varphi^2 + 2\omega r^2 \dd t \dd \varphi  + \dd z^2 \right)
\end{equation*} \cite{adler:1975} 
deutlicher hervor als in \refb{eq:kerrnachadler}.
% viererindizes \mu \nu usw   - 3er lateinische buchstaben NOCH ERKLÄREN
%Riemann Räume
%Riemann Tensoren   DONE
%Einsteingleichungen ableiten (zumindest grob) DONE
%Was sind Geodäten DONE
%Geodäten Gleichung aus Einsteingleichung ableiten - NUR ERWÄHNT
%Christoffelsymbole einführen und erklären DONE
%Kovariante Ableitung erklären DONE
%moderne Tests/Anwendungen der ART
%
%Lösungen der Feldgleichungen
%-Schwarzschild DONE
%-Reissner-Nordström
%-Kerr mit ausführlicherer Ableitung
%-Kerr-Newman NUR ERWÄHNEN
%
%
%%%%%%%%%%%%%%%%%%%%%%%%%%%%%%%%%%%%%%%%%%%%%%%%%%%%%%%%%%%%%%%%%%%%%%%%%%%%%%%%%%%%%%%%%%%%%%%%%%%%%%%%%%%%%%%%%%%%%%%%%%%%%%%%%%%%%%%%%%%%%%%
%\setcounter{footnote}{0}
\setcounter{equation}{0}

%
%################    EOF    #######################
\clearpage

\clearpage
%
% Kapitel über die pseudokomplexen Zahlen
%

\section{Die pseudokomplexen Zahlen}\label{sec:pseudocomplex}
Eine wichtige Grundlage der vorliegen Arbeit bilden die so genannten pseudokomplexen Zahlen\footnote{In der Literatur werden sie auch als hyperbolisch, hyperkomplex, semi- oder split-komplex bezeichnet.}. Sie stellen eine Erweiterung der reellen Zahlen, ähnlich der komplexen Zahlen dar. Sie wurden zuerst von James Cockle in der Literatur erwähnt \cite{cockle:1848}. In \cite{Antonuccio:1993et} findet sich eine gute Einführung in dieses Gebiet, an der sich dieses Kapitel orientieren wird. Wir werden allerdings die Nomenklatur wie im Paper über die pseudokomplexe ART \cite{Hess:2008wd} verwenden, welche sich davon leicht unterscheidet. \\

\subsection{Grundlegende Eigenschaften pseudokomplexer Zahlen}
Eine beliebige pseudokomplexe Zahl lässt sich als
\begin{equation}
	X = x_1 + I x_2
	\label{eq:pseudoczahl}
\end{equation}
schreiben. Im Gegensatz zur imaginären Einheit $i$, mit der komplexen Variablen $z = x +iy$ definiert sind, gilt $I^2=1$. Man kann für die pseudokomplexen Zahlen auch einfache Regeln der Addition und Multiplikation aufstellen
\begin{equation}
	(x_1 + I x_2) + (y_1 +I y_2) = (x_1+y_1) +I(x_2+y_2)
	\label{eq:pseudocadd}
\end{equation}
\begin{equation}
	(x_1 + I x_2) \cdot (y_1 +I y_2) = (x_1y_1 + x_2y_2) + I(x_1y_2 + x_2y_1) \qquad .
	\label{eq:pseudocmult}
\end{equation}
Neben diesen beiden Eigenschaften ist es zweckmäßig die pseudokomplexe Konjugation zu definieren. Die Zahl
\begin{equation}
	X^* = x_1 - Ix_2
	\label{eq:pseudockonj}
\end{equation}
ist die pseudokomplex Konjugierte zu $X = x_1 + I x_2$. Offensichtlich gilt für die Konjugation
\begin{equation}
	(XY)^* = X^*Y^* \qquad (X+Y)^* = X^* + Y^* \qquad .
\end{equation}
Analog zu den komplexen Zahlen lässt sich durch die Konjugation das Betragsquadrat einer pseudokomplexen Zahl definieren
\begin{equation}
	\left|X\right|^2 = X^*X = x_1^2 -x_2^2 \qquad .
\end{equation}
Im Gegensatz zu den gewohnten Eigenschaften des Betragsquadrats ist es hier nicht immer positiv, sehr wohl aber reell. Solange das Betragsquadrat nicht verschwindet, kann man es benutzen um ein Inverses zu $X$ zu definieren
\begin{equation}
	X^{-1} := \frac{X^*}{\left|X\right|^2} \qquad .
\end{equation}
Da das Betragsquadrat auch negativ sein kann, eignet es sich nicht als Norm. Es gibt aber eine andere Möglichkeit eine Norm für die pseudokomplexen Zahlen einzuführen. Dafür benötigt man die Größen
\begin{equation}
	X_\pm = x_1 \pm x_2 \qquad \sigma_\pm = \frac{1}{2}\left(1 \pm I\right) \qquad .
\end{equation}
Mit ihnen lässt sich jede pseudokomplexe Zahl als
\begin{equation}
	X = X_+ \sigma_+ + X_- \sigma_-
\end{equation}
schreiben. Eine Norm für die pseudokomplexen Zahlen ist dann \cite{Antonuccio:1993et}
\begin{equation}
\norm{X} := \sqrt{\left|X_+\right|^2 + \left|X_-\right|^2} = \sqrt{2\left(x_1^2 +x_2^2\right)} \qquad .
\end{equation}
Für unsere weiteren Betrachtungen hat die Norm keine wesentliche Bedeutung und ist hier nur der Vollständigkeit wegen erwähnt. Im Gegensatz dazu spielen die dabei eingeführten Größen $\sigma_\pm$ eine wesentliche Rolle in der pseudokomplexen ART. Sie bilden eine Basis des so genannten Null-Teilers $\mathcal{P}^0$ und erfüllen die wichtigen Relationen
\begin{equation}
	\sigma_\pm^2 = 1 \qquad \sigma_+\sigma_- = 0  \qquad .
	\label{eq:sigmaeig}
\end{equation}
\begin{figure}[ht]	
\begin{center}
\includegraphics[width=0.5\textwidth]{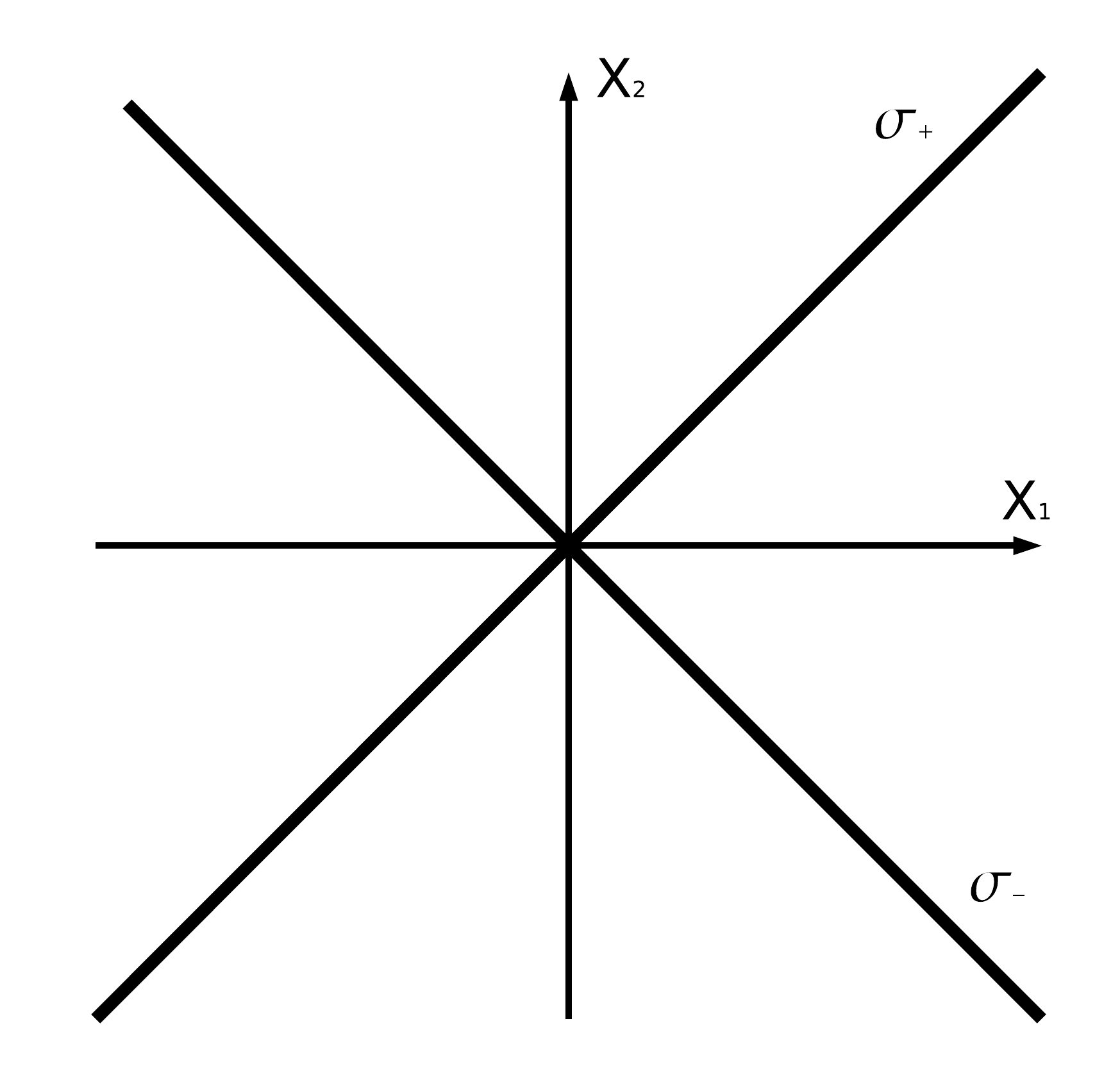}
  \label{fig:pseudonullteiler} 
\begin{flushleft}
\caption{Pseudokomplexe Ebene mit Null-Teiler. Quelle: \cite{Hess:2007pp} }
\end{flushleft}
\end{center}
\end{figure}  \\
 Der Null-Teiler beinhaltet alle pseudokomplexen Zahlen mit Betrag 0 (also gerade die Zahlen ohne Inverses). Im Kapitel \ref{sec:pseudoart} werden wir sehen, warum die $\sigma_\pm$ für uns bedeutend sind. Ein kleines Beispiel, warum man dem Null-Teiler eine besondere physikalische Rolle zuschreiben kann soll hier aber schon erwähnt sein. Wählt man $X = x + It$, wobei $x$ eine räumliche und $t$ eine zeitliche Koordinate sein sollen, so entspricht der Null-Teiler gerade dem Lichtkegel. Man kann also Größen innerhalb des Null-Teilers eine besondere physikalische Bedeutung zuweisen.\\
\subsection{Pseudokomplexe Funktionen}
Nun da wir einige grundlegende Eigenschaften der pseudokomplexen Zahlen kennengelernt haben, können wir Funktionen mit pseudokomplexen Variablen definieren und ihre Eigenschaften untersuchen. Eine der wohl bedeutensten Funktionen in der Physik ist die Exponentialfunktion. Man kann sie - analog zum reellen und komplexen Fall - über die Reihe
\begin{align}
	e^X := \sum_{k=0}^\infty{\frac{X^k}{k!}}
	\label{eq:expfun}
\end{align}
definieren. Auch die Eulerformel gilt in etwas abgewandelter Form
\begin{equation}
	e^{I\varphi} = \cosh\varphi + I \sinh\varphi \qquad .
	\label{eq:eulerformel}
\end{equation}
Dadurch, dass pseudokomplexe Zahlen untereinander kommutieren, lässt sich das Produkt zweier $e$-Funktionen vereinfachen
\begin{equation}
	e^{X}\cdot e^{Y} = e^{X+Y} \qquad .
\end{equation}
Also lässt sich jede pseudokomplexe Exponentialfunktion auftrennen
\begin{equation}
	e^{X} = e^{x_1}\left(\cosh x_2 + I \sinh x_2\right) \qquad .
\end{equation}
Wie bei komplexen Zahlen können wir $e^{I\varphi}$ als Phasenfaktor identifizieren, denn es gilt
\begin{equation}
	\left|e^{I\varphi}\right|^2 = \cosh^2\varphi - \sinh^2\varphi = 1 \qquad .
\end{equation}
Man kann jede pseudokomplexe Funktion in ihren Real- und Pseudoimaginärteil zerlegen
\begin{equation}
	F(x_1 + I x_2) = u(x_1,x_2) + Iv(x_1,x_2) \qquad .
	\label{eq:realim}
\end{equation}
Mit Hilfe von \refb{eq:sigmaeig} können wir außerdem jede in eine Potenzreihe entwickelbare Funktion zerlegen. Dazu stellen wir fest, dass 
\begin{equation}
	X^n = \left(X_+ \sigma_+ +  X_- \sigma_-\right)^n = \left(X_+ \sigma_+\right)^n + \left(X_- \sigma_-\right)^n = X_+^n \sigma_+ + X_-^n \sigma_-
	\label{eq:potaufteil}
\end{equation}
gilt (siehe \cite{Hess:2007pp} ). Es ist jetzt offensichtlich, wie man eine Funktion $F(X)$ in zwei Funktionen, die jeweils nur von $X_+$ beziehungsweise $X_-$ abhängen, aufteilen kann
\begin{eqnarray}
	F(X) &=&F(X_+\sigma_+ + X_-\sigma_-) = \sum_n{a_n \left(X_+ \sigma_+ +  X_- \sigma_-\right)^n} \notag\\
	 &=& \sum_n{a_n X_+^n \sigma_+ + a_n X_-^n \sigma_-} 	= \sum_n{a_n X_+^n} \sigma_+ + \sum_n{a_n X_-^n} \sigma_- \notag\\
	 &=& F(X_+)\sigma_+ + F(X_-)\sigma_-  \qquad .
	 \label{eq:funkaufteil}
\end{eqnarray}
Diese Art der Aufteilung in einen $\sigma_+$- und $\sigma_-$-Anteil wird in Kapitel \ref{sec:pseudoart} wichtig werden. Als eine erste Anwendung sehen wir direkt die Aufteilung der Exponentialfunktion
\begin{equation}
	e^X = e^{X_+}\sigma_+ + e^{X_-}\sigma_- \qquad .
\end{equation}
Für das Produkt zweier beliebiger (entwickelbarer) Funktionen $F$ und $G$ haben wir dann
\begin{equation}
	F(X)G(X) = F(X_+)G(X_+)\sigma_+ + F(X_-)G(X_-)\sigma_-  \qquad .
\end{equation}
\subsubsection*{Die Ableitung pseudokomplexer Funktionen}
Neben den unterschiedlichen Schreibweisen für Funktionen pseudokomplexer Variablen, die wir gerade kennengelernt haben, ist es von großem Interesse zu wissen, wie sich Funktionen verhalten und verändern. Dafür ist es notwendig die pseudokomplexe Ableitung zu definieren. Eine Möglichkeit ist es den Differenzenquotienten
\begin{equation}
	\frac{F(X +\Delta X) - F(X)}{\Delta X}
	\label{eq:diffquot}
\end{equation}
zu betrachten, der für $\Delta X \rightarrow 0$ der tatsächlichen Ableitung entspricht. Problematisch dabei ist allerdings, dass wir durch die pseudokomplexe Größe $\Delta X$ teilen. In \cite{Antonuccio:1993et} ist die Ableitung aus diesem Grund als $ \dfrac{DF}{DX} =F'(X)$ definiert, unter der Voraussetzung, dass
\begin{equation}
	\frac{F(X + \Delta X) - F(X) - F'(X)}{\left\|\Delta X\right\|} \overset{\Delta X \rightarrow 0}{\rightarrow}0
\end{equation}
gilt. Wir werden uns im Folgenden nicht weiter mit den mathematischen Feinheiten der Definition aufhalten und werden uns an die Definition der Ableitung über den Differenzenquotienten \refb{eq:diffquot} halten. Analog zu komplexen Funktionen nennen wir eine überall differenzierbare Funktion holomorph.\\
Man stellt nun fest, dass für eine differenzierbare Funktion F auch folgende Relationen gelten
\begin{equation}
	F'(X) = \lim_{h\rightarrow 0} \frac{F(X+h)- F(X)}{h} = \lim_{h\rightarrow 0} \frac{F(X+Ih)- F(X)}{Ih}  \qquad .
 	\label{eq:pseudodiff}
\end{equation}
Nun können wir durch die Aufteilung der Funktion in Real- und Pseudoimaginärteil einen interessanten Zusammenhang herausfinden
\begin{eqnarray}
	F'(X) &=&\lim_{h\rightarrow 0} \frac{F(X+h)- F(X)}{h} \notag \\
	&=& \lim_{h\rightarrow 0} \frac{u(x_1 +h,x_2) + Iv(x_1+h,x_2) - u(x_1,x_2) -  I v(x_1,x_2)}{h} \notag \\
	&=& \lim_{h\rightarrow 0} \left( \frac{u(x_1 +h,x_2) - u(x_1,x_2)}{h} + \frac{Iv(x_1+h,x_2) -  I v(x_1,x_2)}{h} \right) \notag \\
	&=& \fracpd{u}{x_1} + I \fracpd{v}{x_1}
\end{eqnarray} 
und analog dazu auch
\begin{equation}
	F'(X) = I \fracpd{u}{x_2} + \fracpd{v}{x_2} \qquad .
\end{equation}
Durch Gleichsetzen der Real- und Imaginärteile erhalten wir die pseudokomplexen \textit{Cauchy-Riemann-Gleichungen}
\begin{equation}
	\fracpd{u}{x_1} = \fracpd{v}{x_2} \qquad \fracpd{v}{x_1} = \fracpd{u}{x_2} \qquad .
	\label{eq:pseudocr}
\end{equation}
Eine Funktion, die diese Gleichungen erfüllt ist holomorph. Der Beweis dazu kann beispielsweise in \cite{Antonuccio:1993et} nachgelesen werden. Durch die Einführung des Operators
\begin{equation}
	\frac{D}{DX^*} = \fracpd{}{x_1} - I \fracpd{}{x_2}
\end{equation}
können wir eine Funktion auf Holomorphizität überprüfen. Dazu betrachten wir die Forderung, dass $\frac{DF}{DX^*}$ verschwindet
\begin{equation}
	\frac{DF}{DX^*} = \fracpd{F}{x_1} - I \fracpd{F}{x_2} = \fracpd{u}{x_1} + I \fracpd{v}{x_1} - I \fracpd{u}{x_2} - \fracpd{v}{x_2} \sose 0 \qquad .
\label{eq:wannholomorph}
\end{equation}
Es ergeben sich hier also die pseudokomplexen Cauchy-Riemann-Gleichungen aus der Forderung, dass die Funktion nicht von der pseudokomplex Konjugierten $X^*$ abhängt. Dieser Zusammenhang ist uns von den komplexen Zahlen bereits bekannt. \\
 Ähnlich wie wir in \refb{eq:funkaufteil} eine beliebige pseudokomplexe Funktion aufgeteilt haben, können wir auch deren Ableitung in einen $\sigma_+$- und $\sigma_-$-Anteil zerlegen
\begin{eqnarray}
	\frac{DF}{DX} &=& \lim_{h\rightarrow 0} \frac{F(X+h)- F(X)}{h} \notag \\
	 &=& \lim_{h\rightarrow 0}\left( \frac{F(X_+ +h)- F(X_+)}{h} \sigma_+ + \frac{F(X_- +h)- F(X_-)}{h} \sigma_-  \right) \notag \\
	 &=& \fracpd{F(X_+)}{X_+} \sigma_+ + \fracpd{F(X_-)}{X_-} \sigma_-  \qquad .
	\label{eq:diffaufteil}
\end{eqnarray}
\subsubsection*{Die Integration pseudokomplexer Funktionen}
Nun da wir die Ableitung von pseudokomplexen Funktionen kennen gelernt haben, wenden wir uns deren Integration zu. Wie im Komplexen ist das Integral der Funktion $F$ ein Linienintegral
\begin{equation}
	\int_\gamma{F(X) \text{d}X} = \int_\gamma{[(u + I v) (\text{d}x_1 + I \text{d}x_2)]} = \int_\gamma{[u\text{d}x_1 + v \text{d}x_2 +I(u\text{d}x_2 + v\text{d}x_1)]} \qquad .
	\label{eq:pseudoint}
\end{equation}
Mit Hilfe dieser Definition können wir einen Zusammenhang finden, der dem Cauchy-Theorem im Komplexen entspricht. Dazu betrachten wir das Integral einer pseudokomplexen Funktion entlang einer geschlossenen Kurve
\begin{align}
 \oint_{\partial C} F(X) \dd X &= \oint_{\partial C} F(X)[ \dd x_1 + I \dd x_2  ]= \oint_{\partial C} [F \dd x_1 + I F \dd x_2] \notag \\
 &= \iint_C \left( I \fracpd{F}{x_1} - \fracpd{F}{x_2}\right)  \dd x_1 \dd x_2 = \iint_C \frac{DF}{DX^*} I \dd x_1 \dd x_2  \qquad .
 \label{eq:pseudocauchy}
\end{align}
Hierbei haben wir das Greensche Theorem\footnote{Es stellt einen Spezialfall des Satzes von Stokes in der Ebene dar.} ausgenutzt \cite{Antonuccio:1993et, arnold:2000}. 
Eine ausführlichere Ableitung befindet sich im Anhang ab Gleichung \refb{eq:satzvonstokes}.
 Die letzte Zeile in \refb{eq:pseudocauchy} liefert uns das Analogon zum Cauchy-Theorem. Ist $F$ eine holomorphe Funktion, so verschwindet die Ableitung $\frac{DF}{DX^*}$, wie wir in \refb{eq:wannholomorph} bereits gesehen haben. Im Gegensatz zu komplexen Funktionen sind pseudokomplexe holomorphe Funktionen aber nicht automatisch analytisch (Gegenbeispiel $F(X) = e^{-\frac{1}{X^2}}$  siehe \cite{Antonuccio:1993et}). 
%%%%%%%%%%%%%%%%%%%%%%%%%%%%%%%%%%%%%%%%%%%%%%%%%%%%%%%%%%%%%%%%%%%%%%%%%%%%%%%%%%%%%%%%%%%%%%%%%%%%%%%%%%%%%%%%%%%%%%%%%%%%%%%%%%%%%%%%%%%%%%%
%\setcounter{footnote}{0}
\setcounter{equation}{0}

%
%################    EOF    #######################
%

\clearpage

\clearpage
%
% Kapitel über die pseudokomplexe ART
%

\section{Pseudokomplexe ART}\label{sec:pseudoart}

Wie im Vorwort bereits kurz angesprochen wird in diesem Kapitel eine Aufarbeitung der pseudokomplexen ART nach \cite{Hess:2008wd} stattfinden. Es gab bisher 
bereits viele Versuche die ART weiter zu verallgemeinern \cite{Einstein:1945}. Unter anderem auch mit Hilfe komplexer Zahlen \cite{Einstein:1946, Mantz:2008hm}. Es wurde aber in \cite{KellyMann:1986} gezeigt, dass solche 
Erweiterungen auch automatisch unphysikalische Lösungen liefern und daher nicht in Frage kommen. In Kapitel \ref{sec:pseudocomplex} haben wir schon ein kleines Beispiel
für einen pseudokomplexen Raum mit physikalischer Bedeutung kennengelernt. Hier entsprach der Null-Teiler gerade dem Lichtkegel. Wir werden sehen, dass 
dem Null-Teiler auch noch eine andere besondere physikalische Bedeutung zugewiesen werden kann.\\

Zunächst definieren wir die pseudokomplexe Metrik $g_{\mu\nu}$ als Funktion der pseudokomplexen Koordinaten $X^\lambda = x_1^\lambda + I x_2^\lambda$. Nach Gleichung \refb{eq:funkaufteil} 
können wir sie in die Null-Teiler-Basis entwickeln
\begin{equation}
 g_{\mu\nu}  = g_{\mu\nu}^+ \sigma_+  + g_{\mu\nu}^- \sigma_- \qquad .
 \label{eq:metricpseudo}
\end{equation}
Sie hängt in dieser Form von den $X_{\pm}^\lambda$ ab. Diese lassen sich auch als $X_{\pm}^\lambda = x^\lambda \pm l u^\lambda$ schreiben, wobei $l$ die minimale Länge und $u^\lambda = \frac{\dd x^\lambda}{\dd \tau}$ mit $\tau$ als Eigenzeit ist. Die minimale Länge $l$ wird an dieser Stelle aus Dimensionsgründen eingeführt \cite{Hess:2007pp}. Mit Hilfe der Metrik können wir nun das pseudokomplexe Längenelement definieren %(NEUE ART DER PROJEKTION DIREKT VERWENDEN) war hier noch nicht notwendig
 \begin{align}
	\dd \omega^2 &= g_{\mu\nu}(X) DX^\mu DX^\nu \notag\\
	&= g^+_{\mu\nu}(X_+) DX^\mu_+ DX^\nu_+ \sigma_+ + g^-_{\mu\nu}(X_-) DX^\mu_- DX^\nu_- \sigma_-  \qquad .
	\label{eq:pseudolaenge}
\end{align}
Da die Differentiale der pseudokomplexen Koordinaten kommutieren und die Indizes in \refb{eq:pseudolaenge} nur Summationsindizes sind, die man beliebig umbenennen kann, muss die Metrik symmetrisch sein.\\
Durch die Darstellung der Metrik in \refb{eq:metricpseudo} erhält man zunächst zwei Formulierungen der ART, einmal in $\sigma_+$ und einmal in $\sigma_-$ . Durch die
lineare Unabhängigkeit der $\sigma_\pm$ sind die beiden Formulierungen der ART auch unabhängig. Der Vorteil dieser Unabhängigkeit liegt darin, dass wir
die bekannten  Prinzipien und Ergebnisse aus Kapitel \ref{sec:art} verwenden können. So sind zum Beispiel die Christoffelsymbole der zweiten Art \refb{eq:christoffel2} durch
\begin{equation}
 \crs{\gamma}{\mu}{\nu} = - \Gamma^\gamma_{\nu\mu} = \frac{1}{2} g^{\sigma\gamma} \left( \fracD{g_{\nu\sigma}}{X^\mu} + \fracD{g_{\sigma\mu}}{X^\nu} - \fracD{g_{\mu\nu}}{X^\sigma}\right)
\end{equation}
 gegeben, wobei $\fracD{}{X^\mu}$ die pseudokomplexe Ableitung \refb{eq:pseudodiff} nach der Variablen $X^\mu$ ist. Wie alle Funktionen, mit denen wir arbeiten werden,
 lassen auch sie sich in eine $\sigma_+$- und $\sigma_-$-Komponente entwickeln
\begin{equation}
  \crs{\gamma}{\mu}{\nu} =  \crs{\gamma}{\mu}{\nu}_+ \sigma_+ +  \crs{\gamma}{\mu}{\nu}_- \sigma_- \qquad .
\end{equation}
Das gleiche gilt u.a. auch für die kovariante Ableitung \refb{eq:kovdiffkontra}. \\
An dieser Stelle stellt sich nun die Frage, welche Vorteile die pseudokomplexe Formulierung mit sich bringt. Bisher haben wir nur zweimal die selbe alte ART reproduziert. 
Die entscheidende Neuerung kommt durch die Verbindung der beiden Theorien über eine Änderung des Variationsprinzips. Wie zunächst vorgeschlagen von Schuller \cite{schuller:2003,schullerphd:2003} und dann 
aufgegriffen von Greiner und Hess in \cite{Hess:2008wd,Hess:2007pp} verändern wir das Variationsprinzip \refb{eq:variationspr} indem wir fordern, dass die Variation der Wirkung im Null-Teiler liegt. Dadurch wird die Gleichung \refb{eq:varprinadler} zu
\begin{equation}
 \delta S  = \delta \int R \sqrt{-g} ~\dd^4 x \in \mathcal{P}^0
\end{equation}
und wir erhalten damit die modifizierten Einsteingleichungen für den freien Raum als
\begin{equation}
 R_{\mu\nu} - \frac{1}{2} g_{\mu\nu} R \in \mathcal{P}^0 \qquad .
\end{equation}
Da keines der beiden $\sigma_\pm$ besonders hervorgehoben ist, spielt es keine Rolle ob die Variation proportional zu $\sigma_+$ oder $\sigma_-$ ist. Wir wählen sie proportional zu $\sigma_-$. \\
Für die Beschreibung von Teilchenbahnen fordern wir aber weiterhin das alte Variationsprinzip (siehe Kapitel \ref{sec:art}), damit Geodäten weiterhin extremale Kurven beschreiben
\begin{equation}
 \delta S = \delta \int L \dd \tau = 0 \qquad .
\end{equation}
Durch Ausführen der Variation erhält man die wohlbekannten Euler-Lagrange-Gleichungen
\begin{equation}
 \fracD{}{\tau} \left( \fracD{L}{\dot{X}^\mu} \right) - \fracD{L}{X^\mu} = 0 \qquad .
\end{equation}
Wie vorher führt uns diese Variation dann auf die Geodätengleichung für ein freies Teilchen im Gravitationsfeld
\begin{equation}
 \fracDD{X^\mu}{s} +\crs{\mu}{\sigma}{\gamma} \fracD{X^\sigma}{s}  \fracD{X^\gamma}{s} = 0 \qquad .
\end{equation}

% schreiben über Ansatz mit R=0.

\subsection{Projektion auf die reellen Zahlen}
Bisher sind alle Größen und Funktionen von den pseudokomplexen Variablen abhängig. Um tatsächliche physikalische Observablen zu erhalten, müssen wir noch eine Vorgehensweise definieren, wie man auf reelle Größen abbildet. Das im Artikel über die pseudokomplexe ART \cite{Hess:2008wd} vorgeschlagene Verfahren bringt leider einige Probleme mit sich. Im Laufe des letzten Jahres wurde deshalb ein neues Verfahren entwickelt, welches bisher noch nicht in einer Publikation veröffentlicht wurde. Zuerst betrachten wir aber das Projektionsverfahren aus \cite{Hess:2008wd}.\\

Zerlegt man die Metrik wie in \refb{eq:metricpseudo}, so kann man den Real- und Pseudoimaginärteil der Metrik als
\begin{align}
  &\g^R = \frac{1}{2} \left( \g^+ + \g^- \right) =: \g^0  \notag \\
  &\g^I = \frac{1}{2} \left( \g^+ - \g^- \right) =: h_{\mu\nu}
 \label{eq:projektionalt} 
\end{align}
schreiben. Per Konstruktion kann man nun mit der Metrik $\g^\pm$ die Indizes der $X_\pm^\mu$ absenken und mit ihrer Inversen $g^{\mu\nu}_\pm$ die Indizes von $X^\pm_\mu$ anheben
\begin{align}
 & X_\mu^\pm = x_\mu \pm l u_\mu  = \g^\pm \left(x^\nu \pm l u^\nu \right) = \g^\pm X^\nu_\pm \notag \\
 & X^\mu_\pm = x^\mu \pm l u^\mu  = g^{\mu\nu}_\pm \left(x_\nu \pm l u_\nu \right) = g^{\mu\nu}_\pm X_\nu^\pm  \qquad .
 \label{eq:projektionX}
\end{align}
 Aus \refb{eq:projektionX} erhalten wir die Transformationsvorschriften für $x_\mu$ und $u_\mu$
\begin{align}
 x_\mu = \frac{1}{2} \left( \g^+ + \g^- \right)x^\nu + \frac{1}{2} \left( \g^+ - \g^- \right) l u^\nu \notag \\
 l u_\mu = \frac{1}{2} \left( \g^+ - \g^- \right)x^\nu + \frac{1}{2} \left( \g^+ + \g^- \right) l u^\nu  \qquad .
\end{align}
Zusammen mit \refb{eq:projektionalt}  erhält man
\begin{align}
 x_\mu = \g^0 x^\nu + h_{\mu\nu} l u^\nu \notag \\
 l u_\mu = h_{\mu\nu} x^\nu + \g^0 l u^\nu \qquad .
\end{align}
Diese Gleichungen sind insofern problematisch, da nun $x_\mu$ und $u_\mu$ keine kovarianten Vektoren mehr darstellen (entsprechend dazu sind natürlich auch $x^\mu$ 
und $u^\mu$ keine kontravarianten Vektoren mehr), bzw. die Metriken $\g^0$ und $h_{\mu\nu}$ sind nicht geeignet um die Transformation von Real- und Pseudoimaginärteil getrennt zu beschreiben. Ausgehend von $\g^\pm g^{\mu\lambda}_\pm = \delta^\lambda_\nu $ ergibt sich ein weiteres Problem. Es gilt dann 
\begin{align}
& g^0_{\mu\nu}g_0^{\mu\lambda} + h_{\mu\nu}h^{\mu\lambda} = \delta^\lambda_\nu \notag \\
& g^0_{\mu\nu}h^{\mu\lambda} + g^{\mu\lambda}_0 h_{\mu\nu} = 0  \qquad .
\end{align}
Dadurch ist das Produkt $g^0_{\mu\nu}g_0^{\mu\lambda}$ kein Kronecker-Delta mehr. \\ %hier evtl noch weiter darauf eingehen was x_\mux^\mu ist siehe Report 3

Nun können wir mit dieser Art der Projektion das pseudokomplexe Linienelement \refb{eq:pseudolaenge} betrachten. Da es sich hier um eine Observable handelt, wird in \cite{Hess:2008wd} gefordert, dass
\begin{equation}
 \dd \omega^2 = {\dd \omega^2}^{*} 
 \label{eq:laengenelementreellalt}
\end{equation}
gelten muss. Diese Bedingung liefert uns einen Zusammenhang zwischen den Nullteiler-Komponenten des Längenelements
\begin{equation}
 \g^+ \DD X_+^\mu \DD X_+^\nu = \g^- \DD X_-^\mu \DD X_-^\nu \qquad .
 \label{eq:bedingunglaengenelement}
\end{equation}
Durch Einsetzen der Differentiale $\DD X_\pm^\mu = \dd x^\mu \pm l \dd u^\mu$ und der Definition für $\g^0$ und $h_{\mu\nu}$ \refb{eq:projektionalt} können wir das Linienelement zu
\begin{equation}
 \dd \omega^2 = \g^0 \left( \dd x^\mu \dd x^\nu + l^2 \dd u^\mu \dd u^\nu \right) + 2l h_{\mu\nu} \dd x^\mu \dd x^\nu 
 \label{eq:laengenelementreellalt2}
\end{equation}
umschreiben. Gleichzeitig liefert uns \refb{eq:bedingunglaengenelement} noch die Dispersionsrelation
\begin{equation}
 \frac{1}{2} h_{\mu\nu} \left( \dd x^\mu \dd x^\nu + l^2 \dd u^\mu \dd u^\nu \right) + \frac{1}{2} \g^0 \left( 2l \dd x^\mu \dd u^\nu \right) = 0 \qquad .
 \label{eq:dispersionalt}
\end{equation}
Der letzte Term in \refb{eq:laengenelementreellalt2} ist problematisch, da es sich hierbei um einen Mischterm handelt. %Brandt und Beil (REFERENZ  8-15 aus dem Paper) haben gezeigt, dass man einen solchen Ausdruck nicht mehr umformen %(EVTL WEITER VEREINFACHEN??)kann.
Die Dispersionsrelation \refb{eq:dispersionalt} ist in dieser Form unnötig kompliziert. Sie enthält einen zusätzlichen Term proportional zu $h_{\mu\nu}$ und stellt eine Art erweiterte Orthogonalitätsrelation dar. Durch die neue Art der Projektion werden diese beiden Formeln in einer sehr angenehmen Art vereinfacht und auch das Problem, dass $\g^0$ nicht die Indizes der Koordinaten heben und senken kann, wird behoben werden.
\bigskip\\
%%%%%%%%%%%%%%%%%%%%%%%%%%%%%%%%%%%%%%%%%%%%%%%%%%%%%%%%%%
Die Idee für die neue Art der Projektion lehnt sich an die Behandlung der Elektrodynamik mit komplexen Variablen an. Hier projiziert man die zunächst als komplexe Größen angesetzten elektrischen und magnetischen Felder erst auf reale Größen, bevor man diese in Funktionen einsetzen kann \cite{greiner:1991}. Die Wahl der Felder als zunächst komplexe Größen ist nur ein mathematischer Trick, um die Rechnungen zu vereinfachen. Alle physikalischen Größen bleiben nach wie vor reell. Analog dazu werden wir auch vorgehen und schrittweise Größen auf das Reelle abbilden (siehe dazu auch das Kapitel "`Extracting the physical component of a field"` in \cite{Hess:2007pp}). Zuerst schauen wir uns das Linienelement \refb{eq:pseudolaenge} an und schreiben es als Funktion der $x^\mu$ und $u^\mu$
\begin{equation}
	\dd \omega^2 = \g \left( \dd x^\mu \dd x^\nu + l^2 \dd u^\mu \dd u^\nu  + 2Il \dd x^\mu \dd u^\nu \right) \qquad .
	\label{eq:pseudolaenge2}
\end{equation}
Als nächsten Schritt projizieren wir den Metrischen Tensor $\g$ auf seinen Realteil $\g^{\text{Re}}$. Der Tensor selbst ist eine Funktion von pseudokomplexen Größen, die wir auch zunächst auf ihren Realteil projizieren, um diesen dann in den Tensor einzusetzen. Welche das im Einzelnen sind, werden wir später noch sehen. Nun können wir das Linienelement auf das Reelle abbilden
\begin{align}
	&\dd \omega^2 \rightarrow \g^{\text{Re}} \left( \dd x^\mu \dd x^\nu + l^2 \dd u^\mu \dd u^\nu \right) + 2Il \g^{\text{Re}} \dd x^\mu \dd u^\nu \notag \\
	& \Rightarrow \dd \omega^2_{\text{Re}} = \g^{\text{Re}} \left( \dd x^\mu \dd x^\nu + l^2 \dd u^\mu \dd u^\nu \right) \qquad .
	\label{eq:laengeproj}
\end{align} 
Damit das Längenelement nun reell ist haben wir noch
\begin{equation}
	\g^{\text{Re}} \dd x^\mu \dd u^\nu = 0
	\label{eq:dispersionneu}
\end{equation}
gefordert. Das hat gegenüber der alten Projektion den Vorteil, dass im Längenelement kein Mischterm zwischen $x^\mu$ und $u^\mu$ auftaucht und wir gleichzeitig eine einfachere Dispersionsrelation erhalten. %(LAENGENELEMENT ENTSPRICHT DEM VON BRANDT UND  BEIL NACHSCHLAGEN).
Die Form von $\g^{\text{Re}}$ hängt natürlich von der entsprechenden Metrik ab. Im nächsten Kapitel werden wir sie für die pseudokomplexe Schwarzschildlösung bestimmen. 

\subsection{Die pseudokomplexe Schwarzschildlösung} \label{sec:pseudoschwarz}
Auch hier werden wir zunächst die Lösung aus dem Artikel \cite{Hess:2008wd} vorstellen und dann einige damit verbundene Probleme erläutern. Die Vorgehensweise zur Aufstellung der Schwarzschildmetrik erfolgt ganz analog zu Kapitel \ref{sec:loesungenklassisch}. Zunächst kann man den Ansatz für das Längenelement aus den selben Symmetriebetrachtungen wie im klassischen Fall auf die Form
\begin{equation}
	\dd \omega^2 = e^{\nu(R)} (\DD X^0)^2 - e^{\lambda(R)} (\DD R)^2 - R^2\left( (\DD \theta)^2  + \sin^2\theta (\DD \phi)^2\right)
	\label{eq:schwarzschildlängepseudo}
\end{equation}
bringen. Nun verwenden wir aber statt des klassischen Variationsprinzips \ref{eq:varprinadler} das neue
\begin{equation}
	\delta S  = \delta \int R \sqrt{-g} ~\dd^4 x \in \mathcal{P}^0 \qquad .
	\label{eq:varneu}
\end{equation}
Alle weiteren Rechnungen laufen nun analog zum klassischen Fall und wir erhalten den Ricci-Tensor
\begin{align}
	 \mathcal{R}_{00} &= -\frac{e^{\nu-\lambda}}{2} \left( \nu'' + \frac{\nu'^2}{2} - \frac{\lambda'\nu'}{2} + \frac{2\nu'}{R} \right)  \notag \\
 \mathcal{R}_{11} &= \frac{1}{2} \left( \nu'' + \frac{\nu'^2 }{2} - \frac{\lambda'\nu'}{2}  - \frac{2\lambda'}{R} \right)  \notag \\
 \mathcal{R}_{22} &=  e^{-\lambda} - R \lambda' e^{-\lambda}  -1 + R e^{-\lambda} \left( \frac{\lambda' + \nu'}{2}  \right)  \notag \\
 \mathcal{R}_{33} &=  \sin^2(\vartheta) \left[ e^{-\lambda} - R \lambda'e^{-\lambda} - 1 + R e^{-\lambda}  \left(\frac{\nu' + \lambda'}{2} \right)  \right]  \qquad ,
	\label{eq:riccitensorpseudo}
\end{align}
wobei der Strich nun eine Ableitung nach $R$ bedeutet, z.B. $\nu' = \fracD{\nu}{R}$. Der Unterschied zur klassischen Schwarzschildlösung besteht nun darin, dass die Komponenten des Ricci-Tensors aufgrund der freien Einsteingleichungen \refb{eq:einsteinfrei} nicht mehr Null sein müssen, sondern im Nullteiler liegen. Wie bereits beschrieben, wählen wir diese Funktion proportional zu $\sigma_-$. Wir gehen desweiteren davon aus, dass die lokale Krümmung, beschrieben durch den Krümmungsskalar $\mathcal{R}$, wie im Klassischen verschwindet. Die Einsteingleichungen lauten also % (Die Definition der $\xi$-Funktionen stimmt für die ersten beiden Gleichungen nicht mit ihrer ersten Einführung \refb{eq:xidef} überein, aus Gründen der Übersichtlichkeit werden wir aber darauf verzichten sie hier anders zu benennen.)
\begin{align}
 \mathcal{R}_{00} &= -\frac{e^{\nu-\lambda}}{2} \left( \nu'' + \frac{\nu'^2}{2} - \frac{\lambda'\nu'}{2} + \frac{2\nu'}{R} \right) = -\frac{e^{\nu-\lambda}}{2} \xi_0 \sigma_-  \notag \\
 \mathcal{R}_{11} &= \frac{1}{2} \left( \nu'' + \frac{\nu'^2 }{2} - \frac{\lambda'\nu'}{2}  - \frac{2\lambda'}{R} \right) = \frac{1}{2} \xi_1 \sigma_- \notag \\
 \mathcal{R}_{22} &=  e^{-\lambda} - R \lambda' e^{-\lambda}  -1 + R e^{-\lambda} \left( \frac{\lambda' + \nu'}{2}  \right) = \xi_2 \notag \\
 \mathcal{R}_{33} &=  \sin^2(\vartheta) \left[ e^{-\lambda} - R \lambda'e^{-\lambda} - 1 + R e^{-\lambda}  \left(\frac{\nu' + \lambda'}{2} \right)  \right] = \xi_3 \qquad .
 \label{eq:einsteinpseudo}
\end{align}
Zunächst stellt man fest, dass die linke Seite der ersten drei Gleichungen nur eine Funktion von $R$ ist. Damit dürfen die ersten drei $\xi$ auch ausschließlich von $R$ (nach \refb{eq:funkaufteil} und \refb{eq:sigmaeig} sogar nur von $R_-$) abhängen. Subtraktion der beiden ersten Gleichungen gibt uns wieder einen Zusammenhang zwischen den Funktion $\nu$ und $\lambda$
\begin{equation}
	\nu' + \lambda' = \frac{1}{2} R (\xi_0 -\xi_1) \sigma_- = \frac{1}{2} R_- (\xi_0 - \xi_1) \sigma_- \qquad .
	\label{eq:nulambdaneu}
\end{equation}
Da wir für die Metrik $e^\nu$ und $e^\lambda$ benötigen, integrieren wir diese Gleichung noch und erhalten
\begin{align}
	e^{-\lambda} &= e^{\nu - \int{R_-(\xi_0-\xi_1)/2 \dd R_- \sigma_-}} \notag \\
	 &= e^{\nu_+} \sigma_+ + e^{\nu_- - \int{R_-(\xi_0-\xi_1)/2 \dd R_-}} \sigma_- \qquad .
	\label{eq:nulambdaneu2}
\end{align}
Für große Abstände müssen wir auch hier wieder den flachen Raum mit Minkowski-Metrik - $e^\nu \rightarrow1$ und $e^\lambda \rightarrow 1$ für $r  \rightarrow \infty$ - erhalten. Diese Bedingung liefert uns Einschränkungen an die $\xi$-Funktionen, denn auch der Unterschied zwischen $\nu$ und $\lambda$ muss für große Abstände beliebig klein werden. In \cite{Hess:2008wd} wurde deshalb vorgeschlagen, dass $\xi_0 = \xi_1$ gelten soll, dadurch also der Integrand in \refb{eq:nulambdaneu2} verschwindet und man an dieser Stelle die Lösung $\nu = - \lambda$ wie im Klassischen \refb{eq:nulambdaschwarzschild} erhält. Diese Annahme führt schlussendlich auf quadratische Korrekturen in der Metrik, welche wiederum durch experimentelle Daten sehr unwahrscheinlich sind (siehe dazu die Anmerkungen ab Gleichung \refb{eq:bppn}).
Nun führen wir aber erst die Lösung unter der Annahme, dass $\xi_0 = \xi_1$ und damit $\nu = -\lambda$ ist, fort. Die Gleichung für $\mathcal{R}_{22}$ liefert
\begin{equation}
 \left( R e^{-\lambda} \right)' - 1 = \xi_2 \sigma_-  \qquad .
 \label{eq:r22neu}
\end{equation}
Durch Integration bleibt dann
\begin{equation}
  e^{-\lambda} = 1 - \frac{R_S}{R} + \frac{1}{R} \int \xi_2 \sigma_- \dd R_- \qquad .
\label{eq:ehochminuslambda}
\end{equation}
Die Komponente dieser Gleichung proportional zu $\sigma_+$ ist identisch mit \refb{eq:r22klassisch}, wenn wir $R_S = R_S^+ \sigma_+ + R_S^- \sigma_-$ mit der pseudokomplexen Erweiterung des Schwarzschildradius identifizieren. Für die $\sigma_-$-Komponente bleibt dann
\begin{equation}
 e^{-\lambda_-} = 1 - \frac{R_S^-}{R_-} + \frac{1}{R_-} \int \xi_2 \dd R_- \qquad .
 \label{eq:lambdaneu}
\end{equation}
An dieser Stelle wenden wir uns wieder der Gleichung für $\mathcal{R}_{11}$ zu. Sie lässt sich wie im klassischen Fall \refb{eq:r11klassisch} umschreiben, nur dass hier die rechte Seite der Gleichung nicht verschwindet (genauer gesagt auch nur für die $\sigma_-$-Komponente, der $\sigma_+$-Teil der Gleichung bleibt gleich)
\begin{equation}
  \frac{e^{\lambda}}{R} \left( R e^{-\lambda} \right)'' = \xi_1 \sigma_- \qquad .
\end{equation}
Zusammen mit Gleichung \refb{eq:r22neu} ergibt das
\begin{equation}
 \xi_2  = R_- \xi_1 e^{-\lambda_-}
\end{equation}
und unter Ausnutzung von \refb{eq:lambdaneu} bleibt
\begin{equation}
 \xi_2' = \xi_1 \left( R_- - R_S^- + \int \xi_2 \dd R_- \right) \qquad .
\label{eq:xi1xi2}
\end{equation}
Wie bereits in der Standard-Schwarzschildlösung sind die Gleichungen für $\mathcal{R}_{22}$ und $\mathcal{R}_{33}$ bis auf einen Faktor $\sin^2\theta$ identisch. Daraus können wir direkt
\begin{equation}
 \xi_3 = \sin^2\theta \xi_2 
\end{equation}
folgern. Nun muss noch eine Verbindung zwischen $\xi_1$ und $\xi_2$ hergestellt werden, um \refb{eq:xi1xi2} lösen zu können. Wir greifen hier auf die Forderung, dass der Krümmungsskalar $\mathcal{R}$ verschwinden soll, zurück. Unter Berücksichtigung von \refb{eq:einsteinpseudo} erhält man 
\begin{equation}
 \mathcal{R} = - e^{-\lambda_-} \xi_1  - \frac{2}{R_-^2} \xi_2 = 0 \qquad \Leftrightarrow \qquad \xi_1 = - e^{\lambda_-} \frac{2}{R_-^2} \xi_2 \qquad .
\label{eq:kruemmungsskalarnull}
\end{equation}
Setzt man das zusammen mit \refb{eq:lambdaneu} in \refb{eq:xi1xi2} ein, erhält man eine Differentialgleichung für $\xi_2$
\begin{equation}
 \xi_2' = - \frac{2}{R_-} \xi_2
\end{equation}
welche sich direkt durch 
\begin{equation}
 \xi_2 = \frac{-B}{R_-^2}
\label{eq:xi2}
\end{equation}
lösen lässt. Die genaue Bestimmung der Integrationskonstanten $B$ verschieben wir zunächst. Um Konsistenz mit \cite{Hess:2008wd} zu wahren, führen wir noch eine neue Variable 
\begin{equation}
 \Omega := \int \xi_2 \dd R_-
\label{eq:omegadef}
\end{equation}
ein. Sie ist nur von $\sigma_-$ abhängig und wir können daher $\Omega = \Omega_-$ und $\Omega_+ = 0$ schreiben. Mit der Lösung für $\xi_2$ \refb{eq:xi2} sind wir jetzt in der Lage die modifizierte Schwarzschildlösung aufzustellen. Über \refb{eq:lambdaneu} und mit der Annahme, dass $\nu = -\lambda$ gilt, haben wir die $\sigma_-$-Komponente des metrischen Tensors
\begin{equation}
  g^{-}_{\mu\nu} = \begin{pmatrix} 1 - \frac{R^-_S}{R_-} + \frac{\Omega_-}{R_-} & 0 & 0  & 0 \\ 0 & - \dfrac{1}{1 - \frac{R^-_S}{R_-} +\frac{\Omega_-}{R_-}} & 0 & 0 \\ 0 & 0 & -R_-^2 & 0 \\ 0 & 0 & 0 & -R_-^2 \sin^2\theta
                          \end{pmatrix} \qquad .
\label{eq:gmunuminusschwarzschild}
\end{equation}
Die $\sigma_+$-Komponente bleibt aufgrund des unveränderten Variationsprinzips von der Form her identisch mit der klassischen Lösung \refb{eq:schwarzschildmetrik}, mit der Ausnahme, dass wir noch einen Term $\sim\Omega_+$ hinzugefügt haben, der aber verschwindet
\begin{equation}
 g^{+}_{\mu\nu} = \begin{pmatrix} 1 - \frac{R^+_S}{R_+} +  \frac{\Omega_+}{R_+}& 0 & 0  & 0 \\ 0 & - \dfrac{1}{1- \frac{R^+_S}{R_+} + \frac{\Omega_+}{R_+}} & 0 & 0 \\ 0 & 0 & -R_+^2 & 0 \\ 0 & 0 & 0 & -R_+^2 \sin^2\theta
                          \end{pmatrix} \qquad .
\end{equation}
Die gesamte pseudokomplexe Metrik lässt sich damit und mit \refb{eq:funkaufteil} als
\begin{equation}
  g_{\mu\nu} = \begin{pmatrix} 1 - \frac{R_S}{R} + \frac{\Omega}{R} & 0 & 0  & 0 \\ 0 & - \dfrac{1}{1 - \frac{R_S}{R} +\frac{\Omega}{R}} & 0 & 0 \\ 0 & 0 & -R^2 & 0 \\ 0 & 0 & 0 & -R^2 \sin^2\theta
                          \end{pmatrix} \qquad .
\end{equation}
schreiben. 
Jetzt ist es an der Zeit die im vorherigen Kapitel angesprochene Projektion auf die reelle Zahlenwelt vorzunehmen. Dazu müssen wir zunächst alle Größen im metrischen Tensor auf ihren Realteil projizieren. Für die radiale Koordinate $R$ und die Winkelvariable $\theta$ ist das direkt klar
\begin{align}
 R &\rightarrow \text{Re}(r + I l \dot{r}) = r \notag \\
\theta & \rightarrow \text{Re}(\vartheta + Il \dot{\vartheta}) = \vartheta \qquad . \notag
\end{align}
Die Größe $\Omega$ besteht per Konstruktion nur aus einem $\sigma_-$-Teil und damit ist
\begin{equation}
 \text{Re}(\Omega)  = \frac{1}{2} (\Omega_- + \Omega_+) = \frac{\Omega_-}{2} \qquad .
\end{equation}
Schließlich bleibt noch das pseudokomplexe Äquivalent des Schwarzschildradius' $R_S$. Um als Grenzfall die klassische Schwarzschildlösung zu erhalten, müssen wir hier
\begin{equation}
 \text{Re}(R_S) = r_S
\end{equation}
fordern. Letztendlich haben wir damit dann eine neue Schwarzschildmetrik
\begin{equation}
  g^{\text{Re}}_{\mu\nu} = \begin{pmatrix} 1 - \frac{r_S}{r} + \frac{\Omega_-}{2r} & 0 & 0  & 0 \\ 0 & - \dfrac{1}{1 - \frac{r_S}{r} +\frac{\Omega_-}{2 r}} & 0 & 0 \\ 0 & 0 & -r^2 & 0 \\ 0 & 0 & 0 & -r^2 \sin^2\vartheta
                          \end{pmatrix} \qquad .
\label{eq:schwarzschildmetrikpseudo}
\end{equation}
Die Schreibweise mit $\Omega_-$ statt $B$ verdeckt an dieser Stelle zwar, das der Korrekturterm zur Schwarzschildlösung hier quadratisch ist (mit \refb{eq:xi2} und \refb{eq:omegadef} ist $\Omega = \frac{B}{R_-}$ und damit $\frac{\Omega}{R_-} = \frac{B}{R_-^2}$),  ermöglicht uns aber diese Form später weiter zu verwenden, wenn die Korrekturterme nicht mehr quadratisch sind. 
\bigskip \\
Bis jetzt sind wir noch nicht näher auf die Bestimmung der Konstanten $B$ eingegangen. Zunächst müssen wir aus \refb{eq:schwarzschildmetrikpseudo} im Grenzfall für große Abstände wieder die klassische Schwarzschildlösung erhalten. Das ist allerdings kein Problem, da der Term proportional zu $B$ schnell genug mit dem Radius abfällt. \\
Die zweite Einschränkung, die wir beachten müssen, ist der Wertebereich des Linienelements $\dd s^2$. Es darf nicht negativ werden. Für die Metrik \refb{eq:schwarzschildmetrikpseudo} ergibt das 
\begin{equation}
 \dd s^2 = g_{00} \dd t^2 - \frac{1}{g_{00}} \dd r^2  -r^2 \dd \vartheta^2 - r^2\sin^2\vartheta \dd \varphi^2 \geq 0 \qquad .
\end{equation}
Solange $g_{00} \geq 0$ ist, ist diese Forderung immer erfüllt. Wird $g_{00}$ aber negativ, so wäre das Längenelement für ein ruhendes Testteilchen ebenfalls negativ, was nicht sein kann. In \cite{Hess:2008wd} wird deshalb $g_{00} > 0$ gefordert. Der Grenzfall $g_{00} = 0$ 
\begin{align}
 &1 - \frac{r_S}{r} + \frac{B}{2r^2} = 0 \notag \\
\Leftrightarrow & r^2 - r_S r + \frac{B}{2} = 0 \notag \\
\Rightarrow& r_{1/2} = \frac{r_S}{2} \pm  \sqrt{\frac{r_S^2}{4} - \frac{B}{2}} \qquad .
 \label{eq:einschraenkb}
\end{align}
liefert dann eine einschränkende Bedingung an $B$. Wenn die Gleichung \refb{eq:einschraenkb} zwei Lösungen aufweist, also $B < \frac{r_S^2}{2}$ ist, dann ist $g_{00}$ im Bereich zwischen $r_1$ und $r_2$ negativ. Für den Fall $B = \frac{r_S^2}{2}$ stellt die Kugel mit Radius $r = \frac{r_S}{2}$ eine Fläche unendlicher Rotverschiebung dar. %(EVTL BERÜNDEN WARUM).
 In \cite{Hess:2008wd} wird deshalb gefordert, dass
\begin{equation}
 B > \frac{r_S^2}{2}
\label{eq:bppn}
\end{equation}
sein muss. Durch einen Kommentar von Joachim Reinhardt wurden die Autoren allerdings darauf aufmerksam gemacht, dass eine solche Einschränkung an $B$ experimentellen Ergebnissen widerspricht. Mit Hilfe des Parametrised-Post-Newton-Formalismus' (PPN) lassen sich Abweichungen der Gravitation von der Newtonschen Theorie durch Entwicklung der Metrik beschreiben \cite{PPN:2006}. Dabei ist eine Obergrenze an die Konstante $B$ durch $ \frac{B}{r_S^2} = \beta -1  < 2,3 \cdot 10^{-4}$ festgelegt \cite{PPN:2006}. Der Parameter $\beta$ ist einer der Standard-Parameter des PPN-Formalismus'. In der Allgemeinen Relativitätstheorie nach Einstein entspricht er gerade 1. In unserem Fall lässt er sich relativ leicht durch die "`Robertson-Entwicklung"' bestimmen \cite{fliessbach:2006}. Sie stellt eine Entwicklung der Metrik-Koeffizienten $e^\nu$ und $e^\lambda$ in Potenzen von $\frac{r_S}{r}$ dar
\begin{align}
 e^\nu & = 1 - \frac{r_S}{r} + \frac{1}{2} (\beta -\gamma) \left( \frac{r_S}{r} \right)^2 + \cdots \notag \\
e^\lambda &= 1 +  \gamma  \frac{r_S}{r} + \cdots \qquad .
\label{eq:robertsonentwicklung}
\end{align}
Für die Metrik  \refb{eq:schwarzschildmetrikpseudo} ergibt das
\begin{align}
 e^\nu & = 1 - \frac{r_S}{r} + \frac{B}{2r^2} \notag \\
e^\lambda & = 1 + \frac{r_S}{r} + \cdots \qquad ,
\label{eq:repseudo}
\end{align}
wodurch der Zusammenhang $\gamma = 1$ und damit $\beta-1 = \frac{B}{r_S^2}$  hervortritt. Mit Hilfe der Robertson-Entwicklung berechnet \cite{fliessbach:2006} auch die Periheldrehung
\begin{equation}
 \Delta \phi = \frac{3 \pi r_S}{p} \frac{2 - \beta +2 \gamma}{3} \qquad .
\label{eq:periheldrehung}
\end{equation}
Der Parameter $p$ ist durch $\frac{2}{p} = \frac{1}{r_+} + \frac{1}{r_-}$ gegeben. $r_\pm$ bezeichnen den sonnennächsten Punkt (Perihel) und den sonnenfernsten Punkt (Aphel). Wenn die Parameter $\beta$ und $\gamma$ gleich 1 sind, wie in der Standard-ART, dann ergibt \refb{eq:periheldrehung} eine Periheldrehung von 43'' pro Jahrhundert für Merkur. Für den Fall, dass wir \refb{eq:bppn}  annehmen, erhalten wir mit \refb{eq:periheldrehung} einen Faktor von $5/6$ in der Periheldrehung im Vergleich zum Standardwert.  Das entspricht einer Abweichung von über 7'' pro Jahrhundert bei Merkur.

  % NOCH NACHRECHNEN GETAN! SIEHE DAZU DIE ROBERTSON-ENTWICKLUNG IM FLIESSBACH

\subsubsection*{Ansatz für die Schwarzschild-Metrik mit $\xi_0 \neq \xi_1$}

Betrachtet man das Gleichungssystem \refb{eq:einsteinpseudo}, so stellt man fest, dass es darin 6 unbekannte Größen gibt - die vier $\xi$-Funktionen und die beiden Funktionen $\nu$ und $\lambda$. Die letzten beiden Gleichungen des Systems liefern den Zusammenhang $\xi_3 = \sin^2\theta \xi_2$ zwischen zwei der $\xi$-Funktionen. Zur Lösung des gesamten Gleichungssystems wurden im letzten Abschnitt noch die beiden zusätzlichen Annahmen $\xi_0 = \xi_1$ \refb{eq:r22neu} und $\mathcal{R} = 0$ \refb{eq:kruemmungsskalarnull} verwendet, was uns letztendlich auf quadratische Korrekturen im Vergleich zur klassischen Schwarzschid-Lösung führte \refb{eq:schwarzschildmetrikpseudo}. Lassen wir nun diese beiden Annahmen fallen, können wir die Gleichung \refb{eq:nulambdaneu} zunächst nicht mehr weiter vereinfachen. Löst man \refb{eq:nulambdaneu} nach $\nu'$ auf und setzt in die dritte Gleichung aus \refb{eq:einsteinpseudo} ein, bleibt
\begin{align}
 & \xi_2 = e^{-\lambda_-} \left[ 1 + \frac{R_-}{2} \left(- 2 \lambda_-' + \frac{R_-}{2} (\xi_0 - \xi_1) \right) \right] - 1 \notag \\
\Leftrightarrow & \left( R_- e^{-\lambda_-} \right)' = 1 + \xi_2 - \frac{R_-^2}{4} e^{-\lambda_-} (\xi_0 - \xi_1)
\end{align}
übrig. Integriert ergibt die Gleichung das Pendant zu \refb{eq:lambdaneu}
\begin{equation}
 e^{-\lambda_-} = 1 - \frac{R_S^-}{R_-} + \frac{1}{R_-} \int \xi_2 \dd R_- - \frac{1}{4R_-} \int R_-^2 e^{-\lambda_-} (\xi_0 - \xi_1) \qquad .
\label{eq:ehochminuslambda2}
\end{equation}
An dieser Stelle muss man nun eine Möglichkeit finden, die $\xi$-Funktionen genauer zu bestimmen. Ein möglicher Ansatz dafür ist die Annahme, dass sich die $\xi$-Funktionen durch den Energie-Impuls-Tensor einer idealen Flüssigkeit repräsentieren lassen. Dafür berechnen wir den Krümmungsskalar $\mathcal{R}$ aus den Gleichungen \refb{eq:einsteinpseudo}
\begin{align}
 \mathcal{R} & = -e^{-\nu} \frac{e^{\nu -\lambda}}{2}  \xi_0 - \frac{e^{-\lambda}}{2} \xi_1 - \frac{1}{R^2}\xi_2 - \frac{1}{R^2\sin^2\theta} \xi_3\notag \\
&=  - \frac{e^{-\lambda}}{2} \left( \xi_0 + \xi_1 \right) -\frac{2}{R^2} \xi_2 \qquad .
\label{eq:kruemmskalarpseudo}
\end{align}
Damit können wir wieder in die Einsteingleichungen \refb{eq:einsteinmat} in gemischter Darstellung ${R^\mu}_\nu - \frac{1}{2} {g^\mu}_\nu = - 8 \pi \kappa {T^\mu}_\nu$ gehen
\begin{align}
 \frac{e^{-\lambda}}{4} \left( - \xi_0 + \xi_1 \right) + \frac{\xi_2}{R^2} &=  {\Xi^0}_0 = C {T^0}_0 \notag \\
\frac{e^{-\lambda}}{4} \left( \xi_0 - \xi_1 \right) + \frac{\xi_2}{R^2} &= {\Xi^1}_1 =  C {T^1}_1 \notag \\
\frac{e^{-\lambda}}{4} \left(\xi_0 + \xi_1 \right) & =  {\Xi^2}_2 = C {T^2}_2 \notag \\
\frac{e^{-\lambda}}{4} \left(\xi_0 + \xi_1 \right) & = {\Xi^3}_3 =  C {T^3}_3  \qquad ,
 \label{eq:einsteinmatpseudo}
\end{align}
wobei wir wieder die Konstante $C = - 8 \pi \kappa$ verwenden\footnote{ Die Symbole ${\Xi^\mu}_\nu$ werden erst im nächsten Kapitel wieder benutzt werden}. Aufgrund der sphärischen Symmetrie bringt die letzte Gleichung keine neue Information, denn hier gilt ${T^2}_2 = {T^3}_3$. Addition der ersten beiden Gleichungen liefert
\begin{equation}
 \frac{\xi_2}{R^2} = \frac{1}{2} C \left({T^0}_0 +{T^1}_1  \right) \qquad .
\end{equation}
Für die Summe aus der ersten und dritten Gleichung ergibt sich damit zusammen
\begin{equation}
 e^{-\lambda} \xi_1 = C ( 2 {T^2}_2 - {T^1}_1 + {T^0}_0 )
\end{equation}
und analog für die zweite und dritte Gleichung
\begin{equation}
 e^{-\lambda} \xi_0 = C ( 2 {T^2}_2 + {T^1}_1 - {T^0}_0 ) \qquad .
\end{equation}
Jetzt können wir den Energie-Impuls-Tensor für eine ideale Flüssigkeit (siehe \cite{adler:1975})
\begin{align}
 {T^\mu}_\nu = \begin{pmatrix}
                \rho & 0 & 0 &0 \\
		0 & - p & 0 &0 \\
		0 & 0 & -p & 0 \\
		0 & 0 & 0 & -p
               \end{pmatrix}
\label{eq:energieimpulsidealfluid}
\end{align}
einsetzen, womit  %(XI0 GLEICH XI1 LIEFERT HIER EINEN VERSCHWINDENDEN DRUCK P)
\begin{align}
   \frac{\xi_2}{R^2} &=  4\pi \kappa \left(p -\rho \right) \notag \\
 e^{-\lambda} \xi_1 &= 8\pi \kappa \left(p -\rho \right) \notag \\
 e^{-\lambda} \xi_0 &= 8\pi \kappa \left(3p +\rho \right)
\end{align}
bleibt. Das in \refb{eq:ehochminuslambda2} eingesetzt liefert 
\begin{equation}
 e^{-\lambda_-} = 1 - \frac{R^-_S}{R_-} - \frac{8\pi \kappa}{R_-}  \int R_-^2 \rho \dd R_- \qquad .
\end{equation}
Nun muss noch geklärt werden, über welche Grenzen man hier integriert. Eine Möglichkeit ist es von einem Beobachter im Unendlichen bis zum Radius $R_-$ zu integrieren. Vom Ursprung bis zum Radius $R$ zu integrieren ist an dieser Stelle problematisch, weil die Schwarzschildlösung nur im Außenraum, also materiefreien Raum, einer zentralen Masse gilt. Führt man die Integration nun aus, so bleibt
\begin{equation}
 e^{-\lambda_-} = 1 - \frac{R^-_S}{R_-} + \frac{2\kappa}{R_-}  M_{de}(R_-) \qquad .
\end{equation}
Der Vorzeichenwechsel kommt hier durch die Vertauschung der Integrationsgrenzen zustande. Bei der Größe $M_{de}(R_-)$ handelt es sich um eine Masse. In \cite{Hess:2008wd} hat sich durch den Korrekturterm in der Schwarzschildmetrik \refb{eq:schwarzschildmetrikpseudo} ein abstoßender Effekt für einfallende Materie ergeben. In Analogie dazu wurde der Massenterm $M_{de}(R_-)$ mit dem Index "`$de$"' (für "'dark energy"') versehen. Der Term gibt also die Masse der Dunklen Energie bis zum Radius $R_-$ an. Für Radien, die kleiner als der Schwarzschildradius sind, ist der zusätzliche Term für die Dunkle Energie bereits in der Masse des Objekts (hier repräsentiert durch den Term $R^-_S$) mit inbegriffen. Das heißt, dass jede Masse bereits einen Teil Dunkle Energie enthält, den wir automatisch mitmessen. \\
Falls die Annahme, dass hier eine ideale Flüssigkeit \refb{eq:energieimpulsidealfluid} vorliegt, gerechtfertigt ist und falls es sich hier um Dunkle Energie handelt, muss noch ihre Dichteverteilung gefunden werden. \refb{eq:nulambdaneu2} liefert dann auch $e^{\nu_-}$, wenn man zusätzlich einen Zusammenhang zwischen Dichte $\rho$ und Druck $p$ kennt.

% Paper vorstellen mit Schwarzschildlösung
% auf Probleme hinweisen 
% - projektion FERTIG
% - einschränkungen durch PPN FERTIG (ABER NOCHMAL NACHVOLLZIEHEN)
% -
% Änderungen, die wir uns überlegt hatten
% _ \xi_0 \neq \xi_1
% _
% _
% _ (Joker mit kleinem B) macht gunther
% - Literaturverweis neues Projektionsverfahren Greiner E-Dynamik Buch FERTIG
%%%%%%%%%%%%%%%%%%%%%%%%%%%%%%%%%%%%%%%%%%%%%%%%%%%%%%%%%%%%%%%%%%%%%%%%%%%%%%%%%%%%%%%%%%%%%%%%%%%%%%%%%%%%%%%%%%%%%%%%%%%%%%%%%%%%%%%%%%%%%%%
%\setcounter{footnote}{0}
\setcounter{equation}{0}

%
%################    EOF    ########################

% neuer befehl   \newcommand{\fracD}[2]{\frac{D #1}{D #2}} %  Differential mit großen D

\clearpage

\clearpage
%
% Kapitel mit eigenen Rechnungen und Ergebnissen
%
\section{Verallgemeinerung der Kerr-Metrik} \label{sec:ergebnisse}

\subsection{Kerr-Metrik als komplexe Koordinatenverschiebung der Schwarzschildmetrik} \label{sec:ergebnissekerr1}
In diesem Abschnitt werden wir die Möglichkeiten der Verallgemeinerung des Ansatzes aus Kapitel \ref{sec:kerrklassisch} für die Kerr-Metrik diskutieren. Wie in Kapitel \ref{sec:kerrklassisch} beschrieben, gründet diese Art der Herleitung der Kerr-Metrik auf der Eddington-Form der Schwarzschildmetrik \refb{eq:eddington}. Im letzten Kapitel haben wir die pseudokomplexe Verallgemeinerung der Schwarzschildmetrik vorgestellt \refb{eq:schwarzschildmetrikpseudo}. Allerdings haben wir auch festgestellt, dass eine Metrik dieser Form für quadratische Korrekturterme im Vergleich zur klassischen Schwarzschildmetrik unwahrscheinlich scheint. Die quadratischen Korrekturterme haben ihren Ursprung in den zwei Annahmen $\nu = - \lambda$ und $\mathcal{R}=0$. Daher gehen wir nun von der Form \refb{eq:schwarzschild} aus, um eine Verallgemeinerung für die Eddington-Form der Metrik zu bekommen. Als Ansatz  für die entartete Metrik \refb{eq:degmetric} können wir
\begin{align}
  \g = \eta_{\mu\nu} - C l_\mu l_\nu \notag \\
l_\mu  = \left( T(R), F(R) \vec{X} \right)
\label{eq:degmetric2}
\end{align}
wählen. Hierbei ist $C$ eine beliebige reelle Konstante und $T(R),F(R)$ zwei noch nicht weiter eingeschränkte Funktionen des Abstands $R$. Dieser Ansatz berücksichtigt alle in \refb{eq:schwarzschild} vorhanden Symmetrien und die Zeitunahängigkeit der Metrik. Das zu \refb{eq:eddingtoncart} analoge Linienelement hat dann die Form
\begin{align}
 \dd s^2 = (\dd \bar{X}^0)^2 - (\dd \vec{X})^2 - C & \left[ T^2(\dd \bar{X}^0)^2 + 2TF \dd \bar{X}^0 (X \dd X + Y \dd Y + Z \dd Z) \right. \notag \\
 & \left. +F^2 (X \dd X + Y \dd Y + Z \dd Y)^2 \right] \qquad ,
\end{align}
wobei $(\dd \vec{X})^2 = \dd X^2 + \dd Y^2 + \dd Z^2$ ist. In Kugelkoordinaten ist das Lienienelement dann
\begin{equation}
 \dd s^2 = (\dd \bar{X}^0)^2 - \dd R^2 - R^2 \left(\dd \theta^2 + \sin^2\theta \dd \phi^2 \right) - C \left[ T \dd \bar{X}^0 + r F \dd R \right]^2 \qquad .
\label{eq:linienelementtrafo}
\end{equation}
Um nun die Transformation zwischen $\bar{X}^0$ und $X^0$ zu finden, betrachten wir das Linienelement, welches man durch die Metrik \refb{eq:schwarzschild} erhält
\begin{equation}
 \dd s^2 = e^\nu (\dd X^0)^2 - e^\lambda \dd R^2 - R^2 \left(\dd \theta^2 + \sin^2\theta \dd \phi^2\right) \qquad .
\label{eq:linienelementschwarzschild}
\end{equation}
In Analogie zu \cite{adler:1975} setzen wir die Transformation als Funktion $X^0 = f(\bar{X}^0,R)$ an. Damit folgt
\begin{align}
 \dd X^0 &= \fracpd{f}{\bar{X}^0} \dd \bar{X}^0 + \fracpd{f}{R} \dd R \notag \\
 (\dd X^0)^2 &= \dot{f}^2 (\dd \bar{X}^0)^2 + f'^2 \dd R^2 + 2 \dot{f}f' \dd \bar{X}^0 \dd R
\end{align}
mit den offensichtlichen Abkürzungen $ \dot{f} := \fracpd{f}{\bar{X}^0}$ und $f' := \fracpd{f}{R}$. Setzt man das in \refb{eq:linienelementschwarzschild} ein und vergleicht mit \refb{eq:linienelementtrafo}, so bleiben drei Gleichungen
\begin{align}
 1 - CT^2 &= e^\nu \dot{f}^2 \notag \\
 1 + C R^2 F^2 &= e^\lambda - e^\nu f'^2 \notag \\
-2 C R T F &= 2 \dot{f} f' e^\nu \qquad .
 \label{eq:dreigleichungen}
\end{align}
Aus der dritten erhalten wir direkt
 \begin{align}
T= - \frac{\dot{f}f'e^\nu}{C R F}  \qquad ,
\label{eq:Tbestimmen}
 \end{align}
was wir in die erste einsetzen können
\begin{equation}
 1 - C \frac{\dot{f}^2f'^2 e^{2\nu}}{C^2 R^2 F^2} = e^\nu \dot{f}^2 \qquad .
\end{equation}
Multipliziert man das mit $CR^2F^2$ und sortiert die Terme um, so ergibt sich
\begin{equation}
 CR^2F^2 \left(1- e^\nu \dot{f}^2 \right) = \dot{f}^2f'^2 e^{2\nu} \qquad .
\label{eq:Fbestimmen}
\end{equation}
Hier kann man nun die zweite Gleichung aus \refb{eq:dreigleichungen} umformen und einsetzen
\begin{align}
 &\left(e^\lambda - e^\nu f'^2 -1 \right) \left(1 - e^\nu \dot{f}^2 \right) = \dot{f}^2 f'^2 e^{2\nu} \notag \\
\Leftrightarrow &  \left( e^\lambda -1  \right) \left(e^{-\nu} - \dot{f}^2 \right) = f'^2  \qquad .
\end{align}
Jetzt setzen wir nun in Analogie zur Transformation in \cite{adler:1975} $\dot{f} = 1 $. Solange wir nicht von $\nu = -\lambda$ ausgehen, ist allerdings nicht garantiert, dass die Gleichung
\begin{equation}
 f' = \pm \sqrt{ \left( e^\lambda -1  \right) \left(e^{-\nu} - 1 \right)}
\label{eq:lambdanuproblem}
\end{equation}
eine reelle Lösung besitzt. An dieser Stelle wollen wir aber eine reelle Koordinatentransformation zwischen $X^0$ und $\bar{X}^0$ haben. Unter der Annahme, dass $\xi_0 = \xi_1$ gilt (siehe Kapitel \ref{sec:pseudoart}), gilt aber auch $\nu = - \lambda$ und damit haben wir 
\begin{equation}
 f' = \pm \left( e^{-\nu} -1 \right) \qquad .
\end{equation}
Das Vorzeichen können wir dadurch festlegen, dass wir als Spezialfall die Transformation wie im Abschnitt \ref{sec:kerrklassisch} erhalten. Dazu muss das negative Vorzeichen gewählt werden. \\
Für die Metrik  \refb{eq:schwarzschildmetrikpseudo} erhält man dann
\begin{equation}
 f' = 1 - \frac{1}{1 - \dfrac{R_S}{R} + \dfrac{B}{R^2} } \qquad .
\label{eq:fstrichbestimmt}
\end{equation}
Diese Funktion lässt sich integrieren (siehe z.B. \cite{Bronstein:2008}), sodass man einen Ausdruck für die Transformation erhält
\begin{equation}
 \bar{X}^0 = \frac{(2B -R_S^2)}{\sqrt{4B - R_S^2}}\text{arctan}\left(\frac{2R - R_S}{\sqrt{4B - R_S^2}}\right) - \frac{R_S}{2} \ln \left(B - R_S R +R^2\right) + X^0 \qquad .
\end{equation}
Dieser Ausdruck gilt unter der Einschränkung $B > R_S^2/4$, was erfüllt ist, solange man \refb{eq:einschraenkb} annimmt. Ansonsten gilt \cite{Bronstein:2008}
\begin{equation}
 \bar{X}^0 = -\frac{(2B -R_S^2)}{\sqrt{R_S^2 - 4B}}\text{Artanh}\left(\frac{2R - R_S}{\sqrt{R_S^2 - 4B}}\right) - \frac{R_S}{2} \ln \left(B - R_S R +R^2\right) + X^0 \qquad .
\end{equation}
 Es bleibt nun noch die Bestimmung der Funktionen $T$ und $F$ und der Konstanten $C$. Hierfür ist es noch nicht einmal notwendig, die genaue Transformationsvorschrift zu kennen, es genügt zu wissen, dass sie existiert. \\
Die Gleichung \refb{eq:Fbestimmen} lässt sich nach $F^2$ umstellen, sodass wir 
\begin{equation}
 F^2 = \frac{1}{CR^2} \frac{1}{1-e^\nu} f'^2 e^{2\nu}
\end{equation}
haben. Setzt man nun $f'$ aus \refb{eq:fstrichbestimmt} ein und formt den enstehenden Term noch ein wenig um, so bleibt
\begin{equation}
 F = \pm \frac{1}{\sqrt{C}R^2}\sqrt{R_S R - B} \qquad .
\end{equation}
Zusammen mit \refb{eq:Tbestimmen} können wir auch $T$ bestimmen
\begin{equation}
 T = - \frac{f'e^\nu}{CRF} = \mp \frac{R}{\sqrt{C}} \frac{1}{\sqrt{R_S R -B}} \left(e^\nu -1\right) = \pm \frac{1}{\sqrt{C}R}\sqrt{R_S R - B} \qquad .
\end{equation}
Nun da $T$ und $F$ festgelegt sind, lässt sich ein Ausdruck für die Metrik \refb{eq:degmetric2} finden
\begin{equation}
 \g = \eta_{\mu\nu} - C l_\mu l_\nu = \eta_{\mu\nu} - L_\mu L_\nu \qquad ,
 \label{eq:degmetric3}
\end{equation}
wobei die 
\begin{equation}
 L_\mu = \sqrt{\frac{R_S}{R}- \frac{B}{R^2}} \left( 1, \frac{X}{R} , \frac{Y}{R} ,\frac{Z}{R}\right)
\label{eq:Lmu}
\end{equation}
 eingeführt wurden, um Konsistenz mit der Form für die entartete Metrik \refb{eq:degmetric} nach \cite{adler:1975} zu wahren. Die Form der $L_\mu$ stellt allerdings ein Problem dar. Zunächst stellen wir fest, dass in der Metrik ein Term quadratisch in $L_\mu$ vorkommt. Dieser Term hat in der Standard-Kerrlösung nach \cite{adler:1975} den Vorfaktor $\frac{r_S}{r}$. Hier ist der Vorfaktor durch $\frac{R_S}{R}- \frac{B}{R^2}$ gegeben. Der weitere Lösungsweg von Adler et al. \cite{adler:1975}  - ab Gleichung \refb{eq:degmetric} - beruht darauf, dass sich die Einsteingleichungen als Polynom in $r_S$ auffassen lassen \refb{eq:polynomeinstein}. Problematisch ist nun, dass bereits in den Termen der Ordnung $L_\mu^4$ der Vorfaktor die Form $	\frac{R_S^2}{R^2} + \frac{B}{R^4} - \frac{2 R_S B}{R^3} $ annimmt. Hier treten Mischterme zwischen $R_S$ und $B$ auf. Die Einsteingleichungen lassen sich so nicht mehr in der Form \refb{eq:polynomeinstein} auffassen. Durch den Zusatzterm $- \frac{B}{R^2}$ werden die Gleichungen deutlich komplizierter, so sind die Christoffelsymbole zweiter Art nun
\begin{align}
	\crs{\gamma}{\nu}{\mu} & = \frac{1}{2} g^{\sigma\gamma} [\nu \mu , \sigma] \notag \\
	&= \frac{1}{2} \left(\eta^{\sigma \gamma} - \frac{R_S}{R} \bar{L}^\sigma \bar{L}^\gamma + \frac{B}{R^2} \bar{L}^\sigma \bar{L}^\gamma \right) \cdot \left( R_S \left[ \frac{\nu \mu, \sigma}{R} \right] - B \left[ \frac{\nu \mu, \sigma}{R^2} \right] \right) \notag \\
	&= \frac{1}{2} \left[ \eta^{\sigma\gamma} \left(R_S \left[ \frac{\nu \mu, \sigma}{R} \right] -B \left[ \frac{\nu \mu, \sigma}{R^2} \right] \right) - \frac{R_S^2}{R} \bar{L}^\sigma \bar{L}^\gamma \left[ \frac{\nu \mu, \sigma}{R} \right] \right. \notag \\
	&~~~  \left. + \frac{R_S B}{R} \bar{L}^\sigma \bar{L}^\gamma \left[ \frac{\nu \mu, \sigma}{R^2} \right]	 + \frac{B R_S}{R^2} \bar{L}^\sigma \bar{L}^\gamma	\left[ \frac{\nu \mu, \sigma}{R} \right] - \frac{B^2}{R^2} \bar{L}^\sigma \bar{L}^\gamma \left[ \frac{\nu \mu, \sigma}{R^2} \right]  \right] \qquad ,
\end{align}
mit 
\begin{equation*}
	 \bar{L}^\sigma = \frac{1}{\sqrt{\frac{R_S}{R} - \frac{B}{R^2}}}   L^\sigma
\end{equation*}
 und
 \begin{equation*}
	\left[ \frac{\nu \mu, \sigma}{R} \right] := \left( \frac{\bar{L}_\nu \bar{L}_\sigma}{R} \right)_{|\mu} + \left( \frac{\bar{L}_\sigma \bar{L}_\mu}{R} \right)_{|\nu} -\left( \frac{\bar{L}_\mu \bar{L}_\nu}{R} \right)_{|\sigma} \qquad . 
	\end{equation*}
  Hinzu kommt, dass die rechte Seite in \refb{eq:polynomeinstein} nicht mehr Null ist, sondern Terme aus dem Nullteiler enthält.
 Dieses Problem, zusammen mit der Problematik für $\nu \neq -\lambda$ in Gleichung \refb{eq:lambdanuproblem}, legt nahe einen anderen Ansatz auf der Suche nach einer Erweiterung der Kerr-Lösung zu verwenden.

 %Allerdings geht die Konsistenz mit \refb{eq:degmetric} nicht weit genug, denn die so definierten $L_\mu$ sind durch den quadratischen Korrekturterm immer noch von $R_S$ abhängig. Es gibt keine Möglichkeit eine Konstante aus den $l_\mu$ ganz zu extrahieren, sodass diese nicht mehr davon abhängen. Das ist insofern problematisch, da die weiteren Rechnungen von Adler et al. \cite{adler:1975} - ab Gleichung \refb{eq:degmetric} - darauf beruhen, dass sich die Einsteingleichungen als Polynom in Potenzen der Konstanten $R_S$ schreiben lassen \refb{eq:polynomeinstein}.  %Die Konstante $C$, die Rolle von $R_S$ übernehmen könnte, kürzt sich aus den Termen in \refb{eq:degmetric3}.  \\

 % Diese Form muss bestehen bleiben, da Adler et al. eine Entwicklung in der Konstanten $R_S$ vornehmen (IRGENDWIE GEHT DAS JA NICHT MEHR SO, WENN MAN DAS RS IMMERNOCH IN DEN L DRIN STEHEN HAT. WEITERHIN MÜSSTEN GLEICHUNGEN 7.15 AUS DEM ADLER MODIFIZIERT WERDEN - DIE GLEICHUNGEN WERDEN DADURCH DEUTLICH KOMPLIZIERTER - ES IST ALSO WAHRSCHEINLICH EINFACHER DEN WEG MIT DEN DIFFFORMEN ZU VERFOLGEN). 
%Die so gefundenen $L_\mu$ in \refb{eq:Lmu} gehen für den Fall $B = 0$ in die $l_\mu$ aus \refb{eq:degmetric} über.\\
%Nun müssen wir analog zu den Rechnungen in Kapitel \ref{sec:kerrklassisch} eine Lösung der Gleichungen \refb{eq:laplaceeikonal} bzw. \refb{eq:loesungdeg} finden.

%%%%%%%%%%%%%%%%%%%%%%%%%%%%%%%%%%%%%%%%%%%%%%%%%%%%%%%%%%%%%%%%%%%%%%%%%%%%%%%%%%%%%%%%%%%%%%%%%%%%%%%%%%%%%%%%%%%%%%%%%%%%%%%%%%%%%%%%%%%%%%%
\clearpage
\subsection{Die Kerr-Metrik - Ansatz von Carter} 
Der Ansatz von Carter \cite{carter:1972,plebanski:2006} besteht darin, die Separierbarkeit der Klein-Gordon-Gleichung 
\begin{equation}
 \frac{1}{\Psi} \fracpd{}{x^\alpha} \left( \sqrt{-g}g^{\alpha\beta} \fracpd{\Psi}{x^\beta} \right) - m_0^2\sqrt{-g} = 0
\end{equation}
zu fordern\footnote{Carter verwendet die Signatur (-,+,+,+), Plebanski et al. und wir verwenden (+,-,-,-).}. Zunächst betrachtet er dafür die "`metric co-form"' $\left(\fracpd{}{s}\right)^2 = g^{\alpha\beta} \left(\fracpd{}{x^\alpha}\right) \left(\fracpd{}{x^\beta}\right)$, da man hier die einzelnen Terme der Klein-Gordon-Gleichung gut ablesen kann. Durch Einführung der Variablen $\mu := \cos\vartheta$ wird dann daraus für den Schwarzschildfall
\begin{equation}
 \left(\fracpd{}{s}\right)^2  = -\frac{1}{r^2} \left[ (1-\mu^2) \left(\fracpd{}{\mu}\right)^2  + \frac{1}{1-\mu^2} \left(\fracpd{}{\varphi}\right)^2 
+ \Delta_r \left(\fracpd{}{r}\right)^2  - \frac{Z_r^2}{\Delta_r} \left(\fracpd{}{t}\right)^2  \right] \label{eq:coform1} \qquad .
\end{equation}
Dabei sind bei Plebanski et al. \cite{plebanski:2006} noch die beiden Funktionen $\Delta_r :=  r^2 - r_sr$ und $Z_r := r^2$ festgelegt worden. Carter geht zunächst nicht näher auf die Form von $\Delta_r$ und $Z_r$ ein. Die einzige Einschränkung an die beiden Funktionen besteht darin, dass sie nur von $r$ abhängen dürfen. Die "`metric co-form"' kann somit auch folgendermaßen aussehen
\begin{equation}
  \left(\fracpd{}{s}\right)^2  = -\frac{1}{r^2} \left[ (1-\mu^2) \left(\fracpd{}{\mu}\right)^2  + \frac{1}{1-\mu^2} \left(\fracpd{}{\varphi}\right)^2 
 +e^{-\lambda}r^2 \left(\fracpd{}{r}\right)^2 - e^\nu r^2 \left(\fracpd{}{t}\right)^2  \right] \qquad .
\end{equation}
Also muss man hier für den Fall $\nu \neq -\lambda$ keine Einschränkung machen. Unter Verwendung von \refb{eq:coform1} resultiert dann die Klein-Gordon-Gleichung
\begin{eqnarray}
 \frac{r^2}{Z_r} \left[ - \frac{1}{\Psi} \fracpd{}{\mu} \left( (1-\mu^2) \fracpd{\Psi}{\mu} \right) - \frac{1}{\Psi (1-\mu^2)} \fracppd{\Psi}{\varphi} - m_0^2 r^2\right]&& \notag \\
 - \frac{1}{\Psi} \fracpd{}{r} \left( \frac{r^2 \Delta_r}{Z_r} \fracpd{\Psi}{r} \right) + \frac{1}{\Psi} \frac{r^2 Z_r}{\Delta_r} \fracpd{\Psi}{t} &=& 0 \qquad .
\end{eqnarray}
 Durch eine geeignete Substitution $\Psi = \prod_i \psi_i(x^i)$ ist diese Gleichung separierbar und zerfällt in vier partielle Differentialgleichungen für die jeweiligen Variablen. Hiervon ausgehend soll die "`mectric co-form"' erweitert werden, sodass auch die Kerr-Metrik als Spezialfall enthalten ist.
Gleichzeitig soll aber auch die Separierbarkeit der Klein-Gordon-Gleichung erhalten bleiben. Carter setzt nun für die "`metric co-form"' im Kerr-Fall folgendermaßen an
\begin{eqnarray}
 \left(\fracpd{}{s}\right)^2  &=& -\frac{1}{Z} \left[ \Delta_\mu \left( \fracpd{}{\mu} \right)^2 + \frac{1}{\Delta_\mu} \left( Z_\mu \fracpd{}{t} + Q_\mu \fracpd{}{\varphi}\right)^2 \right] \notag \\
 && - \frac{1}{Z} \left[ \Delta_r \left( \fracpd{}{r} \right)^2 - \frac{1}{\Delta_r} \left( Z_r \fracpd{}{t} + Q_r \fracpd{}{\varphi}\right)^2 \right]  \qquad .
  \label{eq:coformkerr}
\end{eqnarray}
Dieser Ansatz entspricht insofern der Kerr-Metrik, da nur gemischte Terme in $t$ und $\varphi$ auftauchen (siehe hierzu die Boyer-Lindquist-Form der Kerr-Metrik \refb{eq:boyerlindquist} ). Die Funktionen $\Delta_\mu,~ Q_\mu,~ Z_\mu$ und $\Delta_r,~Q_r,~Z_r$ hängen jeweils nur von der indizierten Variablen ab und die Funktion $Z$ ist hier noch nicht näher eingeschränkt. \refb{eq:coform1} geht also als Spezialfall für $\Delta_\mu = 1-\mu^2$, $Z=r^2$, $Q_r=0$, $Z_\mu = 0$ und $Q_\mu = 1$ aus \refb{eq:coformkerr} hervor. Als Determinante erhält Carter dann
\begin{equation}
	\sqrt{-g} = \frac{Z^2}{|Z_r Q_\mu - Z_\mu Q_r|} \qquad .
	\label{eq:detcarter}
\end{equation} 
Die Klein-Gordon-Gleichung hat für diesen Ansatz die Form \cite{plebanski:2006}
\begin{eqnarray}
	\frac{1}{\Psi} \left[ - \fracpd{}{\mu} \left(\frac{\sqrt{-g}}{Z} \Delta_\mu \fracpd{\Psi}{\mu}\right) - \fracpd{}{r} \left(\frac{\sqrt{-g}}{Z} \Delta_r \fracpd{\Psi}{r}\right) - \frac{\sqrt{-g}}{Z} m_0^2 Z \Psi\right] \notag \\
	- \frac{1}{\Psi} \left(Z_\mu \fracpd{}{t} + Q_\mu \fracpd{}{\varphi} \right) \left[ \frac{1}{\Delta_\mu} \frac{\sqrt{-g}}{Z} \left( Z_\mu \fracpd{\Psi}{t} + Q_\mu \fracpd{\Psi}{\varphi}\right) \right]  \notag \\ 
		+ \frac{1}{\Psi} \left(Z_r \fracpd{}{t} + Q_r \fracpd{}{\varphi} \right) \left[ \frac{1}{\Delta_r} \frac{\sqrt{-g}}{Z} \left( Z_r \fracpd{\Psi}{t} + Q_r \fracpd{\Psi}{\varphi}\right) \right] = 0 \qquad .
	\label{eq:kleingordonkerr}
\end{eqnarray}
Um die Separierbarkeit der Gleichung zu garantieren, muss der Faktor $\frac{\sqrt{-g}}{Z}$ ein Produkt einer von $r$ abhängenden Funktion mit einer nur von $\mu$ abhängenden Funktion sein. Plebanski et al. \cite{plebanski:2006} beschreiben die Möglichkeit durch eine Transformation der Variablen $r$ und $\mu$ und Umdefinition der Funktionen $\Delta_\mu$, $\Delta_r$, $Z_\mu$, $Z_r$, $Q_\mu$ und $Q_r$ den Faktor $\sqrt{-g}/Z = 1$ zu wählen. Mit \refb{eq:detcarter} entspricht das der Wahl
\begin{equation}
	Z = Z_r Q_\mu - Z_\mu Q_r \qquad .
	\label{eq:wahlvonz}
\end{equation}
Weiterhin muss $Z$ als Summe von Funktionen, die von $r$ respektive $\mu$ abhängen, darstellbar sein, sonst bereitet der Term proportional zu $m_0 Z \Psi$ Probleme bei der Separierbarkeit. Das ergibt die Bedingung
\begin{equation}
	\fracd{Z_r}{r} \fracd{Q_\mu}{\mu} -\fracd{Z_\mu}{\mu}\fracd{Q_r}{r} = 0 \qquad .
	\label{eq:bedingunganqundz}
\end{equation}
Es gibt nun verschiedene Möglichkeiten diese Bedingung zu erfüllen. Wir wählen an dieser Stelle wie \cite{carter:1972, plebanski:2006} die Konstanz der Funktionen $Q_r$ und $Q_\mu$ und nennen sie $C_r$ und $C_\mu$\footnote{Für weitere Diskussionen dieser Wahl siehe \cite{editorialnotetocarter}.}. 
% Um die Separierbarkeit der Gleichung zu garantieren wählt Carter an dieser Stelle den Faktor $\frac{\sqrt{-g}}{Z} = 1$ und legt damit $Z = Z_r Q_\mu - Z_\mu Q_r$ fest. Wir sind zunächst davon ausgegangen, dass man hier zusätzlich einen Faktor $e^{\frac{\nu + \lambda}{2}}$ - also $Z = e^{\frac{\nu + \lambda}{2}} (Z_r Q_\mu - Z_\mu Q_r) $  - bekommt, weil für den verallgemeinerten Schwarzschildfall $\sqrt{-g} = r^2 e^{\frac{\nu + \lambda}{2}}$ gilt. Daraus resultieren dann aber einige Probleme bei den weiteren Rechnungen. Weiterhin muss $Z$ als Summe von Funktionen, die von $r$ respektive $\mu$ abhängen darstellbar sein, sonst bereitet der Term proportional zu $m_0 Z \Psi$ Probleme bei der Separierbarkeit. Hier kann man nun $Q_\mu$ und $Q_r$ konstant wählen. Wenn man aber von  $Z = e^{\frac{\nu + \lambda}{2}} (Z_r Q_\mu - Z_\mu Q_r) $ ausgeht, muss man diese Forderung ein wenig verändern, sodass $Q_r$ nicht mehr konstant ist. Daraus ergeben sich in den späteren Rechnungen enorm aufwendige Terme (wir hatten die Christophelsymbole mit diesem Ansatz berechnet). Aktuell gehen wir aber davon aus, dass die Freiheiten in den verschiedenen Funktionen ausreichen um $\frac{\sqrt{-g}}{Z} = 1$ zu gewährleisten. Damit kann man wahrscheinlich die Rechnungen von Carter übernehmen und muss nur später in den Einsteingleichungen Quellterme hinzufügen. 
Nach diesen Überlegungen nimmt die Metrik die folgende Form an \\
\begin{align}
	\g &=& \frac{1}{Z} \begin{pmatrix}
	\Delta_r C_\mu^2 - \Delta_\mu C_r^2 & 0 & 0 & \Delta_\mu C_r Z_r - \Delta_r C_\mu Z_\mu \\
	0 & - \frac{Z^2}{\Delta_r} & 0 & 0 \\
	0& 0& - \frac{Z^2}{\Delta_\mu} & 0\\
	\Delta_\mu C_r Z_r - \Delta_r C_\mu Z_\mu & 0 & 0 & \Delta_r Z_\mu^2 - \Delta_\mu Z_r^2	\end{pmatrix} \qquad .
	\label{eq:ansatzcartermetrik}
\end{align} \\
Jetzt, da der Ansatz für die Metrik aufgestellt ist, ist es an der Zeit die Einsteingleichungen zu finden und wenn möglich zu lösen. Carter geht nun anders als Adler et al. \cite{adler:1975} vor, um die Einsteingleichungen aufzustellen. Er berechnet nicht die Christophelsymbole, sondern benutzt ein anderes Verfahren, welches im nächsten Kapitel kurz vorgestellt wird. 

%inverse Version der Carter-Metrik
%\begin{eqnarray}
%g^{\mu\nu} &=& \frac{1}{Z} \begin{pmatrix}
% \frac{Z_r^2}{\Delta_r} - \frac{Z_\mu^2}{\Delta_\mu} & 0 & 0 & \frac{Z_r Q_r}{\Delta_r} - \frac{Z_\mu C_\mu}{\Delta_\mu} \\
% 0 & - \Delta_r & 0 & 0 \\
% 0 & 0 & - \Delta_\mu & 0 \\
%\frac{Z_r Q_r}{\Delta_r} - \frac{Z_\mu C_\mu}{\Delta_\mu}  & 0 & 0 & \frac{Q_r^2}{\Delta_r} - \frac{C_\mu^2}{\Delta_\mu} 
%\end{pmatrix}
%\end{eqnarray}

%%%%%%%%%%%%%%%%%%%%%%%%%%%%%%%%%%%%%%%%%%%%%%%%%%%%%%%%%%%%%%%%%%%%%%%%%%%%%%%%
\subsubsection{Bestimmung der Einsteingleichungen mit Hilfe von Differentialformen} \label{sec:diffformmethode}
Die hier vorgestellte Methode zur Bestimmung der Einsteingleichungen geht auf Cartan \cite{cartan:51} zurück. Misner et al. \cite{misner:1973} erläutern diese Methode recht ausführlich in ihrem Buch. Eine beispielhafte Rechnung, auf die sich auch Carter \cite{carter:1972} bezieht, lässt sich im Anhang von \cite{misner:1963} finden. Eine mathematisch formale Ausführung findet man in \cite{ONeill:1995}. Wir werden hier die Methode kurz vorstellen und die Einsteingleichungen für die Schwarzschildmetrik ausführlich ableiten. Die Methode lässt sich grob in drei Schritte gliedern:
\begin{enumerate}
	\item Man wechselt in ein neues Koordinatensystem, das (mitbewegte) Vierbein oder auch Tetrade (englisch "`tetrad"') genannnt wird. %(STIMMT DAS? JA ABER MAN BETRACHTET AUS RECHENTEHNISCHEN GRÜNDEN DAS DAZU DUALE SYSTEM DER 1-FORMEN)
 Dieses System bringt den Vorteil mit sich, dass die Metrik hier Minkowski-Form annimmt und dadurch einige Rechnungen wesentlich vereinfacht werden können. Das Längenelement ist dann also
	\begin{equation}
	\dd s^2 = \g \dd x^\mu \dd x^\nu = \eta_{\mu\nu} \boldsymbol{\omega}^\mu \boldsymbol{\omega}^\nu \qquad .
	\label{eq:laengenelementtetrad}
	\end{equation}
	Wobei die $\boldsymbol{\omega}^\mu$ Differential- oder auch 1-Formen sind. 
	\item Nun bestimmt man die so genannten Verbindungs-Formen ${\boldsymbol{\omega}^\mu}_\nu$, welche eindeutig\footnote{für den Eindeutigkeitsbeweis siehe z.B. \cite{ONeill:1995}.} durch 	
\begin{eqnarray}
	&	0 = \bd \boldsymbol{\omega}^\mu + {\boldsymbol{\omega}^\mu}_\nu \wedge \boldsymbol{\omega}^\nu \notag \\
	& \bd \g = \boldsymbol{\omega}_{\mu\nu} + \boldsymbol{\omega}_{\nu\mu} 
	\label{eq:verbindungsformen}
\end{eqnarray}
 festgelegt sind. Das Symbol $\wedge$ bezeichnet das "`Dach-"', oder "`äußere"' Produkt mit der wichtigen Eigenschaft $\alpha \wedge \beta = - \beta \wedge \alpha$. Wie häufig in der Physik kann man aufgrund der Eindeutigkeit Lösungen durch Ansätze finden oder "`raten"'. Sobald sie \refb{eq:verbindungsformen} erfüllen, sind sie richtig. Wenn man sich aber mit der Methode vertraut macht, ist es empfehlenswert die Verbindungsformen zunächst direkt zu berechnen. Dies geschieht über
  \begin{align}
	 &\bd \bomega^\alpha = - {c_{|\mu\nu|}}^\alpha \bomega^\mu \wedge \bomega^\nu \notag\\ 
	 &\bomega_{\mu\nu} = \frac{1}{2} \left( c_{\mu\nu\alpha} + c_{\mu\alpha\nu} - c_{\nu\alpha\mu} \right) \bomega^\alpha \qquad ,
	\label{eq:cbestimmen}
	\end{align}
	wobei die hier verwendete Schreibweise eine Summation über $\mu$ und $\nu$ mit $\mu < \nu$ bedeutet \cite{misner:1973}. Bei den ${c_{\mu\nu}}^\alpha$ handelt es sich um so genannte "`Kommutations-Koeffizienten"'. Sie sind über $[\vec{e}_\mu,\vec{e}_\gamma]=: {c_{\mu\nu}}^\alpha \vec{e}_\alpha$ definiert \cite{misner:1973}, wobei die $\vec{e}_\mu$ Basisvektoren sind. Für normale Koordinatenbasisvektoren verschwinden sie, nicht aber für die Tetradenbasis. %%%%%%%%%%%%
  \item Hat man die (in 4 Dimensionen) sechs Verbindungsformen ${\boldsymbol{\omega}^\mu}_\nu$ gefunden, muss man noch die Krümmungsformen
  \begin{equation}
	 {\Theta^\mu}_\nu = \bd {\boldsymbol{\omega}^\mu}_\nu +{\boldsymbol{\omega}^\mu}_\alpha \wedge {\boldsymbol{\omega}^\alpha}_\nu
	\label{eq:kruemmform}
	\end{equation}
	bestimmen. Über den Zusammenhang\footnote{Die Gleichung \refb{eq:kruemmundriemann} findet sich in \cite{misner:1973} ohne Minuszeichen, was daran liegt, dass man bei der Definition des Riemann-Tensors eine Freiheit bei der Wahl des Vorzeichens hat. Dies wirkt in \cite{misner:1973} insofern aus, dass hier die Einsteingleichungen durch $G_{\mu\nu} = + 8 \pi \kappa T_{\mu\nu}$ gegeben sind im Gegensatz zu \refb{eq:einsteinmat}. } \cite{misner:1973}
	\begin{equation}
	 {\Theta^\mu}_\nu = - {\mathcal{R}^\mu}_{\nu|\alpha\beta|}~ \boldsymbol{\omega}^\alpha \wedge \boldsymbol{\omega}^\beta	
	\label{eq:kruemmundriemann}
	\end{equation}
	kann man den Riemann-Tensor ablesen. Durch Kontraktion erhält man den Ricci-Tensor $\mathcal{R}_{\mu\nu} = {\mathcal{R}^\alpha}_{\mu\alpha\nu}$. An dieser Stelle muss noch angemerkt werden, dass die so enthaltenen Tensoren noch in Tetrad-Komponenten aufgeschrieben sind. Um die Koordinaten-Komponenten zu bekommen muss man sie noch zurücktransformieren. Für die Rücktransformation stellt man die $\bomega^\mu$ in Abhängigkeit der Koordinatendifferentialformen $\bd x^a$ als $\bomega^\mu = {\lambda^\mu}_a \dd x^a$ dar. Die Koordinatenkomponenten des Ricci-Tensors erhält man dann über  \cite{carter:1972, feliceclarke:1990}
	\begin{equation}
	 \mathcal{R}_{ab} = \mathcal{R}_{\mu\nu} ~{\lambda^\mu}_a {\lambda^\nu}_b \qquad .
	  \label{eq:ruecktrafo}
	\end{equation}
\end{enumerate} 
\bigskip 
Neben dieser Vorgehensweise um die verschiedenen Krümmungsformen zu bestimmen brauchen wir noch Rechenregeln für die äußere Ableitung. So gelten für eine 0-Form $u$ und eine 1-Form $\boldsymbol{\sigma}$ die folgenden Regeln \cite{misner:1973, ONeill:1995} %(SIEHE MISNER THORNE WHEELER SEITE 350 OBEN, BZW ONEILL ANHANG B)
\begin{align}
	& \bd ^2 u = 0 \notag \\
	& \bd (u \boldsymbol{\sigma}) = \bd u \wedge \boldsymbol{\sigma} + u \bd \boldsymbol{\sigma} \qquad .
	\label{eq:regelnaussereableitung} 
\end{align}
Die erste Zeile in \refb{eq:regelnaussereableitung} ist eine andere Darstellung der in der Physik bekannten Regeln, dass die Divergenz einer Rotation und die Rotation eines Gradienten verschwindet. Dies kann man zum Beispiel durch die Korrespondenz zwischen der äußeren Ableitung und dem Gradienten, bzw. der Divergenz sehen. Für eine Funktion $f$ und einen Vektor $\vec{A}$ im dreidimensionalen euklidischen Raum gilt \cite{arnold:2000, misner:1973}
\begin{equation}
	\bd f = \bomega^1_{\boldsymbol{\text{\bf{grad}}}~f} \quad , \quad \bd \bomega^1_{\vec{A}} = \bomega^2_{\boldsymbol{\text{\bf{rot}}}~\vec{A}} \quad, \quad \bd \bomega^2_{\vec{A}} = (\text{div}~ \vec{A}) \bomega^3 \qquad ,
	\label{eq:aeusserableitung}
\end{equation}
wobei $\bomega^1_{\vec{A}}$ und $\bomega^2_{\vec{A}}$ die zu $\vec{A}$ korrespondierenden 1- beziehungsweise 2-Formen sind und $\bomega^1_{\boldsymbol{\text{grad}}~f}$ die zum Gradienten von $f$ korrespondierende 1-Form ist \cite{arnold:2000}. $\bomega^3$ ist das Volumenelement. Unter Berücksichtung von $\bd^2 = \bd \bd = 0$ folgen {\bf rot}({\bf grad})$=0$ und div({\bf rot})$=0$.
%%%%%%%%%%%%%%%%%%%%
\subsection{Die Einsteingleichungen für die Kerr-Metrik über Differentialformen} \label{sec:ergebnissekerr2}
Mit der im Abschnitt \ref{sec:diffformmethode} beschriebenen Vorgehensweise lassen sich aus dem Ansatz von Carter \refb{eq:ansatzcartermetrik} die Einsteingleichungen ableiten. Die Rechnungen dazu sind länglich und einige Zwischenergebnisse finden sich im Anhang ab Gleichung \refb{eq:tetradkerr}. Der hier aufgeführte Einstein-Tensor, auf den man schließlich kommt, geht auf Carter \cite{carter:1972} zurück und wurde in dieser Form von Plebanski et al. \cite{plebanski:2006} aufgestellt
\begin{align}
 {G^0}_{0} =~&  \frac{1}{2Z} \Delta_{\mu|\mu\mu}  + \frac{1}{Z^2} \Delta_r Z_{r|rr}  + \frac{a^2}{4Z^3} \Delta_\mu \left( {Z_{\mu|\mu}}^2 +{Z_{r|r}}^2 \right)  \notag \\
& - \frac{3}{4Z^3} \Delta_r \left( {Z_{\mu|\mu}}^2 +{Z_{r|r}}^2 \right)  + \frac{a}{2Z^2}\Delta_{\mu|\mu} Z_{\mu|\mu} + \frac{1}{2Z^2} \Delta_{r|r} Z_{r|r} \notag \\
{G^0}_{3} =~ & - \frac{1}{2Z^2} \sqrt{\Delta_r \Delta_\mu} \left(a Z_{r|rr} + Z_{\mu|\mu\mu} \right) \notag \\
{G^1}_{1} =~ &  \frac{1}{2Z} \Delta_{\mu|\mu\mu} + \frac{a^2}{4Z^3} \Delta_\mu \left( {Z_{\mu|\mu}}^2 + {Z_{r|r}}^2\right) - \frac{1}{4Z^3} \Delta_r \left( {Z_{\mu|\mu}}^2 + {Z_{r|r}}^2 \right)  \notag \\
 &  + \frac{a}{2Z^2}\Delta_{\mu|\mu} Z_{\mu|\mu}  + \frac{1}{2Z^2}\Delta_{r|r} Z_{r|r} \notag \\
{G^2}_{2} =~ &   \frac{1}{2Z} \Delta_{r|rr} - \frac{a^2}{4Z^3} \Delta_\mu \left( {Z_{\mu|\mu}}^2 + {Z_{r|r}}^2 \right) + \frac{1}{4Z^3} \Delta_r \left( {Z_{\mu|\mu}}^2 + {Z_{r|r}}^2 \right) \notag \\
 &~- \frac{a}{2Z^2} \Delta_{\mu|\mu} Z_{\mu|\mu} - \frac{1}{2Z^2} \Delta_{r|r} Z_{r|r}   \notag \\
{G^3}_{3} =~ &   \frac{1}{2Z} \Delta_{r|rr} - \frac{a}{Z^2} \Delta_\mu Z_{\mu|\mu\mu}  - \frac{3a^2}{4Z^3} \Delta_\mu \left( {Z_{\mu|\mu}}^2 + {Z_{r|r}}^2 \right) \notag \\
  &~ + \frac{1}{4Z^3} \Delta_r \left( {Z_{\mu|\mu}}^2 + {Z_{r|r}}^2 \right) - \frac{a}{2Z^2} \Delta_{\mu|\mu} Z_{\mu|\mu} - \frac{1}{2Z^2} \Delta_{r|r} Z_{r|r}  \qquad .
\label{eq:einsteintensorkerr}
\end{align}
Durch geeignete Variablentransformationen kann man die Konstanten $C_r = a$ und $C_\mu = 1$ wählen \cite{editorialnotetocarter}.\\ 
Die Kerr-Lösung nach Boyer und Lindquist \cite{boyerlindquist:1967} erhält man hier aus den freien Einsteingleichungen ${G^\mu}_\nu =0$. Zunächst merken \cite{plebanski:2006} an, dass sich die Gleichung für ${G^0_3}$ auf
\begin{equation*}
 a Z_{r|rr} + Z_{\mu|\mu\mu} = 0
\end{equation*}
reduziert, also eine Summe aus zwei Funktionen, die von unterschiedlichen Variablen abhängen. Diese Gleichung lässt sich direkt durch
\begin{equation}
 Z_r = C r^2 + C_1 r + C_2 \qquad , \qquad Z_\mu = -a C \mu^2 + C_3 \mu + C_4
\label{eq:zplebanski}
\end{equation}
lösen. Eingesetzt in ${G^2}_2 - {G^3}_3 = 0$ liefert das nach einigen Umformungen
\begin{equation}
 C_4 = \frac{C_2}{a} - \frac{C_1^2 + C_3^2}{4aC} \qquad .
\label{eq:c4pleb}
\end{equation}
Setzt man das in \refb{eq:zplebanski} ein, so stellt man fest, dass eine Variablentransformationen $\mu = \mu' + \frac{C_3}{2aC}$ und Umdefinition $C_2 = aC_2' + \frac{C_1^2}{4C}$ denselben Effekt haben, wie die Wahl $C_3=0$. Also bleibt
\begin{equation}
 Z_r = C \left( r + \frac{C_1}{2C} \right)^2 + a C_2' \qquad , \qquad Z_\mu = -a C\mu^2 + C_2' \qquad .
\end{equation}
Durch eine weitere Variablentransformationen, diesmal von $r$, lässt sich die Konstante $C_1 =0 $ setzen. Der Faktor $Z = Z_r - a Z_\mu$ hängt nicht von $C_2'$ ab, deshalb wählen \cite{plebanski:2006} $C_2' = aC$. Anschließend lässt sich der Faktor $C$ durch eine Umdefinition von $\varphi$, $\Delta_\mu$ und $\Delta_r$ absorbieren, was  gleichbedeutend mit der Wahl $C=1$ ist und auf
\begin{equation}
 Z_r = r^2 + a^2 \qquad , \qquad Z_\mu = a (1-\mu^2)
\label{eq:zplebanski2}
\end{equation}
führt \cite{plebanski:2006}. Als nächstes betrachten wir ${G^1}_1 + {G^2}_2 = 0$
\begin{equation}
 \frac{1}{2Z} \left( \Delta_{\mu|\mu\mu} + \Delta_{r|rr} \right) = 0 \qquad .
\label{eq:deltaplebansatz}
\end{equation}
Wieder liegt eine Summe aus zwei Funktionen unterschiedlicher Variablen vor. Die Funktionen müssen somit konstant sein und wir erhalten \cite{plebanski:2006}
\begin{equation}
 \Delta_r = E r^2 - r_S r + E_2 \qquad , \qquad \Delta_\mu = - E \mu^2 + E_3 \mu + E_4 \qquad .
\label{eq:deltapleb}
\end{equation}
Um die Konstanten näher zu bestimmen betrachten wir zunächst ${G^2}_2 = 0$. Durch Einsetzen von \refb{eq:zplebanski2} und \refb{eq:deltapleb} und Umformen bleibt
\begin{equation}
 E_4 a^2 - E_2 = 0 \qquad . 
\label{eq:e2e4}
\end{equation}
 Weiterhin fordern \cite{plebanski:2006} einige Einschränkungen an die verbleibenden Konstanten. So muss $E_3 = 0$ sein, denn sonst würde ein Term proportional zu $\mu = \cos \vartheta$ in der Metrik auftauchen. Dieser Term verletzt aber die Symmetrie in Hinsicht auf eine Spiegelung an der Äquatorebene. Um eine Koordinatensingularität an den Polen zu vermeiden fordern \cite{plebanski:2006} $E = 1$. Schließlich muss noch $E_4 = 1$ gesetzt werden, um die Schwarzschildlösung als Grenzfall erhalten zu können. Damit haben wir also
\begin{align}
& Z_r = r^2 + a^2 \quad , \quad Z_\mu = a(1-\mu^2) \quad , \notag \\
& \Delta_r = r^2 -r_S r + a^2 \quad , \quad \Delta_\mu = 1 - \mu^2 \quad , \notag \\
& Z = Z_r - aZ_\mu = r^2 + a^2 \mu^2 \quad .
\label{eq:loespleb}
\end{align}
Eingesetzt in \refb{eq:ansatzcartermetrik} ergibt sich nach einigen Umformungen gerade die Boyer-Lindquist-Darstellung der Kerr-Metrik \refb{eq:boyerlindquist}. \\
\bigskip \\
%%%%%%%%%%%%%%%%%%%%%%% 
Im Gegensatz zu den gerade gezeigten Rechnungen für die klassische Kerr-Lösung, setzen wir die Einsteingleichungen durch das erweiterte Variationsprinzip \refb{eq:varneu} nicht mehr Null, sondern gleich Elementen aus dem Nullteiler (siehe \refb{eq:einsteinpseudo} und \refb{eq:einsteinmatpseudo}). Auf der Suche nach einer Lösung für die Einsteingleichungen ${G^\mu}_\nu = {\Xi^\mu}_\nu$ mit \refb{eq:einsteintensorkerr} bietet es sich nun an, ähnliche Kombinationen wie in \cite{plebanski:2006} zu betrachten
\begin{eqnarray}
 {G^0}_3 : &- \frac{1}{2Z^2} \sqrt{\Delta_{R_-} \Delta_{\mu_-}} (a Z_{R_-|R_- R_-} + Z_{\mu_-|\mu_- \mu_-} ) &= {\Xi^0}_3 \notag \\
{G^1}_1 + {G^2}_2 : & \frac{1}{2Z} (\Delta_{\mu_-|\mu_- \mu_-} + \Delta_{R_-|R_- R_-}) &= {\Xi^1}_1 + {\Xi^2}_2 \notag \\
 {G^2}_2 - {G^3}_3 : & \frac{a}{Z^2} \Delta_{\mu_-} Z_{\mu_-|\mu_- \mu_-} + \frac{a^2}{2Z^3} \Delta_{\mu_-} ( Z_{\mu_-|\mu_- }^2 + Z_{R_-|R_- }^2 )& = {\Xi^2}_2 - {\Xi^3}_3 \notag \\
 {G^0}_0 - {G^1}_1 : & \frac{1}{Z^2} \Delta_{R_-} Z_{R_-|R_- R_-} - \frac{1}{2Z^3} \Delta_{R_-} ( Z_{\mu_-|\mu_- }^2 + Z_{R_-|R_- }^2 ) &= {\Xi^0}_0 - {\Xi^1}_1  \notag\\
 {G^2}_2 : & \frac{1}{2Z} \Delta_{R_-|R_- R_-} - \frac{a^2}{4Z^3} \Delta_{\mu_-} \left( {Z_{\mu_-|\mu_- }}^2 + {Z_{R_-|R_- }}^2 \right) - \frac{a}{2Z^2} \Delta_{\mu_-|\mu_- } Z_{\mu_-|\mu_- }  \notag \\
 &~+ \frac{1}{4Z^3} \Delta_{R_-} \left( {Z_{\mu_-|\mu_- }}^2 + {Z_{R_-|R_- }}^2 \right) - \frac{1}{2Z^2} \Delta_{R_-|R_- } Z_{R_-|R_- }   & = {\Xi^2}_2 \qquad . \notag \\
~\label{eq:systemkerr}
\end{eqnarray}
%\begin{eqnarray}
% {G^0}_3 : &- \frac{1}{2Z^2} \sqrt{\Delta_r \Delta_\mu} (a Z_{r|rr} + Z_{\mu|\mu\mu} ) &= {\Xi^0}_3 \notag \\
%{G^1}_1 + {G^2}_2 : & \frac{1}{2Z} (\Delta_{\mu|\mu\mu} + \Delta_{r|rr}) &= {\Xi^1}_1 + {\Xi^2}_2 \notag \\
% {G^2}_2 - {G^3}_3 : & \frac{a}{Z^2} \Delta_\mu Z_{\mu|\mu\mu} + \frac{a^2}{2Z^3} \Delta_\mu ( Z_{\mu|\mu}^2 + Z_{r|r}^2 )& = {\Xi^2}_2 - {\Xi^3}_3 \notag \\
% {G^0}_0 - {G^1}_1 : & \frac{1}{Z^2} \Delta_r Z_{r|rr} - \frac{1}{2Z^3} \Delta_r ( Z_{\mu|\mu}^2 + Z_{r|r}^2 ) &= {\Xi^0}_0 - {\Xi^1}_1  \notag\\
% {G^2}_2 : & \frac{1}{2Z} \Delta_{r|rr} - \frac{a^2}{4Z^3} \Delta_\mu \left( {Z_{\mu|\mu}}^2 + {Z_{r|r}}^2 \right) - \frac{a}{2Z^2} \Delta_{\mu|\mu} Z_{\mu|\mu}  \notag \\
% &~+ \frac{1}{4Z^3} \Delta_r \left( {Z_{\mu|\mu}}^2 + {Z_{r|r}}^2 \right) - \frac{1}{2Z^2} \Delta_{r|r} Z_{r|r}   & = {\Xi^2}_2 
%\label{eq:systemkerr}
%\end{eqnarray}
\bigskip \\
Der erste Ansatz zur Lösung dieses Systems bestand darin, die ${\Xi^\mu}_\nu$ aus \refb{eq:einsteinmatpseudo} mit Hilfe der Transformationsformeln \refb{eq:transformtetrad}, \refb{eq:transformtensor} in Tetradenkomponenten zu schreiben
\begin{align}
 {\Xi^0}_0 & = \frac{Z_{R_-}}{Z} {\Xi^t}_t - \frac{a_- Z_{\mu_-}}{Z} {\Xi^\varphi}_\varphi   \notag \\
 {\Xi^1}_1 & = {\Xi^r}_r  \notag \\
 {\Xi^2}_2 & = {\Xi^\mu}_\mu  \notag \\
 {\Xi^3}_3 & = \frac{ Z_{R_-}}{Z} {\Xi^\varphi}_\varphi -\frac{a_- Z_{\mu_-}}{Z} {\Xi^t}_t  \notag \\
 {\Xi^3}_0 & = \frac{a_- Z_{R_-}}{Z} \frac{\sqrt{\Delta_{\mu_-}}}{\sqrt{\Delta_{R_-}}} \left( {\Xi^t}_t - {\Xi^\varphi}_\varphi \right)  \qquad .
\end{align}
Die Tetradenindizes sind hier durch Ziffern und die Koordinatenindizes durch die jeweilige Variable gegeben. Anschließend wurden die ${\Xi^\mu}_\nu$ aus \refb{eq:einsteinmatpseudo} mit \refb{eq:xi2} und \refb{eq:kruemmungsskalarnull} bestimmt - also mit den Annahmen, dass $\xi_0 = \xi_1$ und $\mathcal{R} = 0$ für die Schwarzschildmetrik ist (siehe dazu Kapitel \ref{sec:pseudoschwarz}). Daraus resultieren die folgenden ${\Xi^\mu}_\nu$
\begin{align}
  {\Xi^0}_0 &= \frac{- Z_{R_-} - a_- Z_{\mu_-}}{Z} \frac{B}{R_-^4} \notag\\
  {\Xi^1}_1 &= - \frac{B}{{R_-}^4} \notag\\
  {\Xi^2}_2 &= \frac{B}{{R_-}^4} = - {\Xi^1}_1 \notag\\
  {\Xi^3}_3 &= \frac{ Z_{R_-} + a_- Z_{\mu_-}}{Z} \frac{B}{R_-^4} = - {\Xi^0}_0 \notag\\
  {\Xi^3}_0 &= - \frac{2 a_- Z_{R_-}}{Z} \frac{\sqrt{\Delta_{\mu_-}}}{\Delta_{R_-}} \frac{B}{R_-^4} \qquad .
  \label{eq:ximunu}
\end{align}
 An dieser Stelle lässt sich der Lösungsweg von Plebanski et al. \cite{plebanski:2006} leider nicht mehr weiter verwenden. Die erste Gleichung aus \refb{eq:systemkerr} wird mit \refb{eq:ximunu}
\begin{equation}
a_- Z_{{R_-}|{R_-}{R_-}} + Z_{{\mu_-}| {\mu_-} {\mu_-}}  = - 4 a_- B \frac{Z_{R_-}}{\sqrt{\Delta_{R_-}}^3 R_-^4 } \left( Z_{R_-} - a_- Z_{\mu_-} \right) \qquad .
\end{equation}
Es tritt hier auf der rechten Seite ein Produkt aus einer Funktion von ${R_-}$ und einer Funktion von ${\mu_-}$ auf. Die einzige Möglichkeit die Gleichung in dieser Form zu lösen, besteht in der Annahme, dass $Z_{\mu_-}$ eine Konstante ist. Diese Annahme würde aber dazu führen, dass man die klassische Kerr-Lösung, wie sie aus \refb{eq:loespleb} folgt, nicht mehr reproduzieren kann. \\
\bigskip \\
%%%%%%%%%%%%%%%%%%%%%%%%%%%%
Den zweiten Ansatz, um eine Lösung für \refb{eq:systemkerr} zu finden, bildet die Annahme, dass die fünf verschiedenen ${\Xi^\mu}_\nu$ beliebige Funktionen von $R_-$ und $\mu_-$ sind und zunächst kein Zusammenhang zu den $\xi_\mu$ der Schwarzschild-Lösung besteht. Durch geeignete Wahl dieser Funktionen lässt sich \refb{eq:systemkerr} derart lösen, dass man eine Verallgemeinerung der Metrik \refb{eq:gmunuminusschwarzschild} erhält. Die erste Annahme treffen wir durch ${\Xi^0}_3 = 0$. Damit sind 
\begin{equation}
 Z_{R_-} = C R_-^2 + C_1 R_- + C_2 \qquad , \qquad Z_{\mu_-} = -a_- C \mu_-^2 + C_3 \mu_- + C_4 \qquad ,
\end{equation}
wie in \refb{eq:zplebanski}. Mit der Wahl ${\Xi^2}_2  - {\Xi^3}_3  = 0 $ erhält man dann \refb{eq:c4pleb} und die daran anschließenden Umformungen können genauso übernommen werden, sodass wie in \refb{eq:zplebanski2}
\begin{equation}
 Z_{R_-} = R_-^2 + a_-^2 \qquad , \qquad Z_{\mu_-} = a_- (1-\mu_-^2)
\end{equation}
bleibt. Die vierte Gleichung in \refb{eq:systemkerr} liefert dann ${\Xi^0}_0 - {\Xi^1}_1 = 0$. Bisher sind die Ergebnisse identisch mit denen aus \cite{carter:1972, plebanski:2006}. Durch die Annahme ${\Xi^1}_1 + {\Xi^2}_2 = 0$ erhält man analog zu Gleichung \refb{eq:deltaplebansatz} 
\begin{equation}
 \Delta_{{R_-}|{R_-}{R_-}} + \Delta_{{\mu_-}|{\mu_-}{\mu_-}} = 0 \qquad , 
\end{equation}
was durch
\begin{equation}
 \Delta_{R_-} = E R_-^2 - R^-_S R_- + E_2  \qquad , \qquad \Delta_{\mu_-} = - E \mu_-^2 + E_3 \mu_- + E_4
\end{equation}
gelöst wird. Eingesetzt in die lezte Gleichung aus \refb{eq:systemkerr} bleibt 
\begin{equation}
 E_2 - a_-^2 E_4 = Z^2 {\Xi^2}_2 \qquad .
\end{equation}
Setzt man nun auch $ {\Xi^2}_2 = 0$, entspricht das Gleichung \refb{eq:e2e4}, was dann auf die klassische Kerr-Lösung führt. Die Wahl ${\Xi^2}_2 = \frac{B}{Z^2}$ ergibt eine Änderung dieser Lösung, die im Grenzfall $a \rightarrow 0$ in \refb{eq:gmunuminusschwarzschild} übergeht. Die Bestimmung der restlichen Konstanten erfolgt hier analog zu den Betrachtungen nach Gleichung \refb{eq:e2e4}. Schließlich haben wir damit
\begin{align}
& Z_{R_-} = R_-^2 + a_-^2 \quad , \quad Z_{\mu_-} = a_-(1-\mu_-^2) \quad , \notag \\
& \Delta_{R_-} = R_-^2 -R^-_S R_- + a_-^2 + B \quad , \quad \Delta_{\mu_-} = 1 - \mu_-^2 \quad , \notag \\
& Z = Z_{R_-} - a_- Z_{\mu_-} = R_-^2 + a_-^2 \mu_-^2 \quad .
\label{eq:loessystemkerr}
\end{align}
Eingesetzt in \refb{eq:ansatzcartermetrik} zusammen mit $\mu = \cos \theta$ haben wir damit den zu $\sigma_-$ proportionalen Teil der Metrik
\begin{align}
g^-_{00} &= \frac{ R_-^2 - R^-_S R_- + B + a_-^2 \cos^2 \theta_-}{R_-^2 + a_-^2 \cos^2\theta_-} \notag \\
g^-_{11} &= - \frac{R_-^2 + a_-^2 \cos^2 \theta_-}{R_-^2 - R^-_S R_- + a_-^2 + B} \notag \\
g^-_{22} &= - R_-^2 - a_-^2 \cos^2 \theta_-  \notag \\
g^-_{33} &= - (R_-^2 +a_-^2 )\sin^2 \theta_- - \frac{a_-^2 \sin^4\theta_-(R^-_S R_- - B)}{R_-^2 + a_-^2 \cos^2 \theta_-}  \notag \\
g^-_{03} &= \frac{-a_- \sin^2 \theta_- ~R^-_S R_- + a_- B \sin^2 \theta_- }{R_-^2 + a_-^2 \cos^2\theta_-}   \qquad .
\end{align}
Die $\sigma_+$-Komponenten entsprechen denen der klassischen Kerr-Lösung. Durch anschließende Projektion auf reelle Größen, wie in Kapitel \ref{sec:pseudoart} beschrieben, erhält man die Metrik
\begin{align}
 g^{\text{Re}}_{00} &= \frac{ r^2 - r_S r + \frac{B}{2} + a^2 \cos^2 \vartheta}{r^2 + a^2 \cos^2\vartheta} \notag \\
g^{\text{Re}}_{11} &= - \frac{r^2 + a^2 \cos^2 \vartheta}{r^2 - r_S r + a^2 + \frac{B}{2}} \notag \\
g^{\text{Re}}_{22} &= - r^2 - a^2 \cos^2 \vartheta  \notag \\
g^{\text{Re}}_{33} &= - (r^2 +a^2 )\sin^2 \vartheta - \frac{a^2 \sin^4\vartheta (r_S r - \frac{B}{2})}{r^2 + a^2 \cos^2 \vartheta}  \notag \\
g^{\text{Re}}_{03} &= \frac{-a \sin^2 \vartheta ~ r_S r + a \frac{B}{2} \sin^2 \vartheta }{r^2 + a^2 \cos^2\vartheta}   \qquad . 
\label{eq:kerrpseudo}
\end{align}
Im Grenzfall $a = 0 $ geht sie gerade in \refb{eq:schwarzschildmetrikpseudo} und für den Fall $B = 0$ in die klassische Kerr-Lösung \refb{eq:boyerlindquist} über. \bigskip \\
Es stellt sich nun noch die Frage, ob man den Parameter $a$ wie für die klassische Kerr-Metrik mit dem Drehimpuls der Zentralmasse in Verbindung bringen kann. Dazu gehen wir analog zu \cite{adler:1975} vor und entwicklen das Längenelement linear in $\frac{a}{r}$, womit
\begin{align}
 \dd s^2 = &\left( 1- \frac{r_S}{r} + \frac{B}{2 r^2} \right) \dd t^2 - \frac{1}{1- \frac{r_S}{r} + \frac{B}{2 r^2}} \dd r^2 - r^2 \dd \vartheta^2 - r^2 \sin^2\vartheta \dd \varphi^2 \notag \\
& +2 a \sin^2 \vartheta \left(- \frac{r_S}{r} +\frac{B}{2r^2} \right)
\end{align}
bleibt. Der nächste Schritt besteht darin, dieses Längenelement in isotrope Koordinaten umzuschreiben. Um sie zu erhalten, vergleicht man das Längenelement der Schwarzschildmetrik \refb{eq:schwarzschildmetrikpseudo} mit
\begin{equation}
\dd s^2 = \left(1 - \frac{r_S}{r} + \frac{B}{2r^2} \right) \dd \hat{t}^2 - \lambda^2(\rho) \left[ \dd \rho^2 + \rho^2 \left( \dd \hat{\vartheta}^2 + \sin^2 \hat{\vartheta} \dd \hat{\varphi}^2 \right) \right] \qquad , 
\label{eq:linienelemententwickelt}
\end{equation}
und erhält dadurch
\begin{equation}
 \lambda^2 \dd \rho^2 = \frac{\dd r^2}{1 - \frac{r_S}{r} + \frac{B}{2r^2}} \quad \text{und}  \quad \lambda^2= \frac{r^2}{\rho^2} \qquad .
\end{equation}
Zusammen ergibt sich eine Differentialgleichung für die Abstandsvariablen $\rho$ und $r$ analog zu \cite{adler:1975}
\begin{equation}
 \frac{\dd \rho}{\rho} = \frac{\dd r}{\sqrt{r^2 - r_S r + \frac{B}{2}}} \qquad .
\end{equation}
Die Lösung dieser Gleichung ist durch
\begin{equation}
 \sqrt{r^2 - r_S r +\frac{B}{2}} + r - \frac{r_s}{2} = 2 \rho
\end{equation}
gegeben. Nach einigen algebraischen Umformungen \cite{adler:1975} erhält man die Umkehrrelation
\begin{equation}
 r = \rho \left( 1+  \frac{r_S}{4 \rho}\right)^2 - \frac{B}{8 \rho} \qquad .
\end{equation}
Jetzt kann man \refb{eq:linienelemententwickelt} umschreiben. Für die folgenden Betrachtungen ist nur der Mischterm in $\varphi$ und $t$ relevant. Er lautet (im Weiteren wird auf die Unterscheidung zwischen $\vartheta$ und $\hat{\vartheta}$ u.ä. verzichtet)
\begin{equation}
 2 a \sin^2\vartheta \left( - \frac{\frac{r_S}{\rho}}{ \left( 1+ \frac{r_S}{4\rho} \right)^2 - \frac{B}{8\rho^2}}  + \frac{\frac{B}{2\rho^2}}{\left[ \left( 1+ \frac{r_S}{4\rho} \right)^2 - \frac{B}{8\rho^2} \right]^2} \right) \dd \varphi \dd t \qquad .
\end{equation}
Diesen Ausdruck entwickeln wir nun linear in $\frac{1}{\rho}$, wodurch
\begin{equation}
 -2 a \sin^2\vartheta \frac{r_S}{\rho} \dd \varphi \dd t
\end{equation}
übrig bleibt. Das können wir mit den Rechnungen von Lense und Thirring \cite{adler:1975, Lense:1918zz} 
\begin{equation}
 \dd s^2 = \left(1 - \frac{r_S}{\rho} \right) \dd t^2 - \left( 1+ \frac{r_S}{r} \right)\left[ \dd \rho^2 + \rho^2 \left( \dd \vartheta^2 + \sin^2 \vartheta \dd \varphi^2 \right)\right] + \frac{4 \kappa J}{c^3 \rho} \sin^2\vartheta \dd \varphi \dd t 
\end{equation}
vergleichen und damit den Parameter $a$ mit dem Drehimpuls $J$ durch
\begin{equation}
  a = - 2 \frac{\kappa J}{r_S c^3}
\end{equation}
identifizieren. Wie gerade gezeigt wurde, lässt sich der Parameter $a$ auch für die pseudokomplexe Erweiterung der Kerr-Metrik noch mit dem Drehimpuls in Verbindung bringen. \bigskip \\
Die klassische Kerr-Metrik weist einige besonders hervorgehobene Hyperflächen auf. Für die Schwarzschildlösung kennen wir schon die Oberfläche, welche durch den Schwarzschildradius festgelegt wird. Analog erhalten wir hier eine Fläche unendlicher Rotverschiebung, indem wir den Metrikkoeffizienten $g_{00} = 0$ setzen. Das ist gleichbedeutend mit
\begin{equation}
 r^2 -r_S r +a^2 \cos^2\vartheta + \frac{B}{2} = 0 \qquad ,
\end{equation}
was die Lösungen
\begin{equation}
 r_{0} = \frac{r_S}{2} - \sqrt{\frac{r_S^2}{4} - a^2 \cos^2\vartheta - \frac{B}{2}} \quad \text{und} \quad r_{\infty} = \frac{r_S}{2} + \sqrt{\frac{r_S^2}{4} - a^2 \cos^2\vartheta - \frac{B}{2}}
\label{eq:rinfty}
\end{equation}
 hat. Adler et al. \cite{adler:1975} zeigen dass die zu $r_\infty$ korrespondierende Fläche allerdings in beide Richtungen passierbar ist. Die zum Schwarzschildradius gehörende Fläche lässt sich nur in einfallender Richtung durchqueren. Diese Eigenschaft lässt sich anhand der Norm des zur Fläche gehörenden Normalenvektors $n_\alpha$ untersuchen. Ist die Norm kleiner als Null, so können physikalische Beobachter die Fläche in beide Richtungen durchqueren. Ist sie größer oder gleich Null, so ist die Fläche nur in eine Richtung durchlässig. Im Folgenden untersuchen wir die Flächen, deren Normalenvektor eine verschwindende Norm hat. Sie sollen axialsymmetrisch und zeitunabhängig sein und können so durch eine Funktion 
\begin{equation}
 u(r,\vartheta) = \const
\end{equation}
beschrieben werden \cite{adler:1975}. Der zugehörige Normalenvektor ist durch
\begin{equation}
 n_\alpha = \left(0 , \fracpd{u}{r} , \fracpd{u}{\vartheta}, 0 \right)
\end{equation}
gegeben \cite{adler:1975}. Wir setzen nun die Norm $n_\alpha n^\alpha$ dieses Vektors gleich Null und erhalten
\begin{equation}
\left(r^2 - r_S r + a^2 + \frac{B}{2} \right)  \left( \fracpd{u}{r}\right)^2 + \left( \fracpd{u}{\vartheta}\right)^2 = 0 \qquad .
\end{equation}
Diese Gleichung lässt sich durch einen Produktansatz $u = R(r) \Theta(\vartheta)$ lösen
\begin{equation}
- \left(r^2 - r_S r + a^2 + \frac{B}{2} \right)  \left( \frac{\fracpd{R}{r}}{R}\right)^2 = \left( \frac{\fracpd{\Theta}{\vartheta}}{\Theta}\right)^2 \qquad .
\end{equation}
Beide Seiten dieser Gleichung sind jeweils von einer anderen Variable abhängig und damit müssen sie gleich einer Konstanten sein, die wir hier in Analogie zu \cite{adler:1975} $\lambda$ nennen. Daraus folgt
\begin{equation}
 \Theta = A e^{\sqrt{\lambda} \vartheta}.
\end{equation}
Diese Lösung ist allerdings nicht zyklisch in $\vartheta$ und kann somit keine Fläche beschreiben, es sei denn man setzt $\lambda = 0$ und damit $\Theta = \const$. Dadurch verbleibt die Gleichung für $R$
\begin{equation}
 \left(r^2 - r_S r + a^2 + \frac{B}{2} \right)  \left( \frac{\fracpd{R}{r}}{R}\right)^2 = 0
\end{equation}
 Der Fall $\fracpd{R}{r} = 0$ ist nicht weiter interessant und deshalb betrachten wir die Lösung von $\left(r^2 - r_S r + a^2 + \frac{B}{2} \right) = 0$
\begin{equation}
 r_\pm = \frac{r_S}{2} \pm \sqrt{\frac{r_S^2}{4} - a^2 - \frac{B}{2}} \qquad .
\label{eq:rplusminus}
\end{equation}
Bis auf den Faktor $\frac{B}{2}$ stimmen \refb{eq:rinfty} und  \refb{eq:rplusminus}  mit den Ergebnissen aus \cite{adler:1975} überein. Wählen wir allerdings die Konstante $B$ wie in \cite{Hess:2008wd} vorgeschlagen (das entspricht \refb{eq:bppn}), werden die Wurzelterme in  \refb{eq:rinfty} und  \refb{eq:rplusminus} imaginär. Es treten dann also keine besonders ausgezeichneten Hyperflächen mehr auf, ganz analog zur in Kapitel \ref{sec:pseudoart} diskutierten Schwarzschildlösung.
\setcounter{equation}{0}

%
%################    EOF    #######################
%

%\boldsymbol{}  für omega usw
% äußere Ableitung
%\newcommand{\bd}{\textbf{d}}
% fettes omega
%\newcommand{\bomega}{\boldsymbol{\omega}}
\clearpage

\clearpage
%
% Fazit
%
\section{Fazit} 

Die zu Beginn der Arbeit formulierten Ziele wurden erreicht. So ist es mit \refb{eq:kerrpseudo} gelungen, wenn auch mit einigen vereinfachenden Annahmen, eine Verallgemeinerung der Kerr-Metrik mit Hilfe des pseudokomplexen Formalismus' zu finden. Diese Lösung entspricht der bereits in \cite{Hess:2008wd} veröffentlichten Metrik \refb{eq:schwarzschildmetrikpseudo} im Grenzfall $a = 0$. 
Sie ist allerdings mit einigen stark einschränkenden Annahmen aufgestellt worden und stellt die wahrscheinlich einfachste pseudokomplexe Verallgemeinerung der Kerr-Metrik dar. Auch ist die Form von \refb{eq:schwarzschildmetrikpseudo} - und damit auch \refb{eq:kerrpseudo} - durch die in Kapitel \ref{sec:pseudoart} diskutierten experimentellen Befunde äußerst fragwürdig geworden.\\
Bisher ist es jedoch noch nicht gelungen eine zu \refb{eq:schwarzschildmetrikpseudo} äquivalente Lösung zu finden, ohne die in \cite{Hess:2008wd} getroffenen Annahmen ( $\xi_0 = \xi_1$ und $\mathcal{R} = 0$ ) zu treffen. Der letzte Teil des dritten Kapitels gibt hier einen kurzen Abriss über die aktuellen Versuche eine solche Lösung zu finden. Zusammen mit der Komplexität der Gleichungen \refb{eq:systemkerr} stellt die Lösung \refb{eq:kerrpseudo} aber ein zufriedenstellendes Ergebnis dar. \\
Ziel zukünftiger Arbeiten kann nun zum einen die Bestimmung der Funktionen $\xi_\mu$ für die Schwarzschildmetrik sein, ohne die Annahmen $\xi_0 = \xi_1$ und $\mathcal{R} = 0$ zu treffen. Zum anderen müssen diese Ergebnisse dann herangezogen werden, um andere Lösungen von \refb{eq:systemkerr} zu finden, ohne die vereinfachenden Annahmen aus Kapitel \ref{sec:ergebnissekerr2} zu machen.\bigskip\\
Die Möglichkeit durch ein verändertes Variationsprinzip zusätzliche Terme in die Einsteingleichung einführen zu können, bringt sehr interessante Folgen mit sich. Einige davon konnten bereits durch experimentelle Daten widerlegt werden. Trotzdem bieten die verbleibenden Möglichkeiten genug Raum für ausführliche Forschungen. Dabei muss in Zukunft auch ein Auge auf mögliche Tests der neuen Theorien geworfen werden.

\setcounter{equation}{0}

%########## EOF##################
\clearpage

\clearpage

\pagestyle{empty}
\renewcommand\theequation{A.\arabic{equation}}  %Nummerierung der Gleichung in der Form:
																													%Gleichung
\renewcommand\thefootnote{\arabic{footnote}}  %Nummerierung der Fußnoten in der Form:
																															%Kapitel.Fußnote																													
%%%%%%%

\begin{center}
\huge Anhang
\end{center}

\section*{Ableitung der Einsteingleichungen mit Hilfe des Cartan-Formalismus'}
Hier finden sich die wichtigsten Zwischenschritte, um die Einsteingleichungen mit Hilfe des in Abschnitt \ref{sec:diffformmethode} vorgestellten Formalismus' aufzustellen. Zunächst am einfacheren Beispiel der Schwarzschildmetrik und anschließend für die Kerr-Metrik. 
\subsection*{Für die Schwarzschild-Metrik}
Am Anfang müssen wir die vier 1-Formen $\boldsymbol{\omega}^\mu$ des mitbewegten Systems, mit denen $\dd s^2 = \eta_{\mu\nu} \boldsymbol{\omega}^\mu \boldsymbol{\omega}^\nu$ gilt, finden. 
\begin{align}
 & \boldsymbol{\omega}^0 = e^{\nu/2} \bd t \notag \\
 & \boldsymbol{\omega}^1 = e^{\lambda/2} \bd r \notag \\
 & \boldsymbol{\omega}^2 = r \bd \vartheta \notag \\
 & \boldsymbol{\omega}^3 = r \sin \vartheta \bd \varphi \qquad .
\end{align}
Jetzt können wir die $\bd \bomega$ bestimmen
\begin{align}
 & \bd \bomega^0 = - \frac{\nu'}{2} e^{-\lambda/2} \bomega^0 \wedge \bomega^1 = - \frac{\nu'}{2} e^{(\nu-\lambda)/2} \bd t \wedge \bomega^1 \notag \\
 & \bd \bomega^1 = 0 \notag \\
 & \bd \bomega^2 = \frac{1}{r} e^{-\lambda/2} \bomega^1 \wedge \bomega^2 = - e^{-\lambda/2} \bd \vartheta \wedge \bomega^1 \notag \\
 & \bd \bomega^3 = -\sin \vartheta e^{-\lambda/2} \bd \varphi \wedge \bomega^1 - \cos \vartheta \bd \varphi \wedge \bomega^2 \qquad .
\end{align}
Daraus können wir die Verbindungsformen \refb{eq:verbindungsformen} ablesen
\begin{align}
 & {\bomega^0}_1 =  \frac{\nu'}{2} e^{(\nu-\lambda)/2} \bd t \notag \\
 & {\bomega^2}_1 =  e^{-\lambda/2} \bd \vartheta \notag \\
 & {\bomega^3}_1 =  \sin \vartheta e^{-\lambda/2}  \bd \varphi \notag \\
 & {\bomega^3}_2 =  \cos \vartheta \bd \varphi  \qquad .
 \label{eq:verbschwarzschild}
\end{align}
Die beiden nicht aufgeführten Verbindungsformen ${\bomega^0}_2$ und ${\bomega^0}_3$ verschwinden. \\ % An dieser Stelle merken Misner et al. \cite{misner:1973} an, dass man die Verbindungsformen nicht als Funktion der $\bomega^\mu$ schreiben sollte. Ansonsten fehlen beim nächsten Schritt Terme, bzw. es können welche hinzu kommen. Dies liegt an der unterschiedlichen Wirkung der äußeren Ableitung auf Funktionen, beziehungsweise 1-Formen. \\
Mit Bestimmung der Verbindungsformen ist der zweite Schritt aus \ref{sec:diffformmethode} abgeschlossen und man muss nun mit Hilfe von \refb{eq:kruemmform} die Krümmungsformen bestimmen. Dazu bilden wir zunächst die Ableitungen von \refb{eq:verbschwarzschild}
\begin{align}
 & \bd {\bomega^0}_1 = e^{-\lambda} \left( \frac{\nu''}{2} + \frac{\nu'}{2} \frac{\nu'-\lambda'}{2} \right) \bomega^1 \wedge \bomega^0 \notag \\
 & \bd {\bomega^2}_1 =  - \frac{\lambda'}{2r} e^{-\lambda} \bomega^1 \wedge \bomega^2 \notag \\
 & \bd {\bomega^3}_1 =  \frac{\cot \vartheta}{r^2} e^{-\lambda/2} \bomega^2 \wedge \bomega^3 - \frac{\lambda'}{2r} e^{-\lambda} \bomega^1 \wedge \bomega^3 \notag \\
 & \bd {\bomega^3}_2 = - \frac{1}{r^2} \bomega^2 \wedge \bomega^3 
 \label{eq:diffverbschwarzschild}
\end{align}
und dann schließlich die Krümmungsformen
\begin{align}
 & {\Theta^0}_1 = e^{-\lambda} \left( \frac{\nu''}{2} + \frac{\nu'}{2}\frac{\nu'-\lambda'}{2} \right) \bomega^1 \wedge \bomega^0 \qquad & {\Theta^1}_2 = \frac{\lambda'}{2r} e^{-\lambda} \bomega^1 \wedge \bomega^2 \notag \\
 & {\Theta^0}_2 = -\frac{\nu'}{2r} e^{-\lambda} \bomega^0 \wedge \bomega^2  \qquad & {\Theta^1}_3 = \frac{\lambda'}{2r} e^{-\lambda} \bomega^1 \wedge \bomega^3 \notag \\
 & {\Theta^0}_3 = -\frac{\nu'}{2r} e^{-\lambda} \bomega^0 \wedge \bomega^3  \qquad & {\Theta^2}_3 =  \frac{1}{r^2}\left( 1- e^{\lambda} \right) \bomega^2 \wedge \bomega^3  \qquad .
 \label{eq:kruemmformschwarzschild} 
\end{align}
Damit haben wir unser Ziel fast erreicht, denn man kann nun die Komponenten des Riemann-Tensors ablesen
\begin{align}
& {\mathcal{R}^0}_{101} =  e^{-\lambda} \left( \frac{\nu''}{2} + \frac{\nu'}{2}\frac{\nu'-\lambda'}{2} \right) \notag \\
& {\mathcal{R}^2}_{323} = - \frac{1}{r^2}\left( 1- e^{\lambda} \right) \notag \\
& {\mathcal{R}^1}_{313} = -\frac{\lambda'}{2r} e^{-\lambda}  \notag \\
& {\mathcal{R}^1}_{212} = -\frac{\lambda'}{2r} e^{-\lambda}  \notag \\
& {\mathcal{R}^0}_{202} =  \frac{\nu'}{2r} e^{-\lambda}\notag \\
& {\mathcal{R}^0}_{303} =  \frac{\nu'}{2r} e^{-\lambda}
\end{align}
und über Kontraktion erhält man den Ricci-Tensor
\begin{align}
& \mathcal{R}_{11} = {\mathcal{R}^0}_{101} + {\mathcal{R}^2}_{121} + {\mathcal{R}^3}_{131} = \frac{e^{-\lambda}}{2} \left( \nu''  + \frac{\nu'^2}{2} - \frac{\nu'\lambda'}{2} - \frac{2\lambda'}{r} \right) \notag \\
& \mathcal{R}_{00} = \frac{e^{-\lambda}}{2} \left( \nu''  + \frac{\nu'^2}{2} - \frac{\nu'\lambda'}{2} + \frac{2\nu'}{r} \right) \notag \\
& \mathcal{R}_{22} =  \frac{e^{-\lambda}}{2r} (\nu' -\lambda' ) + \frac{1}{r^2} (e^{-\lambda} -1) \notag \\
& \mathcal{R}_{33} = R_{22} \qquad .
\end{align}
Der Ricci-Tensor hier entspricht  dem Ricci-Tensor aus Kapitel \ref{sec:art} für die Schwarzschildmetrik, allerdings hier noch in Tetradenkomponenten. Durch die im dritten Punkt beschriebene Rücktransformation \refb{eq:ruecktrafo} erhält man gerade den Ricci-Tensor aus \refb{eq:schwarzschildricci}. 

%gemischter Darstellung $R_{\mu\nu}^{hier} = {R^\mu}_{\nu~ARTKapitel}$. (DAS LIEGT DARAN, DASS MAN HIER NOCH TETRADENKOMPONENTEN HAT.DIE RÜCKTRAFO BIS AUF EIN VORZEICHEN! IST KLAR). 
\subsection*{Für die Kerr-Metrik}
Aus dem Ansatz der Metrik \refb{eq:ansatzcartermetrik} nach Carter lässt sich direkt die Tetrade
\begin{align}
 \bomega^0 & = \sqrt{\frac{\Delta_r}{Z}} \left(C_\mu \bd t - Z_\mu \bd \varphi \right) \notag \\
 \bomega^1 & = \sqrt{\frac{Z}{\Delta_r}} \bd r \notag \\
 \bomega^2 & = \sqrt{\frac{Z}{\Delta_\mu}} \bd \mu \notag \\
 \bomega^3 & = \sqrt{\frac{\Delta_\mu}{Z}} \left( C_r \bd t - Z_r \bd \varphi \right) 
\label{eq:tetradkerr}
\end{align}
ablesen. Mit ihrer Hilfe lassen sich nun die Einsteingleichungen in Tetradenkomponenten bestimmen. Zunächst bestimmen wir die äußeren Ableitungen der Tetrade
\begin{align}
	 \bd\bomega^0 & = \left[ \frac{C_\mu}{2} Z_{r|r} \frac{\sqrt{\Delta_r}}{\sqrt{Z}^3} - \frac{\Delta_{r|r}}{2 \sqrt{\Delta_r} \sqrt{Z}} \right] \bomega^0 \wedge \bomega^1 \notag \\
	 & + \frac{C_r}{2} \frac{\sqrt{\Delta_\mu}}{\sqrt{Z}^3} Z_{\mu|\mu} \bomega^0 \wedge \bomega^2 + C_\mu \frac{\sqrt{\Delta_r}}{\sqrt{Z}^3} Z_{\mu|\mu} \bomega^2 \wedge \bomega^3 \notag \\
 \bd\bomega^1 & = \frac{C_r Z_{\mu|\mu}}{2} \frac{\sqrt{\Delta_\mu}}{\sqrt{Z}^3} \bomega^1 \wedge \bomega^2 \notag \\
 \bd\bomega^2 & = \frac{C_\mu Z_{r|r}}{2} \frac{\sqrt{\Delta_r}}{\sqrt{Z}^3} \bomega^1 \wedge \bomega^2 \notag \\
 \bd\bomega^3 & =  \left[  \frac{C_r}{2} Z_{\mu|\mu} \frac{\sqrt{\Delta_\mu}}{\sqrt{Z}^3} + \frac{\Delta_{\mu|\mu}}{2 \sqrt{\Delta_\mu}\sqrt{Z}} \right] \bomega^2 \wedge \bomega^3 \notag \\
 & + \frac{C_\mu}{2} \frac{\sqrt{\Delta_r}}{\sqrt{Z}^3} Z_{r|r} \bomega^1 \wedge \bomega^3 + C_r \frac{\sqrt{\Delta_\mu}}{\sqrt{Z}^3} Z_{r|r} \bomega^0 \wedge \bomega^1 \qquad .
\end{align}
Die Verbindungsformen bestimmt man nun mit den Gleichungen \refb{eq:cbestimmen} 
\begin{align}
 {\bomega^0}_1 &= \left( \frac{\Delta_{r|r}}{2 \sqrt{\Delta_r}\sqrt{Z}} - \frac{C_\mu}{2}  \frac{\sqrt{\Delta_r}}{\sqrt{Z}^3}Z_{r|r} \right) \bomega^0 + \frac{C_r}{2} \frac{\sqrt{\Delta_\mu}}{\sqrt{Z}^3} Z_{r|r} \bomega^3 \notag \\
 {\bomega^0}_2 &= - \frac{C_r}{2} \frac{\sqrt{\Delta_\mu}}{\sqrt{Z}^3} Z_{\mu|\mu} \bomega^0 + \frac{C_\mu}{2} \frac{\sqrt{\Delta_r}}{\sqrt{Z}^3} Z_{\mu|\mu} \bomega^3  \notag \\
 {\bomega^0}_3 &= \frac{C_r}{2} \frac{\sqrt{\Delta_\mu}}{\sqrt{Z}^3} Z_{r|r} \bomega^1 - \frac{C_\mu}{2} \frac{\sqrt{\Delta_r}}{\sqrt{Z}^3} Z_{\mu|\mu} \bomega^2	\notag \\
 {\bomega^1}_2 &= - \frac{C_r}{2}  \frac{\sqrt{\Delta_\mu}}{\sqrt{Z}^3} Z_{\mu|\mu} \bomega^1 - \frac{C_\mu}{2}  \frac{\sqrt{\Delta_r}}{\sqrt{Z}^3}Z_{r|r} \bomega^2  \notag \\
 {\bomega^1}_3 &= \frac{C_r}{2} \frac{\sqrt{\Delta_\mu}}{\sqrt{Z}^3} Z_{r|r} \bomega^0 - \frac{C_\mu}{2} \frac{\sqrt{\Delta_r}}{\sqrt{Z}^3} Z_{r|r} \bomega^3  \notag \\
 {\bomega^2}_3 &= \frac{C_\mu}{2} \frac{\sqrt{\Delta_r}}{\sqrt{Z}^3} Z_{\mu|\mu} \bomega^0 - \left( \frac{C_r}{2}  \frac{\sqrt{\Delta_\mu}}{\sqrt{Z}^3} Z_{\mu|\mu} + \frac{\Delta_{\mu|\mu}}{2 \sqrt{\Delta_\mu}\sqrt{Z}} \right) \bomega^3
\end{align}
Die Berechnung der Krümmungsformen \refb{eq:kruemmform} ist der wahrscheinlich aufwendigste Teil der Herleitung. Sie lauten
\begin{align}
 {\Theta^0}_1  = & \bomega^0 \wedge \bomega^1 \left[ - \frac{\Delta_{r|rr}}{2Z} + C_\mu \frac{\Delta_{r|r} Z_{r|r}}{Z^2} + \frac{C_\mu}{2} \frac{\Delta_r}{Z^2} Z_{r|rr} - C_\mu^2 \frac{\Delta_r}{Z^3} Z_{r|r}^2 \right. \notag \\
 & \qquad \qquad \qquad \left. - \frac{C_r^2}{4} \frac{\Delta_\mu}{Z^3} Z_{\mu|\mu}^2 + \frac{3C_r^2}{4} \frac{\Delta_\mu}{Z^3} Z_{r|r}^2  \right]	\notag \\ 
 & \bomega^1 \wedge \bomega^3 \left[ \frac{C_r}{2} \frac{\sqrt{\Delta_\mu \Delta_r}}{Z^2} Z_{r|rr} - \frac{C_r C_\mu}{4} \frac{\sqrt{\Delta_\mu \Delta_r}}{Z^3} \left( Z_{r|r}^2 + Z_{\mu|\mu}^2 \right)  \right] \notag \\
  & \bomega^2 \wedge \bomega^3 \left[ \frac{C_\mu}{2} \frac{\Delta_{r|r}}{Z^2} Z_{\mu|\mu} + \frac{C_r}{2} \frac{\Delta_{\mu|\mu}}{Z^2} Z_{r|r} + \left( C_r^2 \Delta_\mu - C_\mu^2 \Delta_r \right) \frac{Z_{\mu|\mu} Z_{r|r}}{Z^3} \right] \notag \\
 {\Theta^0}_2  = & \bomega^0 \wedge \bomega^2 \left[  \frac{C_r^2}{2} \frac{\Delta_\mu}{Z^3} Z_{\mu|\mu}^2 + \frac{C_r}{4} \frac{\Delta_{\mu|\mu}}{Z^2} Z_{\mu|\mu} + \frac{C_r}{2} \frac{\Delta_\mu}{Z^2} Z_{\mu|\mu\mu} \right. \notag \\
 &\qquad \qquad \left. - \frac{C_\mu}{4} \frac{\Delta_{r|r}}{Z^2} Z_{r|r} + \frac{C_\mu^2}{4} \frac{\Delta_r}{Z^3} \left( Z_{r|r}^2 -Z_{\mu|\mu}^2 \right) \right]	\notag \\
 & \bomega^1 \wedge \bomega^3 \left[ - \frac{C_\mu^2}{2} \frac{\Delta_r}{Z^3}Z_{\mu|\mu} Z_{r|r} + \frac{C_\mu}{4} \frac{\Delta_{r|r}}{Z^2} Z_{\mu|\mu} + \frac{C_r^2}{2} \frac{\Delta_\mu}{Z^3} Z_{\mu|\mu} Z_{r|r} \right. \notag \\
 &\qquad \qquad \left. + \frac{C_r}{4} \frac{\Delta_{\mu|\mu}}{Z^2} Z_{r|r}  \right] \notag \\
 & \bomega^2 \wedge \bomega^3 \left[ \frac{C_\mu C_r}{4} \frac{\sqrt{\Delta_\mu \Delta_r}}{Z^3} \left( Z_{\mu|\mu}^2 + Z_{r|r}^2 \right)  + \frac{C_\mu}{2} \frac{\sqrt{\Delta_\mu \Delta_r}}{Z^2} Z_{\mu|\mu\mu} \right] \notag \\
 {\Theta^0}_3  = & \bomega^1 \wedge \bomega^2 \left[ \left( \frac{C_\mu^2 \Delta_r - C_r^2 \Delta_\mu}{2} \right) \frac{Z_{r|r} Z_{\mu|\mu}}{Z^3} - \frac{C_r}{4} \frac{\Delta_{\mu|\mu}}{Z^2} Z_{r|r} - \frac{C_\mu}{4} \frac{\Delta_{r|r}}{Z^2} Z_{\mu|\mu} \right]	\notag \\
 & \bomega^0 \wedge \bomega^3 \left[ - \frac{C_\mu}{4} \frac{\Delta_{r|r}}{Z^2} Z_{r|r} + \frac{C_r}{4} \frac{\Delta_{\mu|\mu}}{Z^2} Z_{\mu|\mu} \right. \notag \\
 & \qquad \qquad \left. + \frac{C_\mu^2}{4} \frac{\Delta_r}{Z^3} \left( Z_{r|r}^2 - Z_{\mu|\mu}^2 \right)  + \frac{C_r^2}{4} \frac{\Delta_\mu}{Z^3} \left( Z_{\mu|\mu}^2 - Z_{r|r}^2 \right) \right] \notag 
 \end{align}
 \begin{align} 
 {\Theta^1}_2  = & \bomega^1 \wedge \bomega^2 \left[ \frac{C_r}{4} \frac{\Delta_{\mu|\mu}}{Z^2} Z_{\mu|\mu} + \frac{C_r}{2} \frac{\Delta_\mu}{Z^2} Z_{\mu|\mu\mu} + \frac{C_r^2}{2} \frac{\Delta_\mu}{Z^3} Z_{\mu|\mu}^2 \right. \notag \\
 & \qquad \qquad \left. - \frac{C_\mu}{4} \frac{\Delta_{r|r}}{Z^2} Z_{r|r} - \frac{C_\mu}{2} \frac{\Delta_r}{Z^2} Z_{r|rr} + \frac{C_\mu^2}{2} \frac{\Delta_r}{Z^3} Z_{r|r}^2 \right]	\notag \\
  & \bomega^0 \wedge \bomega^3 \left[ \frac{C_\mu}{4} \frac{\Delta_{r|r}}{Z^2} Z_{\mu|\mu} - \frac{C_\mu^2}{2} \frac{\Delta_r}{Z^3} Z_{\mu|\mu} Z_{r|r} \right. \notag \\ & \qquad \qquad \left. + \frac{C_r}{4} \frac{\Delta_{\mu|\mu}}{Z^2} Z_{r|r} + \frac{C_r^2}{2} \frac{\Delta_\mu}{Z^3} Z_{\mu|\mu} Z_{r|r} \right] \notag \\
 {\Theta^1}_3  = & \bomega^0 \wedge \bomega^1 \left[ \frac{C_r C_\mu}{4} \frac{\sqrt{\Delta_\mu \Delta_r}}{Z^3}\left( Z_{\mu|\mu}^2 + Z_{r|r}^2 \right) - \frac{C_r}{2} \frac{\sqrt{\Delta_\mu \Delta_r}}{Z^2} Z_{r|rr} \right]	\notag \\
 & \bomega^0 \wedge \bomega^2 \left[ \left(  \frac{C_\mu^2 \Delta_r - C_r^2 \Delta_\mu}{2} \right) \frac{Z_{\mu|\mu} Z_{r|r}}{Z^3} - \frac{C_r}{4} \frac{\Delta_{\mu|\mu}}{Z^2} Z_{r|r} - \frac{C_\mu}{4} \frac{\Delta_{r|r}}{Z^2} Z_{\mu|\mu} \right] \notag \\
 & \bomega^1 \wedge \bomega^3 \left[ \frac{C_\mu^2}{2} \frac{\Delta_r}{Z^3} Z_{r|r}^2 - \frac{C_\mu}{2} \frac{\Delta_r}{Z^2} Z_{r|rr} - \frac{C_\mu}{4} \frac{\Delta_{r|r}}{Z^2} Z_{r|r} \right. \notag \\
 & \qquad \qquad \left. + \frac{C_r^2}{4} \frac{\Delta_\mu}{Z^3} \left( Z_{\mu|\mu}^2 - Z_{r|r}^2 \right) + \frac{C_r}{4} \frac{\Delta_{\mu|\mu}}{Z^2} Z_{\mu|\mu} \right] \notag \\
 {\Theta^2}_3  = & \bomega^0 \wedge \bomega ^1 \left[ - \frac{C_\mu}{2} \frac{\Delta_{r|r}}{Z^2} Z_{\mu|\mu} + \left(C_\mu^2 \Delta_r - C_r^2 \Delta_\mu \right) \frac{Z_{\mu|\mu} Z_{r|r}}{Z^3} - \frac{C_r}{2} \frac{\Delta_{\mu|\mu}}{Z^2} Z_{r|r} \right] \notag \\	
 & \bomega^0 \wedge \bomega^2 \left[ \frac{C_\mu}{2} \frac{\sqrt{\Delta_\mu \Delta_r}}{Z^2} Z_{\mu|\mu\mu} - \frac{C_\mu C_r}{4} \frac{\sqrt{\Delta_\mu \Delta_r}}{Z^3} \left( Z_{\mu|\mu}^2 + Z_{r|r}^2 \right) \right] \notag \\
 & \bomega^2 \wedge \bomega^3 \left[ \frac{3C_\mu^2}{4} \frac{\Delta_r}{Z^3} Z_{\mu|\mu}^2 - \frac{C_\mu^2}{4} \frac{\Delta_r}{Z^3} Z_{r|r}^2 - C_r^2 \frac{\Delta_\mu}{Z^3} Z_{\mu|\mu}^2 \right. \notag \\ 
 & \qquad \qquad \left. -C_r \frac{\Delta_{\mu|\mu}}{Z^2} Z_{\mu|\mu} - \frac{C_r}{2} \frac{\Delta_\mu}{Z^2} Z_{\mu|\mu\mu} - \frac{\Delta_{\mu|\mu\mu}}{2Z} \right]
\end{align}

Mit dem Zusammenhang \refb{eq:kruemmundriemann} haben wir damit auch den Riemann-Tensor bestimmt. Den Einstein-Tensor erhält man über
\begin{align}
 {G^0}_0 &= - \left( {R^{12}}_{12} + {R^{23}}_{23} +{R^{13}}_{13} \right) \notag \\
 {G^1}_1 &= - \left( {R^{02}}_{02} + {R^{03}}_{03} +{R^{23}}_{23} \right) \notag \\
 {G^0}_1 &=  {R^{02}}_{12} + {R^{03}}_{13} \notag \\
 {G^1}_2 &=  {R^{10}}_{20} + {R^{13}}_{23}
\end{align}
und ähnliche Permutationen der Indizes \cite{misner:1973}. Daraus ergibt sich dann der Einstein-Tensor \refb{eq:einsteintensorkerr} von Plebanski et al. \cite{plebanski:2006}. \\
Neben dem Einstein-Tensor fehlt nun noch der entsprechende Part aus dem Nullteiler, um die Einsteingleichungen lösen zu können. Hierfür müssen die ${\Xi^\mu}_\nu$ aus \refb{eq:einsteinmatpseudo} in Tetradenkomponenten transformiert werden. Zunächst stellen wir dazu die Tetrade als Funktion der Koordinatendifferentiale und umgekehrt dar. Wir schreiben $\bomega^\mu = {\lambda^\mu}_a \bd x^a$ und $\bd x^a = {\lambda_\mu}^a \bomega^\mu$, mit
\begin{align}
 {\lambda^0}_t = \frac{\sqrt{\Delta_r}}{\sqrt{Z}} C_\mu  \qquad & \qquad {\lambda_0}^t = \frac{Z_r}{\sqrt{Z \Delta_r}}  \notag \\
 {\lambda^0}_\varphi = - \frac{\sqrt{\Delta_r}}{\sqrt{Z}} Z_\mu \qquad & \qquad {\lambda_3}^t = - \frac{Z_\mu}{\sqrt{Z \Delta_\mu}}  \notag \\
 {\lambda^1}_r = \frac{\sqrt{Z}}{\sqrt{\Delta_r}} \qquad & \qquad {\lambda_1}^r = \frac{\sqrt{\Delta_r}}{\sqrt{Z}}  \notag \\
 {\lambda^2}_\mu = \frac{\sqrt{Z}}{\sqrt{\Delta_\mu}} \qquad & \qquad {\lambda_2}^\mu = \frac{\sqrt{\Delta_\mu}}{\sqrt{Z}}  \notag \\
 {\lambda^3}_t = \frac{\sqrt{\Delta_\mu}}{Z} C_r \qquad & \qquad {\lambda_3}^\varphi =  - \frac{C_\mu}{\sqrt{Z \Delta_\mu}} \notag \\
 {\lambda^3}_\varphi = - \frac{\sqrt{\Delta_\mu}}{\sqrt{Z}} Z_r \qquad & \qquad {\lambda_0}^\varphi = \frac{C_r}{\sqrt{Z \Delta_r}} \qquad .
\label{eq:transformtetrad}
\end{align}
Die Tetradenkomponenten sind hier mit Ziffern und die Koordinatenkomponenten mit der jeweiligen Koordinate gekennzeichnet. Ein Tensor in gemischter Darstellung transformiert sich nun wie folgt \cite{feliceclarke:1990}
\begin{equation}
 {T^{\mu \nu \cdots}}_{\gamma \delta \cdots} = {T^{ij \cdots}}_{rs \cdots} {\lambda^\mu}_i {\lambda^\nu}_j\cdots {\lambda_\gamma}^r {\lambda_\delta}^s \cdots \qquad .
\label{eq:transformtensor}
\end{equation}

\clearpage
\section*{Satz von Stokes für pseudokomplexe Integrale}
Für eine Differentialform $\bomega$  lässt sich der Satz von Stokes\footnote{Dieser Satz tritt in verschiedenen Formen und Ausführungen auf und wird deshalb auch Satz von Newton-Leibniz-Gauss-Green-Ostrogradskii-Stokes-Poincaré genannt \cite{arnold:2000}.} wie folgt formulieren \cite{arnold:2000}
\begin{equation}
	\int_{\partial C} \bomega = \int_C \bd \bomega \qquad .
	\label{eq:satzvonstokes}
\end{equation}
$\partial C$ stellt hier den Rand eines beliebigen Teils $C$ des Raums, in dem die Integration stattfindet, dar. Um nun \refb{eq:pseudocauchy} zu beweisen, wählen wir die 1-Form $\bomega := F \bd x_1 + I F \bd x_2$. Damit können wir $\bd \bomega$ berechnen
\begin{align}
	\bd \bomega &= \bd F \wedge \bd x_1 + I \bd F \wedge \bd x_2 \notag \\
	&= \left( \fracpd{F}{x_1} \bd x_1 + \fracpd{F}{x_2} \bd x_2 \right) \wedge \bd x_1 + I \left( \fracpd{F}{x_1} \bd x_1 + \fracpd{F}{x_2} \bd x_2 \right) \wedge \bd x_2 \notag \\
	&= \fracpd{F}{x_2} \bd x_2 \wedge \bd x_1 + I \fracpd{F}{x_1} \bd x_1 \wedge \bd x_2 \notag \\
	&= \left(I \fracpd{F}{x_1} - \fracpd{F}{x_2}\right) \bd x_1 \wedge \bd x_2 \qquad .
	\label{eq:domegafuerstokes}
\end{align}
An dieser Stelle ist noch anzumerken, dass in der Physik häufig das $\wedge$ zwischen $\bd x_1$ und $\bd x_2$ weggelassen wird. Dafür muss man aber immer auf die Orientierung der von $C$ und $\partial C$ achten.
%%%%%%%%%%%%%%%%%%%%%%%%%%%%%%%%%%%%%%%%%%%%%%%%%%%%%%%%%%%%%%%%%%%%%%%%%%%%%%%%%%%%%%%%%%%%%%%%%%%%%%%%%%%%%%%%%%%%%%%%%%%%%%%%%%%%%%%%%%%%%%%%
\setcounter{footnote}{0}
\setcounter{equation}{0}
\clearpage
%\renewcommand{\baselinestretch}{1.5}\normalsize    %Zeilenabstand
%\clearpage\input{Kapitel/anhang.tex}\clearpage
%
\clearpage
%
% Danksagung
%
%evtl ohne Nummerierung

\thispagestyle{empty}

\section*{Danksagung}

Möglich geworden ist diese Arbeit durch Prof. Dr. Dr. h.c. mult. Walter Greiner, der den Kontakt zu Prof. Dr. Peter O. Hess hergestellt hat. Beide waren meinem Kommilitonen Gunther Caspar und mir während der Arbeit durch rege Diskussionen und ständigen Kontakt zur Hilfe. Trotz der großen räumlichen Entfernung zwischen Prof. Dr. Peter O. Hess und uns konnten wir via Skype und Email regelmäßig korrespondieren. Beide Betreuer waren immer bemüht unsere Probleme ernst zu nehmen. Ich hatte während der Ausarbeitung dieser Arbeit dadurch nie das Gefühl nur ein Arbeitsbeschaffungsprojekt durchzuführen, dessen einziges Ziel es ist einem Studenten einen Abschluss zu ermöglichen. Vielmehr konnte ich durch diese Arbeit direkt an Grundlagenforschung teilhaben. \\
Bei Prof. Dr. Dr. h.c. mult. Walter Greiner möchte ich mich auch für seine unkomplizierte Art bedanken, mit der er den neuen Bestimmungen des Masterstudiums begegnet ist. Dadurch war es mir möglich den organisatorischen Aufwand gering zu halten und mich auf den wissenschaftlichen Teil der Arbeit zu konzentrieren. Bei organisatorischen Fragen darf hier auch Laura Quist nicht vergessen werden. Als Institutssekrätärin war sie uns Studenten immer wieder eine große Hilfe.\bigskip \\
Eine große Hilfe in vielerlei Hinsicht war für mich auch die Zusammenarbeit mit Gunther Caspar. Es macht eindeutig mehr Spaß sich gemeinsam in ein Thema einzuarbeiten. Dies gilt insbesondere, wenn man wie in unserem Fall, auf einige Semester gemeinsames Studium zurückblicken kann. In unzähligen Gesprächen haben wir über verschiedenste - auch nichtphysikalische - Themen diskutiert. \bigskip \\
Vielen Dank an Mirko Schäfer für das Korrekturlesen der Arbeit.
\bigskip \\
Schließlich möchte ich mich noch bei meiner Familie und meiner Freundin für die Unterstützung während meiner Arbeit bedanken.

%
%################    EOF    #######################
\clearpage
%\listoffigures
%\listoftables

\renewcommand{\baselinestretch}{1.0}\normalsize    %Zeilenabstand
\bibliographystyle{alphadin}
\bibliography{Literatur}
%
%Erklärung der eigenständigen Arbeit
\clearpage
\renewcommand{\baselinestretch}{1.5}\normalsize    %Zeilenabstand
\begin{center}
 \Huge Erklärung
\end{center}
Ich versichere hiermit, dass ich diese Masterarbeit selbstständig verfasst und keine anderen als die angegebenen Quellen und Hilfsmittel benutzt habe. \\
Ich bin damit einverstanden, dass meine Masterarbeit öffentlich einsehbar ist und der wissenschaftlichen Forschung zur Verfügung steht.
\bigskip \\ \bigskip \\
Frankfurt am Main, den 23. März 2011 
\flushright Thomas Schönenbach

\end{document}